\documentclass[manuscript,screen,nonacm]{acmart}

\AtBeginDocument{%
  }

\acmJournal{CSUR}

\usepackage{multirow,multicol, makecell, booktabs}
\usepackage{longtable}
\usepackage{wasysym}
\usepackage{hyperref}
\usepackage{cleveref}
\usepackage{booktabs}

\usepackage{enumitem}
\newlist{myenum}{enumerate}{2} 
\setlist[myenum,1]{label=\arabic*), ref=\thesubsection.\arabic*}
\setlist[myenum,2]{label={\alph*.}, ref=\themyenumi.\alph*}

\usepackage{relsize}

\usepackage{array}
\newcolumntype{C}[1]{>{\centering\arraybackslash}p{#1}}
\newcolumntype{?}{!{\vrule width 1pt}}

\usepackage{graphicx}

\newcommand*\rot[1]{\rotatebox{90}{#1}}

\usepackage{tikz}
\usepackage{lipsum,adjustbox}
\usetikzlibrary{trees,calc,positioning}

\begin{document}

\title{A challenge-based survey of e-recruitment recommendation systems}

\author{Yoosof Mashayekhi}
\email{yoosof.mashayekhi@ugent.be}
\orcid{0000-0002-8993-0751}
\affiliation{%
  \institution{IDLAB - Department of Electronics and Information Systems (ELIS), Ghent University}
  \city{Ghent}
  \country{Belgium}
  \postcode{9000}
}

\author{Nan Li}
\email{nan.li@ugent.be}
\orcid{0000-0002-9963-7794}
\affiliation{%
  \institution{IDLAB - Department of Electronics and Information Systems (ELIS), Ghent University}
  \city{Gent}
  \country{Belgium}
  \postcode{9000}
}

\author{Bo Kang}
\email{bo.kang@ugent.be}
\orcid{0000-0002-9895-9927}
\affiliation{%
  \institution{IDLAB - Department of Electronics and Information Systems (ELIS), Ghent University}
  \city{Gent}
  \country{Belgium}
  \postcode{9000}
}

\author{Jefrey Lijffijt}
\email{jefrey.lijffijt@ugent.be}
\orcid{0000-0002-2930-5057}
\affiliation{%
  \institution{IDLAB - Department of Electronics and Information Systems (ELIS), Ghent University}
  \city{Gent}
  \country{Belgium}
  \postcode{9000}
}

\author{Tijl De Bie}
\email{tijl.debie@ugent.be}
\orcid{0000-0002-2692-7504}
\affiliation{%
  \institution{IDLAB - Department of Electronics and Information Systems (ELIS), Ghent University}
  \city{Gent}
  \country{Belgium}
  \postcode{9000}
}
\renewcommand{\shortauthors}{Mashayekhi et al.}

\begin{abstract}
E-recruitment recommendation systems recommend jobs to job seekers and job seekers to recruiters. The recommendations are generated based on the suitability of the job seekers for the positions as well as the job seekers' and the recruiters' preferences. Therefore, e-recruitment recommendation systems could greatly impact job seekers' careers. Moreover, by affecting the hiring processes of the companies, e-recruitment recommendation systems play an important role in shaping the companies' competitive edge in the market. Hence, the domain of e-recruitment recommendation deserves specific attention. Existing surveys on this topic tend to discuss past studies from the algorithmic perspective, e.g., by categorizing them into collaborative filtering, content based, and hybrid methods. This survey, instead, takes a complementary, challenge-based approach, which we believe might be more practical to developers facing a concrete e-recruitment design task with a specific set of challenges, as well as to researchers looking for impactful research projects in this domain. We first identify the main challenges in the e-recruitment recommendation research. Next, we discuss how those challenges have been studied in the literature. Finally, we provide future research directions that we consider promising in the e-recruitment recommendation domain.
\end{abstract}



\maketitle

\section{Introduction}\label{sec_intro}

With the ever-increasing use of the world wide web, many people seek jobs on e-recruitment platforms \cite{paparrizos2011machine}. These platforms, such as LinkedIn\footnote{\url{https://www.linkedin.com/}}, usually provide recommendations for job seekers to apply to several jobs and for recruiters to select suitable job seekers for their job positions \cite{88/kenthapadi2017personalized,530/geyik2018talent}.

The recommendation in e-recruitment is an important subfield of recommendation systems. Recommending the proper job seekers to recruiters could increase the efficiency of the hiring process, and recommending the right jobs to job seekers could have a positive impact on job seekers' career paths; on the other hand, low quality recommendations that poorly match job seekers with vacancies do not only cost time and effort of both recruiters and job seekers but also could have a negative impact on the labor market, companies' competitiveness, and people's lives in the long run. Hence, the domain of recommendation in e-recruitment requires specific attention.

In this study, we review the papers in the past decade about e-recruitment recommendation systems. Existing surveys \cite{7/deRuijt2021,259/freire2021recruitment} on e-recruitment recommendation systems usually focus on categorizing papers based on their methods such as collaborative filtering, content based, hybrid, etc. The range of challenges that these different methods address, on the other hand, has been less central to these prior surveys. Therefore, in this survey we focus on the challenges for e-recruitment recommendation systems and how those challenges have been studied in the literature.

We believe the challenge-based approach used in this survey is useful both for developers of e-recruitment recommendation systems and for researchers in the field. Indeed, developers will typically look for solutions to the practical challenges that naturally pose themselves in the design of their e-recruitment recommendation system, so in their design process the challenges will typically come before the possible algorithmic approaches. For researchers, our challenge-based approach may help in identifying the most impactful research problems of the domain and the proposed solution approaches to address them that have already been attempted. Moreover, open challenges and future research directions are also discussed to provide more insight for future research in this domain.

\textbf{Terminology}. Different entities could be recommended in e-recruitment recommendation systems. The e-recruitment recommendation systems could be categorized into three groups based on the \textbf{entities being recommended}: \textit{job recommendation}, \textit{job seeker recommendation}, and \textit{reciprocal recommendation}. In the rest of the paper, we use the term \textbf{e-recruitment recommendation} to refer to all recommendation systems in this research area.

Unless otherwise stated, the terms \textbf{user} and \textbf{item} can refer to job seekers, job positions, or recruiters, depending on the context: users receive the recommended lists, and items are the entities recommended to users. Throughout this paper, the terms job, job posting, job position, vacancy, and opening are used interchangeably to refer to a job vacancy. The terms recruiter or employer are also used interchangeably to refer to the person responsible for a job position. CVs and resumes denote the textual content of job seekers. We refer to all features and textual content of the users (job seekers or job postings) by the term user profile. Since different terms are used for the job/job seeker recommendation in the literature, we also use phrases such as matching job seekers with job positions (e.g., \cite{116/lavi2021consultantbert,132/zhao2021embedding}), person-job fit (e.g., \cite{478/qin2020enhanced,483/le2019towards}), and recommendation in e-recruitment (e.g., \cite{277/freire2020framework,332/chenni2015content}) to denote the same concept of recommendation in e-recruitment here.

\textbf{Contributions}. This survey will provide an overview of the literature in the past decade (from 2012 onwards) on e-recruitment recommendation systems. It contains the following contributions:
\begin{itemize}
\item Underscoring the importance of a survey on this topic, we list and discuss some important specific characteristics of e-recruitment recommendation systems that make it clear why they require a dedicated approach.
\item We identify and briefly discuss eight challenges that were frequently addressed by research papers covered in this survey, and where appropriate explain how they are the result of specific characteristics of e-recruitment recommendation systems.
\item For each of these challenges, we discuss the papers that have specifically targeted it, and we briefly discuss their approaches.
\item We provide future research directions and discuss the challenges that have been investigated less in recent years.
\item We present a structured overview of the collected 123 papers in Table~\ref{tab:all_papers} in the Appendix. The available properties of each paper in Table~\ref{tab:all_papers} are the recommendation type based on the recommended entities (job, job seeker, reciprocal), recommendation method type, and the challenges that the paper has addressed.
\item We maintain a website\footnote{\url{https://aida-ugent.github.io/e-recruitment-recsys-challenges/}} containing the content of Table~\ref{tab:all_papers} along with paper metadata (e.g. venue, url, authors, etc.) and summaries of the selected papers. We hope this can further facilitate the future research in e-recruitment recommendation systems.
\end{itemize}

For the rest of this section, we first discuss more in detail how our survey complements the existing surveys in Section \ref{sect:diff survey}. Next, we describe how the papers were collected and filtered in Section \ref{sect:methodology}. Finally, Section~\ref{sect:structure} presents the structure of this survey.

\subsection{Differences with previous recent surveys}\label{sect:diff survey}
The two recent surveys on e-recruitment recommendation systems \cite{7/deRuijt2021,259/freire2021recruitment} organized the literature differently from the present survey. 
The work by Freire and de Castro \cite{259/freire2021recruitment} focused on method types, data sources, and assessment methods. The work by de Ruijt and Bhulai \cite{7/deRuijt2021} gave an in-depth discussion about the e-recruitment recommendation system methods with a focus on categorizing hybrid and ensemble hybrid methods. Although de Ruijt and Bhulai \cite{7/deRuijt2021} explored some aspects and challenges of e-recruitment recommendation systems such as large scale, ethical, and reciprocal aspects, their discussion on those challenges and aspects is brief and limited.

Since the type of recommendation methods is well discussed in previous papers, this aspect is not the focus of the present study. Given the limitations of previous surveys, we focus on the challenges in e-recruitment recommendation systems and discuss the solutions that have been proposed for those challenges from a technical point of view. Our survey is valuable in that we emphasize the distinguishing nature of e-recruitment and organize the literature with respect to the special difficulties and challenges in e-recruitment recommendation.

\subsection{Literature search methodology}\label{sect:methodology}
We crawled data from dblp\footnote{\url{https://dblp.org/}} using ten keywords: \{`job recommender',
        `job recommendation',
        `job matching',
        `e-recruitment',
        `e-recruiting',
        `online recruitment',
        `person-job fit',
        `vacancy recommendation',
        `candidate recommendation', 
        `occupation recommendation'\} and as a result, 515 papers were collected.
We only kept papers published after (including) 2012, with at least five citations if published before (including) 2019. Papers that do not recommend actual jobs or job seekers (e.g., papers recommending a job type) were removed as well. This approach resulted in 99 papers in total. We further collected 24 papers from industry leaders and known experts from top conferences and journals. In total, 123 papers are kept for further examination.

\subsection{Structure of the survey}\label{sect:structure}
The structure of the rest of the paper is as follows. In Section \ref{sect:e-recruitment_overview}, we discuss the properties that distinguish e-recruitment recommendation systems from other recommendation systems. Section \ref{sect:challenges} contains our findings, in which Section \ref{sec_problem_challenges} gives a bird's eye view of all the challenges identified in this survey, Section \labelcref{sect:data quality,sect:hetero data,sect:cold start,sect:user preference,sect:interpretability,sect:specific obj,sect:bias,sect:large scale} address the different challenges respectively, and Section \ref{sect:remaining papers} briefly talks about the remaining papers not covered in the challenge sections. Finally, Section \ref{sec_conclusion} concludes our findings and discusses the limitations of this survey, open challenges and future directions.

\section{Specific characteristics and properties of e-recruitment recommendation systems}\label{sect:e-recruitment_overview}
In this section, we discuss the differences between e-recruitment and traditional recommendation systems. Although many challenges and characteristics are common between an e-recruitment recommendation system and a traditional one, such as e-commerce or a movie recommender, certain aspects set e-recruitment recommendation systems apart:

\begin{enumerate}

\item \textbf{One worker, one job (OWOJ)}\label{job_vs_trad_matching}: At a certain period of time, a person can only work at one or a few jobs, and also companies hire one or a few employees for a job posting \cite{541/borisyuk2017lijar}. Moreover, job seekers and job positions are mostly available for a limited time and become inactive after they are employed or filled. In contrast, in a traditional recommender, the same items can be recommended to many users, and users consume several items. The e-recruitment recommendation systems have to consider this aspect in the recommendation. First, the number of recommendations for each job/job seeker may have to be kept relatively small since only one or a few of them can succeed. Moreover, job seekers/jobs usually compete with each other for the same jobs/job seekers. Hence, the recommendation of a job at which others have a higher chance of success could be less interesting. This competition aspect should ideally be taken into consideration in generating the recommendations.

\item \textbf{Two-sided (TS)}\label{job_vs_trad_tow_sided}: In traditional recommendation systems, the success of a recommendation usually depends on the action of one user. For example, in e-commerce a recommendation is successful if the user decides to buy a product. However, in e-recruitment recommendation systems, the ultimate success of a recommendation depends on whether it results in employment. The actions by one user, such as applying for a job position by a job seeker could only show the interest of the job seeker to the job position, while the success of the recommendation also depends on the recruiter of the job posting who makes an offer for the job. Hence, e-recruitment recommendation systems have multiple stakeholders (e.g., job seekers and employers).

\item \textbf{Suitability as well as preference (SP)}\label{job_vs_trad_suitability}: While users' preferences play an important role in all recommendation systems, e-recruitment recommendation systems recommend jobs/job seekers based on suitability and skills as well \cite{3/GuptaG14}. One way to define suitability and user preference is as follows. \emph{Suitability} represents the degree of matchness between a job seeker and job position based on typically but not exclusively knowledge, skills, diplomas, and years of experience of the job seekers and the job position requirements. User \emph{preference}, on the other hand, represents one's inclination towards certain items. For example, a job seeker might be suitable for several positions, but prefer to work for a specific company for various reasons such as higher salary, social connections, etc. In addition, a recruiter often has to pick one job seeker among multiple equally suitable job seekers based on the preferences such as social connections, personality, etc. Hence, the suitability of a job seeker for a job and their preferences will in general not be equal, which poses specific challenges to e-recruitment recommendation systems.

\item \textbf{Multi-faceted (MF)}\label{job_vs_trad_multi-faceted}: In e-recruitment recommendation systems, both suitability and preference are, in fact, dependent on many different facets with different data types. For a job seeker, their previous job history, diplomas, seniority, interests, skills, location, social fit to the job environment, etc. could be relevant for an e-recruitment recommendation system. For a job posting, its required skills, required diplomas, seniority, location, organizational culture, etc. might be available and could be used in an e-recruitment recommendation system. Hence, the nature of data available in the e-recruitment domain is usually multi-faceted and requires specific attention in designing e-recruitment recommendation systems.

\item \textbf{High-stakes (HS)}\label{job_vs_trad_high_stakes}: E-recruitment is a high risk domain because it can have a long-term impact on people's careers and hence, their career fulfillment. Moreover, it plays an important role in shaping the companies' competitive edge in the market. E-recruitment is even defined as one of the high-risk domains according to the EU's AI act (proposal) \cite{eu-aiact-2022}. Hence, considering fairness and trustworthiness aspects is more essential in e-recruitment recommendation systems compared to the traditional ones.



\item \textbf{Short interaction history (SIH)}\label{job_vs_trad_short_interaction_history}: In e-recruitment recommendation systems, job seekers only interact with the system while they are seeking a new job, and they will probably stop using it after they are employed. Moreover, new job positions appear and disappear frequently \cite{535/guo2017howinte}. In contrast, in a traditional recommendation system users and items often have a long history within the system.


\end{enumerate}

\section{Survey structured according to challenges faced in the development of e-recruitment recommendation systems}\label{sect:challenges}
In this survey, we identify some challenges in e-recruitment recommendation systems that have been addressed by studies in recent years. Although there would be many other challenges in the e-recruitment recommendation domain, we focus on the most common ones here.

We first list the main challenges in e-recruitment recommendation systems and describe each of the challenges in Section~\ref{sec_problem_challenges}. Next, we introduce the methods that have been proposed to deal with each of the challenges in
Section \labelcref{sect:data quality,sect:hetero data,sect:cold start,sect:user preference,sect:interpretability,sect:specific obj,sect:bias,sect:large scale}. Finally, we discuss the papers that are not included in the sections covering challenges in Section~\ref{sect:remaining papers}. Moreover, in each section, we provide a visual overview of the problems and solutions (Fig.~\ref{fig_data_quality} to Fig.~\ref{fig_large_scale}). They contain the solutions that we \textbf{observed} in the literature. Of course, other solutions that have not yet been described in the literature may exist.

\subsection{A preview of the challenges}\label{sec_problem_challenges}
\begin{myenum}
    \item \textbf{Data quality}\label{problem_data_quality}: E-recruitment recommendation systems often have a plethora of data sources, including interactions and textual data from the job seekers (CVs) and job postings (job descriptions). There are many relevant facets in the available data (MF aspect \ref{sect:e-recruitment_overview}.\ref{job_vs_trad_multi-faceted}), but with variable quality. Moreover, some facets, e.g. skills, might be implicit and need to be extracted from unstructured data. Some common issues about dealing with such data are:
    
    \begin{myenum}
        \item \textbf{Data cleaning and preprocessing}. \label{data_quality:preprocessing} Recommendation systems usually use features extracted from textual data, which is usually noisy. Hence, data cleaning preprocessing are necessary and crucial for better feature extraction and downstream tasks.
        
        \item \textbf{Semantic gap}. \label{data_quality:noisy} The textual data is usually written by different people, and different terms are often used to address the same concept. This semantic gap results in poor semantic matching.
        
        \item \textbf{Skill extraction}. \label{data_quality:skill_extraction} Although many facets might be implicit and need to be extracted with carefully designed methods, we focus on skills, which are the most important feature in matching job seekers with job postings. Using job seekers' skills and the job postings' required skills is necessary for increasing the performance of e-recruitment recommendation systems. Hence, skill extraction from the textual data is another challenging task in the e-recruitment recommendation systems.
        
        \item \textbf{Multi-linguality}. \label{data_quality:multilingual} In some countries/platforms, job seekers' resumes and job descriptions are written in several languages. In such cases, e-recruitment recommendation systems should support multiple languages for the textual content.
        
        \item \textbf{Data sparsity}. \label{data_quality:sparsity} Many recommendation systems suffer from data sparsity issues, e-recruitment recommendation is no exception (SIH aspect \ref{sect:e-recruitment_overview}.\ref{job_vs_trad_short_interaction_history}). The reason is that job seekers may only use the system a few times and then leave the platform forever after a successful job-hunting; the same is true for vacant job positions: new jobs might appear on a daily basis but disappear quickly after receiving satisfying applications.

    
    \end{myenum}

    \item \textbf{Heterogeneous data, and multiple interaction types and data sources}\label{problem_heterogeneous_data}:
    E-recruitment recommendation systems could use more data sources compared to many other kinds of recommendation systems, as they might have access to job seekers' previous work experiences, interviews, the textual content of their resumes/job descriptions, skills, and preferences (MF aspect \ref{sect:e-recruitment_overview}.\ref{job_vs_trad_multi-faceted}). The availability of unstructured, semi-structured and structured data makes e-recruitment recommendation systems have to deal with the heterogeneous nature of data.
    
In addition, there are also many interaction types in the recommendation systems between job seekers and job postings, e.g., view, click, apply, chat, favorite, like, and comment. Using different interaction types between job seekers and job postings could be both a challenge and an opportunity in the development of e-recruitment recommendation systems.
    
    Moreover, recommendation systems could also make use of other data sources besides job market related data, such as job seekers' and job postings' information in social networks, blogs, etc.
    
    
    \item \textbf{Cold Start}\label{problem_cold_start}:
    The cold start problem in recommendation systems refers to the problem of recommending to new users or recommending new items with few or no interactions. This problem might be more acute for e-recruitment recommendation systems than the traditional ones since new jobs tend to appear and disappear frequently (SIH aspect \ref{sect:e-recruitment_overview}.\ref{job_vs_trad_short_interaction_history}). The jobs usually disappear after a successful match, and new jobs with the same title are often posted as new items. In contrast, the products with the same name in traditional recommenders are usually treated as the same item, and only their availability changes over time (in cases such as movie recommenders, the product is always available).
    
Using data other than interactions could often alleviate the cold start problem in recommendation systems. Hence, it is helpful to have the many facets available in the job seekers' and job postings' profiles (MF aspect \ref{sect:e-recruitment_overview}.\ref{job_vs_trad_multi-faceted}).
    
Also note that in e-recruitment recommendation systems terms, there are user (job seeker or job) cold start and item (job or job seeker) cold start problems. In job recommendation, user cold start refers to job seeker cold start and item cold start refers to job cold start, and it is the other way around in job seeker recommendation.
    
    \item \textbf{User preferences as well as suitability}\label{problem_suitability_preference}:
To find the best matches between job seekers and vacancies, it is crucial to use the knowledge and skills of the job seekers and the requirements of job positions. However, users' preferences are equally important for a personalized recommendation system (SP aspect \ref{sect:e-recruitment_overview}.\ref{job_vs_trad_suitability}). Moreover, users' preferences might change over time, which should be taken into consideration by the recommendation systems.
    
    
    \item \textbf{Interpretability and explainability}\label{problem_explainability}:
    Providing explainable recommendations and designing interpretable models are important in e-recruitment recommendation systems (HS aspect \ref{sect:e-recruitment_overview}.\ref{job_vs_trad_high_stakes}). Job seekers could benefit from explanations of their recommendations since important career decisions will depend on their choices. Moreover, providing explainable results helps design user-friendly applications for job-seekers and recruiters.
    
    \item \textbf{Specific objectives}\label{problem_specific objectives}:
    E-recruitment recommendation systems usually have a multi-objective nature, since they need to satisfy multiple stakeholders, including job seekers, recruiters, and service providers (TS aspect \ref{sect:e-recruitment_overview}.\ref{job_vs_trad_tow_sided}). 
    
In addition, e-recruitment recommendation systems could have specific objectives, such as balancing the number of recommendations each job seeker/job posting receive or recommending items with a high chance of success regarding the competitors (OWOJ aspect \ref{sect:e-recruitment_overview}.\ref{job_vs_trad_matching}), or avoiding false positives to make sure that users wouldn't be bothered by too many spams.
    
    \item \textbf{Bias and fairness}\label{problem_bias_and_fairness}:
    Recommendation systems suffer from all kinds of well-known biases, some of which have raised societal and ethical concerns. Providing fair recommendations in e-recruitment is even more essential than the other types since e-recruitment is a high-stakes domain (HS aspect \ref{sect:e-recruitment_overview}.\ref{job_vs_trad_high_stakes}). It is crucial to mitigate biases for job seekers, such as gender bias, as well as biases regarding the job postings, such as recency bias (recent job postings may be more popular).

    
    
    \item \textbf{Large scale}\label{problem_item_large_scale}: The ever-increasing amounts of data bring the pressing challenge of scalability to the e-recruitment recommendation systems. More specifically, large scale data may cause issues in both training and inference phases: in each phase, there could be issues with speed and storage/memory consumption.
    
\end{myenum}


\subsection{Data quality}\label{sect:data quality}
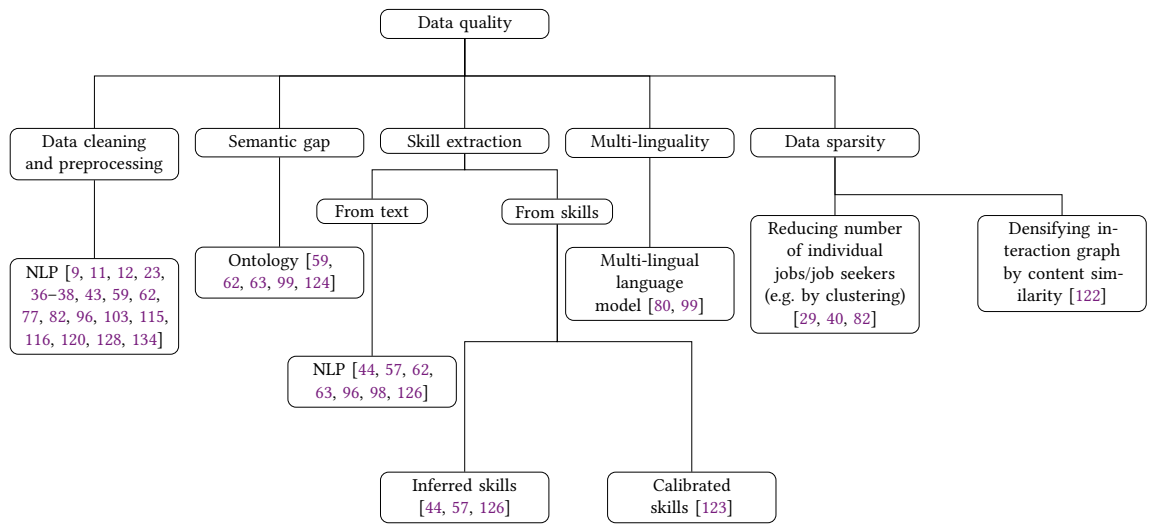
\begin{figure*}[t]
\centering
\begin{adjustbox}{width=\linewidth}
\begin{tikzpicture}[
every node/.style = {draw, rounded corners, text width=27mm, align=center, anchor=north},
node_medium/.style = {draw, rounded corners, text width=20mm, align=center, anchor=north},
node_small/.style = {draw, rounded corners, text width=17mm, align=center, anchor=north},
level distance = 18mm,
sibling distance = 32mm,
edge from parent fork down
                        ]
 \node {Data quality}
    child{ node {Data cleaning and preprocessing}
        child{ node {NLP \cite{164/GuoAH16,16/MughaidOHAA19,19/ElsaftyRB18,27/DiabyVL14,29/DiabyV14,41/Valverde-Rebaza18,58/bellini2020guapp,102/LeeHK16,126/habous2021fuzzy,137/khatua2020matching,150/xu2018matching,167/SchmittCS16,288/MareeKB19,454/diaby2013toward,501/boukari2020huntalent,517/apaza2021job,521/roy2020machine,84/BansalSA17,117/RodriguezC19}}}
        }
    child{ node {Semantic gap}
        child{ node (semantic_gap_ontology) {Ontology \cite{10/ShishehchiB19,52/Mentec0HR21,126/habous2021fuzzy,164/GuoAH16,174/HauffG15}}}
        }
    child { node {Skill extraction} 
        child [level distance = 10mm]{ node [node_small] (from_text) {From text}
        	child[level distance = 25mm]{node  {NLP \cite{57/GugnaniM20,174/HauffG15,288/MareeKB19,477/menacer2021interpretable,211/FaliagkaIKRSTT14,126/habous2021fuzzy,130/smith2021skill}}}
        	}
		child[level distance = 10mm]{ node [node_small] (from_skills) {From skills}
        	child[level distance = 45mm]{node (related_skills) {Inferred skills \cite{57/GugnaniM20,130/smith2021skill,211/FaliagkaIKRSTT14}}}
        	child[level distance = 45mm]{node [right =of related_skills] (select_from) {Calibrated skills \cite{524/shi2020salience}}}
        	}
        }
    child{ node {Multi-linguality}
        child{ node {Multi-lingual language model \cite{52/Mentec0HR21,116/lavi2021consultantbert}}
        }
        }
    child{ node (data_sparsity) {Data sparsity}
        child{ node [below=of data_sparsity] {Reducing number of individual jobs/job seekers (e.g. by clustering) \cite{102/LeeHK16,506/ChenZDGHWW18,509/DongLZBL17}}
        	}
        child{ node [below right=of data_sparsity] {Densifying interaction graph by content similarity \cite{44/ShalabyAKPAQZ17}}
        	}
        }

        ;
\end{tikzpicture}
\end{adjustbox}
\caption{An overview of the \textit{data quality} challenge}
\label{fig_data_quality}
\end{figure*}

Since most e-recruitment recommendation systems use interactions as well as textual data (resumes and job descriptions) to model the user profile or to construct features, various data quality issues affect the quality of recommendations. Most issues in this section are about textual data quality since the facets available in e-recruitment (MF aspect \ref{sect:e-recruitment_overview}.\ref{job_vs_trad_multi-faceted}) are sometimes hidden in free text . We briefly discuss different approaches for each data quality issue discussed in Section~\ref{problem_data_quality}. An overview of this section which includes the categories of the data quality issues and the corresponding solutions in the literature, is presented in Fig.~\ref{fig_data_quality}.

\textbf{Data cleaning and preprocessing} (Section~\ref{data_quality:preprocessing}). E-recruitment recommendation systems usually use textual content to acquire features for job seekers and job descriptions, which could further be used in recommendation methods. However, the textual contents are usually written by different people and are noisy. Therefore, data cleaning and data preprocessing for textual data are crucial for providing high quality recommendations.

Although most approaches using textual content have to do some data cleaning and preprocessing, we only discuss the works that have explicitly focused on NLP techniques to deal with such issues. The data cleaning and preprocessing usually involve common NLP techniques such as tokenization, removing stop words, stemming, and lemmatization \cite{164/GuoAH16,16/MughaidOHAA19,19/ElsaftyRB18,27/DiabyVL14,29/DiabyV14,41/Valverde-Rebaza18,58/bellini2020guapp,102/LeeHK16,126/habous2021fuzzy,137/khatua2020matching,150/xu2018matching,167/SchmittCS16,288/MareeKB19,454/diaby2013toward,501/boukari2020huntalent,517/apaza2021job,521/roy2020machine,84/BansalSA17,117/RodriguezC19}.

\textbf{Semantic gap} (Section~\ref{data_quality:noisy}). 
Since the textual data is written by different people, e-recruitment recommendation systems suffer from a semantic gap between contents from different sources, such as resumes and job descriptions. Different terms might have been used to refer to the same concept. Moreover, the same term could have different meanings depending on the context.

Although most papers that use language models or learn representations of textual data can alleviate the semantic gap to some degree, we only discuss the papers that explicitly focus on this issue. The most common approach that is employed in the literature to tackle the semantic gap is to map skills/concepts to the nodes in an ontology (by exploiting a language model, using Named Entity Recognition (NER), using Named Entity Disambiguation (NED), etc.) and to use the shared nodes to refer to the same skills/concepts \cite{10/ShishehchiB19,52/Mentec0HR21,126/habous2021fuzzy,164/GuoAH16,174/HauffG15}.


\textbf{Skill extraction} (Section~\ref{data_quality:skill_extraction}). 
E-recruitment recommendation systems mostly match job seekers with job postings based on their expertise and skills. Since job seekers' profiles and job descriptions are often available as free text with no structure, skill extraction from the textual data is important for some e-recruitment recommendation systems. Some papers have employed NLP techniques such as n-gram tokenization \cite{288/MareeKB19}, NER \cite{57/GugnaniM20,174/HauffG15,288/MareeKB19,477/menacer2021interpretable}, part-of-speech tagging (PoS tagging) \cite{57/GugnaniM20}, using skill dictionaries or ontology \cite{57/GugnaniM20,174/HauffG15,211/FaliagkaIKRSTT14,288/MareeKB19,126/habous2021fuzzy}, or other techniques (e.g., using the context of a skill term, called skill headwords) \cite{130/smith2021skill} to extract skills from the text. Job seekers' and job postings' skills have also been expanded using skill similarities or relations provided by word embedding models (e.g., word2vec) \cite{57/GugnaniM20, 130/smith2021skill}, and by domain specific ontologies or skill taxonomies \cite{211/FaliagkaIKRSTT14}. Given the extracted skills for job seekers and job postings by an in-house skill tagger in LinkedIn, Shi et al. \cite{524/shi2020salience} selected skills for job postings considering the market supply (enough job seekers having that skill) of the skills and also the importance of each skill in a job posting.





\textbf{Multi-linguality} (Section~\ref{data_quality:multilingual}). Some e-recruitment recommendation systems are multi-lingual, i.e., the textual content of resumes and job descriptions could be in multiple languages. Moreover, matching resumes and job descriptions with different languages results in cross-linguality challenges. Such issues have been studied in \cite{52/Mentec0HR21,116/lavi2021consultantbert}, where a multi-lingual language model was used to support multiple languages. Lavi et al. \cite{116/lavi2021consultantbert} designed a Siamese architecture to fine-tune the multi-lingual Bert using the historical data of recruiters' interactions with candidates.
   
\textbf{Data sparsity} (Section~\ref{data_quality:sparsity}). 
E-recruitment recommendation systems often suffer from data sparsity issues (SIH aspect \ref{sect:e-recruitment_overview}.\ref{job_vs_trad_short_interaction_history}) due to the fact that similar job positions are usually considered as separate entities. Moreover, job seekers often stop using the platform after being employed. Although most approaches that use content in the recommendation could alleviate the data sparsity issue to some extent (e.g. \cite{527/bian2020learning}), we only discuss the works that study data sparsity explicitly.

One approach that has been studied to cope with the data sparsity issue is to reduce the number of distinct job positions by splitting a job position into a job title and a company name \cite{102/LeeHK16} or by clustering similar job positions \cite{506/ChenZDGHWW18,509/DongLZBL17}. Another approach designed by Shalaby et al. \cite{44/ShalabyAKPAQZ17} is to densify the graph of jobs, which is created based on interactions, by adding content similarity links between the entities (job seekers and job positions). The recommendations are then generated using this graph of jobs.


\begin{figure*}[t]
\centering
\begin{adjustbox}{width=0.31\linewidth}
\begin{tikzpicture}[
every node/.style = {draw, rounded corners, text width=27mm, align=center, anchor=north},
level distance = 18mm,
sibling distance = 32mm,
edge from parent fork down
                        ]
 \node {Heterogeneous data}
    child{ node {Computing similarity scores between fields \cite{164/GuoAH16,106/coelho2015hyred,402/lu2013recommender,110/malherbe2014field,77/LiuROX17,90/LiuORSTX16,117/RodriguezC19,129/elgammal2021matching}}
        }
    child{ node {Learning embeddings for each field \cite{525/he2021self,526/he2021finn,132/zhao2021embedding,523/luo2019resumegan}}
        }
        ;
\end{tikzpicture}
\end{adjustbox}
\begin{adjustbox}{width=0.33\linewidth}
\begin{tikzpicture}[
every node/.style = {draw, rounded corners, text width=27mm, align=center, anchor=north},
level distance = 18mm,
sibling distance = 32mm,
edge from parent fork down
                        ]
 \node {Multiple interaction types}
    child{ node {Conversion to ratings \cite{25/ZhangC16}}
        }
    child{ node {Weighting interaction type}
        child{ node {Muti-task \cite{131/fu2021beyond}}
        }
        child{ node {Sample importance weighting \cite{539/volkovs2017content}}
        }
		}
        ;
\end{tikzpicture}
\end{adjustbox}
\begin{adjustbox}{width=0.31\linewidth}
\begin{tikzpicture}[
every node/.style = {draw, rounded corners, text width=27mm, align=center, anchor=north},
level distance = 18mm,
sibling distance = 32mm,
edge from parent fork down
                        ]
 \node {External data sources}
    child{ node {Friends' features \cite{27/DiabyVL14,159/ChalaF17}}
        }
    child{ node {Others \cite{184/BollingerHM12,211/FaliagkaIKRSTT14,212/FaliagkaTT12}}
		}
        ;
\end{tikzpicture}
\end{adjustbox}

\caption{An overview of the \textit{heterogeneous data, and multiple interaction types and data sources} challenge}
\label{fig_hd_mit_ds}
\end{figure*}
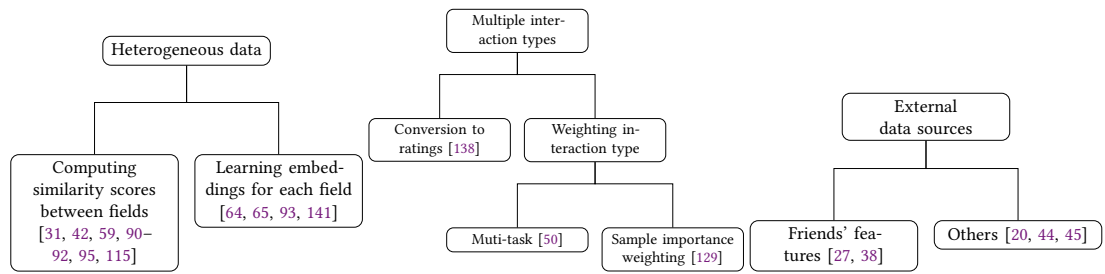

\subsection{Heterogeneous data, and multiple interaction types and data sources}\label{sect:hetero data}
E-recruitment recommendation systems could use the heterogeneous data of job seekers and job postings, including location, textual resume/job description, skills, etc. (MF aspect \ref{sect:e-recruitment_overview}.\ref{job_vs_trad_multi-faceted}). Moreover, different types of behavioral data are available, where using such data is challenging in recommendation systems. In addition, job seekers' and job positions' data could be enriched by their information from external sources. We briefly discuss the papers dealing with these three aspects that are also described in Section~\ref{problem_heterogeneous_data}. An overview of this section is presented in Fig.~\ref{fig_hd_mit_ds}.


Since resumes and job descriptions are among the most important data sources for e-recruitment, it is necessary to carefully use them as well as the behavioral data. Job seeker profiles, resumes, and job descriptions sometimes have several fields with different data types. Hence, the \textbf{heterogeneous nature of the data} should be considered in designing recommendation systems in e-recruitment.

Many papers use features with different types in a recommendation algorithm (e.g., decision trees, deep neural networks, etc.) either directly or by some feature representation techniques such as one-hot encoding, word embedding, etc. (e.g., \cite{75/martinez2018recommendation,478/qin2020enhanced}). However, some methods are explicitly designed to work with heterogenous data. Hence, we mostly focus on those papers. Some studies have combined the similarity scores between the same fields (e.g., education, work experience, etc.) of resumes and job postings \cite{164/GuoAH16,106/coelho2015hyred,402/lu2013recommender,77/LiuROX17,90/LiuORSTX16,117/RodriguezC19,129/elgammal2021matching} or between all fields in resumes and job postings \cite{110/malherbe2014field}. Learning embeddings for each of the fields/data sources of job seeker profiles and job postings, and using the interactions of those embeddings to match job seekers with job postings is another approach employed to deal with heterogeneous data \cite{525/he2021self,526/he2021finn,132/zhao2021embedding,523/luo2019resumegan}. More specifically, Zhao et al. \cite{132/zhao2021embedding} provided recommendations based on the fused embeddings of job seekers and jobs, where they combine the embeddings learned from the textual content, job-skill information graph, and geolocation data. In the deep neural networks proposed in \cite{525/he2021self,526/he2021finn}, the embeddings for the same fields/field types of resumes and job postings were learned by their inner interactions. In \cite{525/he2021self}, a multi-head self-attention module was then applied to the embeddings for different fields as the field outer interaction module. In \cite{523/luo2019resumegan}, different embeddings are learned for different fields of job seekers by their interactions in the neural network. Finally, the learned embeddings were passed to a multi-layer perceptron to compute the matching score between a resume and a job posting \cite{525/he2021self,526/he2021finn,523/luo2019resumegan}.


Moreover, there could be \textbf{multiple types of interactions} between a job seeker and a job position, such as click, apply, like, favorite, invite, interview, hire, etc., where some of them are initiated by the job seeker and some by recruiters. Zhang and Cheng \cite{25/ZhangC16} transformed the implicit feedback (click, bookmark, reply, and click) into ratings and proposed a two-stage ensemble method for generating the recommendations. Fu et al. \cite{131/fu2021beyond} proposed a deep neural network to capture the dynamic preferences of the job seekers and recruiters by learning a multi-task objective of their behavioral data (e.g., click, apply, chat, invite, match). Volkovs et al. \cite{539/volkovs2017content} proposed a content-based recommendation system considering different interaction types as positive with different weights for sampling and used XGBoost to optimize the binary classification loss.

To find a better match between job seekers and vacancies, information other than skills such as personality and traits has also been found to be useful. Some studies have tried to use auxiliary information gathered from \textbf{external data sources} such as friends' features in social networks \cite{27/DiabyVL14,159/ChalaF17} and personal websites \cite{184/BollingerHM12,211/FaliagkaIKRSTT14,212/FaliagkaTT12} to build more comprehensive profiles and improve the recommendations.

\subsection{Cold start}\label{sect:cold start}
As discussed in Section~\ref{problem_cold_start}, cold start in recommendation systems refers to the problem of recommending to new users or items with no or few interaction data. This problem could be more acute for e-recruitment recommendation systems because job opening positions are usually treated as distinct items even if they have the same job title and description, and hence those job openings would be treated as new items (SIH aspect \ref{sect:e-recruitment_overview}.\ref{job_vs_trad_short_interaction_history}). E-recruitment recommenders could suffer from both job seeker cold start and job cold start problems. 

\begin{figure}[t]
\centering
\begin{adjustbox}{width=0.35\linewidth}
\begin{tikzpicture}[
every node/.style = {draw, rounded corners, text width=27mm, align=center, anchor=north},
level distance = 18mm,
sibling distance = 32mm,
edge from parent fork down
                        ]
 \node {Cold start}
    child{ node {Recommending based on the interactions of similar jobs/job seekers \cite{77/LiuROX17,90/LiuORSTX16,38/NigamRSW19,506/ChenZDGHWW18,519/hong2013dynamic,25/ZhangC16,44/ShalabyAKPAQZ17,535/guo2017howinte,538/bianchi2017content,160/SchmittGCS17,102/LeeHK16}}
        }
    child{ node {Recommending based on the features of jobs and job seekers \cite{44/ShalabyAKPAQZ17,3/GuptaG14,167/SchmittCS16,516/yagci2017aranker,538/bianchi2017content,78/yang2017combining,520/sato2017explor,535/guo2017howinte, 537/lian2017practical, 539/volkovs2017content}
          }}  ;
\end{tikzpicture}
\end{adjustbox}

\caption{An overview of the \textit{cold start} challenge}
\label{fig_cold_start}
\end{figure}
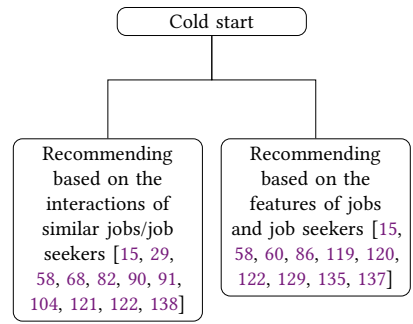

Using content to provide recommendations could alleviate the cold start problem. In the e-recruitment domain, many facets are often available for this purpose (MF aspect \ref{sect:e-recruitment_overview}.\ref{job_vs_trad_multi-faceted}). Hence, the papers with content based approaches or methods that use features based on the content could deal with the cold start problem to some extent. However, we only discuss the papers that explicitly address the cold start problem. The papers dealing with cold start follow two general approaches: recommending using the interactions of similar jobs/job seekers or predicting the recommendation score based on job seekers' and jobs' features. Some papers also employ both approaches to deal with the cold start problem. An overview of this section, including the solutions proposed by recent studies for the cold start problem, is presented in Fig.~\ref{fig_cold_start}.

Two approaches have been used in the literature that recommend based on the \textbf{interactions of similar jobs/job seekers}. First, to compute the matching scores between jobs and new job seekers, some studies find similar job seekers to the new ones based on content features and then use the known (e.g., previously interacted) matching scores between them and the jobs \cite{77/LiuROX17,90/LiuORSTX16,38/NigamRSW19,506/ChenZDGHWW18,519/hong2013dynamic}. In papers \cite{77/LiuROX17,90/LiuORSTX16}, jobs are recommended to new graduate students based on the job offers of similar graduates. In another study by Chen et al. \cite{506/ChenZDGHWW18}, a context-aware multi-arm bandit was employed for generating job recommendations, where the job recommendation scores for new job seekers were computed based on the interaction history of similar job seekers. This method could also deal with the job cold start in case of job seeker recommendation due to the symmetric nature of their model architecture. Second, to compute the matching scores between new jobs and job seekers, some studies find jobs with similar content to the new ones and use the known (e.g., previously interacted) matching scores between them and the job seekers \cite{25/ZhangC16,44/ShalabyAKPAQZ17,535/guo2017howinte,538/bianchi2017content,38/NigamRSW19,160/SchmittGCS17,102/LeeHK16}. 

Some studies \textbf{predict the matching scores between job seekers and jobs using their features} to deal with the cold start problem (e.g., using a machine learning method or a scoring function). The job categories that new job seekers are interested in are predicted using job seekers' textual content \cite{44/ShalabyAKPAQZ17} or attributes \cite{3/GuptaG14} and are further exploited to provide job recommendations. Other papers have provided recommendations based on job seekers' and jobs' content, which tackle both job seeker cold start ad job cold start problems \cite{167/SchmittCS16,516/yagci2017aranker,538/bianchi2017content,78/yang2017combining,520/sato2017explor} (Although many content-based methods could tackle the cold start problem with the same approach, here we only cite the papers that have explicitly addressed the cold start problem). Besides features extracted from job seekers' and jobs' content, several studies \cite{535/guo2017howinte, 537/lian2017practical, 539/volkovs2017content, 520/sato2017explor} also extracted features for job seekers based on the jobs they have interacted with before. Hence, they can deal with the job cold start problem.

\begin{figure*}[t]
\centering
\begin{adjustbox}{width=0.45\linewidth}
\begin{tikzpicture}[
every node/.style = {draw, rounded corners, text width=27mm, align=center, anchor=north},
level distance = 18mm,
sibling distance = 32mm,
edge from parent fork down
                        ]
 \node {User preferences}
    child{ node {Behavioral interactions, e.g.,  \cite{60/wang2020session,56/LacicRKDCL20,76/ReusensLBS17,145/YanLSZZ019,44/ShalabyAKPAQZ17}}
        }
    child{ node {Explicit preferences \cite{39/GutierrezCCHGV19,495/slama2021novel}}
        }
    child{ node {Preference model \cite{49/BiancofioreNSNP21,51/BiancofioreNSNP21,58/bellini2020guapp,115/bills2021looking,3/GuptaG14}}}  ;
\end{tikzpicture}
\end{adjustbox}
\begin{adjustbox}{width=0.45\linewidth}
\begin{tikzpicture}[
every node/.style = {draw, rounded corners, text width=27mm, align=center, anchor=north},
level distance = 18mm,
sibling distance = 32mm,
edge from parent fork down
                        ]
 \node {Dynamic preferences}
    child{ node {Neural architectures (e.g. LSTM) \cite{38/NigamRSW19,23/LiuSKZN16,131/fu2021beyond}}
        }
    child{ node {Time-dependent features \cite{23/LiuSKZN16}}}  
    child{ node {Time-dependent loss function \cite{23/LiuSKZN16}}}
          ;
\end{tikzpicture}
\end{adjustbox}

\caption{An overview of the \textit{user preferences as well as suitability} challenge}
\label{fig_user_preference}
\end{figure*}
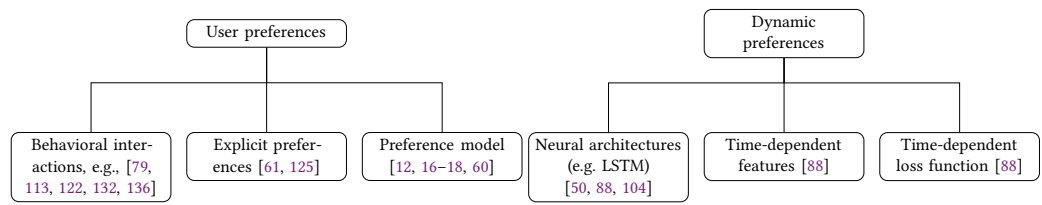

\subsection{User preferences as well as suitability}\label{sect:user preference}
Although considering user preferences is important in all recommendation systems, e-recruitment recommendation systems should also consider suitability in generating the recommendations, i.e. matching job seekers with job postings based on the similarity of their skills and requirements (SP aspect \ref{sect:e-recruitment_overview}.\ref{job_vs_trad_suitability}). Since matching based on the suitability of job seekers for job positions has been the main focus of e-recruitment recommendation systems, we discuss the studies focusing on capturing user preference. Suitability is usually captured by matching the requirements of a job position with the skills and other feature of the job seekers, while preference is often captured by other factors in the profiles of job seekers an job postings, such as location, interests, etc., or by behavioral interactions. As discussed in Section~\ref{problem_suitability_preference}, job seekers' preferences might change over time and modeling the dynamic preferences is also a challenging task in e-recruitment recommendation. In this section, we first discuss the methods explicitly modeling user preferences either based on explicit preferences in user profiles or using a preference model. Next, we present the approaches targeting the dynamic nature of user preferences. An overview of this section is presented in Fig.~\ref{fig_user_preference}.


\textbf{Behavioral interactions} between job seekers and job postings, such as click, apply, invite, etc., can show the user preferences to some extent. Hence, E-recruitment recommendation systems that use such behavioral interactions in their method are considering user preferences in generating recommendations (e.g., \cite{60/wang2020session,56/LacicRKDCL20,76/ReusensLBS17,145/YanLSZZ019,44/ShalabyAKPAQZ17}).

 
Another way that user preferences are taken into consideration in recommendation is by using \textbf{explicit preferences} specified in the user profile (e.g., interests, location, etc.) or in a dashboard. Guti\'errez et al. \cite{39/GutierrezCCHGV19} designed a dashboard for job seekers to visualize and explore available vacancies based on their preferences. A fuzzy-based recommendation was proposed by Slama and Darmon \cite{495/slama2021novel} that matches job seekers with job postings based on their fuzzy preferences in their profiles. Although many studies use such features in the recommendation, we only discuss the papers that explicitly focus on the user preferences.

User preferences are sometimes not obtained directly but rather through a \textbf{preference model}. Some studies learn such models from explicit feedback \cite{49/BiancofioreNSNP21,51/BiancofioreNSNP21,58/bellini2020guapp,115/bills2021looking}. Bills and Ng \cite{115/bills2021looking} proposed a matching model aimed at adults with Autism. They asked both job seekers and employers some questions to form the preference vectors for both sides and used them in the Gale-Shapley stable matching algorithm \cite{gale2013college} to provide the recommendations. Another way that user preferences are modeled is by using implicit feedback and content. Gupta and Garg \cite{3/GuptaG14} designed preference matrices for different job seeker groups generated from historical data and used them in their hybrid recommender.

To consider the \textbf{dynamic} nature of user preferences, the change in user preferences is usually captured through the interactions in time. Nigam et al. \cite{38/NigamRSW19} employed a recurrent neural network (Bidirectional LSTM) with the attention mechanism to capture users' change of preferences over time. Liu et al. \cite{23/LiuSKZN16} proposed an ensemble recommendation system with three different recommenders. Observing the fact that users tend to re-interact with items, a time reweighted linear ranking model was designed to compute the matching score of a job seeker and a job posting based on the frequency of their previous interactions. The time-dependent weights were learned by optimizing a smoothed hinge loss. Next, a temporal matrix factorization algorithm was designed by introducing a time-related loss term to consider the time of the interactions. Finally, an encoder-decoder model based on LSTM was employed to model the sequence of job seekers-jobs interactions. Fu et al. \cite{131/fu2021beyond} proposed a person-job fit model to transform job seekers and jobs heterogeneous dynamic preferences (preferences based on different interactions such as click, apply, chat, invite, and match) into a unified preference space. First, job seekers and jobs were encoded using hierarchical LSTM. Next, their dynamic preferences were captured through a Dynamic Multi-Key Value Memory Network. This network has a global key matrix for each interaction type (along with their attention weights) and a memory matrix for each job seeker/job preferences. Finally, to transfer the preferences from auxiliary behavior (interaction types other than match) to the matching task, the parameters were learned by a multi-task objective, which is the weighted sum of loss for each interaction type.

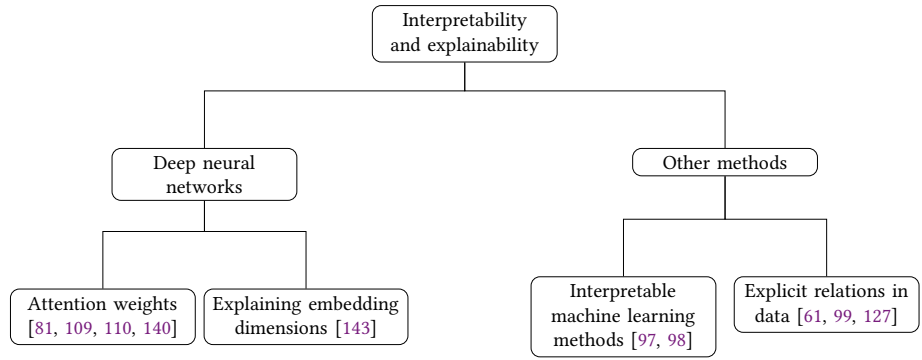
\begin{figure*}[t]
\centering
\begin{adjustbox}{width=0.8\linewidth}
\begin{tikzpicture}[
every node/.style = {draw, rounded corners, text width=27mm, align=center, anchor=north},
level distance = 18mm,
sibling distance = 82mm,
edge from parent fork down
                        ]
 \node {Interpretability and explainability}
    child{ node {Deep neural networks}
             child[sibling distance = 32mm]{ node {Attention weights \cite{478/qin2020enhanced,485/qin2018enhancing,475/zhang2021explainable,483/le2019towards}}
        }
             child[sibling distance = 32mm]{ node[text width=30mm] {Explaining embedding dimensions \cite{469/zhu2018person}}
        }
}
    child{ node {Other methods}
             child[sibling distance = 32mm]{ node {Interpretable machine learning methods \cite{477/menacer2021interpretable,75/martinez2018recommendation}}
        }    
             child[sibling distance = 32mm]{ node {Explicit relations in data \cite{39/GutierrezCCHGV19,53/UpadhyayA0F21,52/Mentec0HR21}}
        }    
    }  ;
\end{tikzpicture}
\end{adjustbox}

\caption{An overview of the \textit{interpretability and explainability} challenge}
\label{fig_explainability}
\end{figure*}

\subsection{Interpretability and explainability}\label{sect:interpretability}
Interpretability often refers to the model transparency and the ability to understand why and how the model generates the predictions. On the other hand, explainability often refers to the ability to explain the predictions in human terms, even for complex models. However, interpretability and explainability have been often been used interchangeably, and we also use the two terms interchangeably in this section. As described in Section~\ref{problem_explainability}, providing explanations for recommendations in e-recruitment is a challenging and important task since the recommendations affect people's future careers and explanations help them make more insightful decisions (HS aspect \ref{sect:e-recruitment_overview}.\ref{job_vs_trad_high_stakes}). We briefly discuss different approaches proposed in the literature to achieve interpretability and explainability for e-recruitment recommendations in the rest of this section, which includes using methods to provide explainability in deep neural network models, using interpretable machine learning methods, and using explicit relations in data to provide explainability. An overview of the approaches that address interpretability and explainability is presented in Fig.~\ref{fig_explainability}.

One way \textbf{explainability is addressed in the deep neural models} that use resumes and job descriptions for person-job fit prediction is to visualize the attention weights. The attention weights could show the importance of different words, sentences, or any part of the resume/job description in the resume/job description \cite{478/qin2020enhanced,485/qin2018enhancing} and also their importance in matching with the target job description/resume words, sentences, or any part of it \cite{475/zhang2021explainable,478/qin2020enhanced, 483/le2019towards, 485/qin2018enhancing}. Another way to address explainability in deep neural models is proposed by Zhu et al. \cite{469/zhu2018person}. For each dimension in the final representation of resumes and jobs resulting from the deep model, high-frequency words were gathered from other resumes and jobs that have high values for that dimension. Hence, a level of explainability was provided for each job posting or resume.

Another approach by which explainability is provided in the literature is by applying \textbf{interpretable machine learning methods} such as decision trees to human-readable features \cite{477/menacer2021interpretable,75/martinez2018recommendation}. 

In other studies, explainability is provided using \textbf{explicit relations in data}. In \cite{39/GutierrezCCHGV19} a dashboard was provided to view the job seekers' affinity with the required skills for the jobs that are recommended. In \cite{53/UpadhyayA0F21}, recommendations were generated using a knowledge graph together with a template for explainability, where the template was then completed using the nodes in the knowledge graph. Mentec et al. \cite{52/Mentec0HR21} provide explanations by the similarity of job seekers' and job postings' skills using an skill ontology.

\begin{figure*}[t]
\centering
\begin{adjustbox}{width=\linewidth}
\begin{tikzpicture}[
every node/.style = {draw, rounded corners, text width=27mm, align=center, anchor=north},
level distance = 18mm,
sibling distance = 62mm,
edge from parent fork down
                        ]
 \node {Specific objectives}
    child{ node {Multiple stakeholders}
             child{ node (reciprocal) {Reciprocal recommenders}
	             child{ node (reciprocal_labeled) [text width=100mm, below=of reciprocal]{Recommending based on the labeled data of both sides used for training \cite{153/MaheshwaryM18,138/LiFTPHC20,331/salazar2015case,116/lavi2021consultantbert, 131/fu2021beyond, 145/YanLSZZ019, 469/zhu2018person, 478/qin2020enhanced, 485/qin2018enhancing, 525/he2021self, 538/bianchi2017content, 537/lian2017practical, 535/guo2017howinte, 520/sato2017explor, 539/volkovs2017content, 526/he2021finn, 527/bian2020learning, 483/le2019towards, 484/bian2019domain, 479/jiang2020learning, 181/MineKO13, 475/zhang2021explainable, 523/luo2019resumegan, 528/cardoso2021matching, 477/menacer2021interpretable, 124/WangJP21,65/liu2019tripartite, 132/zhao2021embedding, 516/yagci2017aranker,402/lu2013recommender,137/khatua2020matching}}
        }
	             child{ node [text width=100mm, right=of reciprocal_labeled] {Recommending based on the features of jobs and job seekers or some inference rules \cite{288/MareeKB19,184/BollingerHM12,159/ChalaF17,115/bills2021looking,130/smith2021skill,166/RaczSS16,495/slama2021novel,277/freire2020framework,331/salazar2015case,156/arita2017gber,129/elgammal2021matching,46/AlmalisTKS15,106/coelho2015hyred,126/habous2021fuzzy}}
        }
        }
}
    child{ node {OWOJ aspect \ref{sect:e-recruitment_overview}.\ref{job_vs_trad_matching}}
             child{ node {Stable matching \cite{115/bills2021looking}}
        }    
             child{ node {Job redistribuition \cite{541/borisyuk2017lijar}}
        }    
    }
    child{ node {Other objectives}
             child{ node {Specific objective functions \cite{78/yang2017combining}}
        }
}    
      ;
\end{tikzpicture}
\end{adjustbox}

\caption{An overview of the \textit{specific objectives} challenge}
\label{fig_specific_objectives}
\end{figure*}
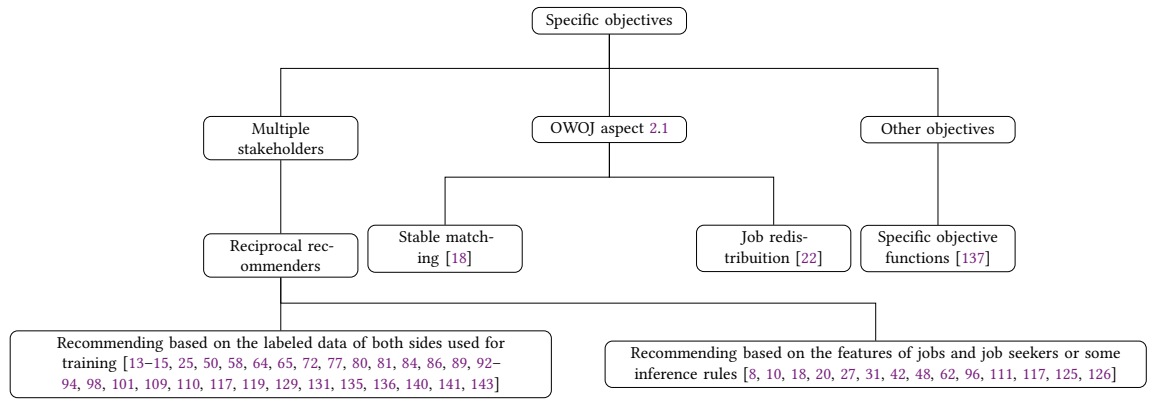

\subsection{Specific objectives}\label{sect:specific obj}
E-recruitment recommendation systems usually should satisfy multiple stakeholders, such as employers, job seekers, and sometimes the recommendation platform, which benefits from matching job seekers with jobs (TS aspect \ref{sect:e-recruitment_overview}.\ref{job_vs_trad_tow_sided}). The platforms' benefits are often included in the job seeker's and employers' benefits since job seekers' and employers' satisfaction also leads to more revenue for the recommendation platform. Hence, most studies try to improve the recommendations for job seekers and employers. In addition, some studies have considered specific objectives for e-recruitment recommendation systems (e.g., OWOJ aspect \ref{sect:e-recruitment_overview}.\ref{job_vs_trad_matching}). We briefly discuss the papers dealing with such issues that are also described in Section~\ref{problem_specific objectives}. An overview of this section is presented in Fig.~\ref{fig_specific_objectives}.

Since \textbf{reciprocal recommenders} recommend job seekers to job postings and vice versa, they usually consider the benefits of job seekers and employers at the same time. Some studies use historical interactions between job seekers and employers that show the interests of both sides for training. The labeled data for such methods usually includes interview and recruitment data \cite{138/LiFTPHC20,116/lavi2021consultantbert,131/fu2021beyond,145/YanLSZZ019,469/zhu2018person, 478/qin2020enhanced,485/qin2018enhancing,525/he2021self,538/bianchi2017content,537/lian2017practical,535/guo2017howinte,520/sato2017explor,539/volkovs2017content,526/he2021finn,527/bian2020learning,483/le2019towards,484/bian2019domain,479/jiang2020learning,181/MineKO13,475/zhang2021explainable,523/luo2019resumegan,528/cardoso2021matching,477/menacer2021interpretable,124/WangJP21,516/yagci2017aranker,65/liu2019tripartite,132/zhao2021embedding}, actions such as favorite or click data by both job seekers and recruiters \cite{402/lu2013recommender,331/salazar2015case}, or manually annotated data \cite{153/MaheshwaryM18,137/khatua2020matching}. On the other hand, some methods compute the matching degree of a job seeker and a job posting based on the similarity of their contents, skills or other features, or by some inference rules \cite{288/MareeKB19,184/BollingerHM12,174/HauffG15,159/ChalaF17,115/bills2021looking,130/smith2021skill,166/RaczSS16,495/slama2021novel,277/freire2020framework,331/salazar2015case,156/arita2017gber,129/elgammal2021matching,46/AlmalisTKS15,106/coelho2015hyred,126/habous2021fuzzy}, which could recommend jobs to job seekers and vice versa with this approach.

Other than the reciprocal nature of recommendation in e-recruitment, some studies have tried to consider the fact that in the job market, for a fixed period of time, each job seeker is hired for one (or a few) job position and vice versa (\textbf{OWOJ} aspect \ref{sect:e-recruitment_overview}.\ref{job_vs_trad_matching}). 
A stable matching algorithm was employed in \cite{115/bills2021looking} to find recommendations for job seekers and recruiters considering this aspect. Moreover, a job application redistribution at LinkedIn was proposed in \cite{541/borisyuk2017lijar} to prevent job postings from receiving too many or too few applications. To achieve this goal, the job recommendation scores were penalized or boosted based on the predicted number of applications using a dynamic forecasting model.

\textbf{Other objectives} have also been investigated for e-recruitment recommendation systems. One of such objectives for e-recruitment systems is to prevent job seekers from receiving spam. To address this issue, false positives were penalized harshly in the hybrid job recommendation proposed by Yang et al. \cite{78/yang2017combining}.

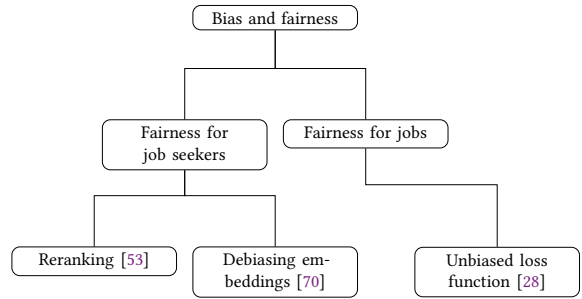
\begin {figure}
\centering
\begin{adjustbox}{width=0.5\linewidth}
\begin{tikzpicture}[
every node/.style = {draw, rounded corners, text width=27mm, align=center, anchor=north},
level distance = 18mm,
sibling distance = 32mm,
edge from parent fork down
                        ]
 \node {Bias and fairness}
    child{ node {Fairness for job seekers}
             child{ node {Reranking \cite{546/geyik2019fairness}}
        }
             child{ node (debiasing) {Debiasing embeddings \cite{531/islam2021debias}}
        }
}
    child{ node {Fairness for jobs}
             child{ node [right=of debiasing] {Unbiased loss function \cite{64/chen2019correcting}}
        }    
    }  ;
\end{tikzpicture}
\end{adjustbox}

\caption{An overview of the \textit{bias and fairness} challenge}
\label{fig_fairness}
\end{figure}

\subsection{Bias and fairness}\label{sect:bias}
The problems related to bias and fairness in AI have gained more attention in recent years.
Since e-recruitment affects people's career choices, it is crucial to consider the fairness aspects of the recommendations (HS aspect \ref{sect:e-recruitment_overview}.\ref{job_vs_trad_high_stakes}): e-recruitment is even defined as one of the high-risk domains according to the EU's AI act (proposal) \cite{eu-aiact-2022}. Realizing the limitation of pure algorithmic debiasing methods, some researchers have argued that mitigating bias and unfairness in e-recruitment deserves an interdisciplinary point of view involving legal and ethical considerations (\cite{544/raghavan2020miti, 545/sanchez2020whatdoes}). Wang et al. \cite{543/wang2022dohumans} addressed the limitation of current debiasing technology by conducting an online user study showing that biased recommendations are preferred by job seekers, which indicates that human bias should be addressed from new perspectives or new technology.

From a technical point of view, fairness concerns may exist on both sides \cite{abdollahpouri2020multistakeholder}, namely for job seekers and also for job postings, since recommendation in e-recruitment is multi-stakeholder. Some examples of such biases and fairness concerns are job seekers' racial or gender discrimination \cite{li2022fairness}, popularity bias \cite{abdollahpouri2020addressing}, selection bias \cite{64/chen2019correcting}, etc. Moreover, fairness concerns exist for both users and items in e-recruitment recommendation systems, e.g., job seekers with a certain sensitive attribute might not be recommended for specific jobs and also might not receive specific jobs in their recommendations.  We briefly discuss the papers dealing with fairness issues, which are also described in Section~\ref{problem_bias_and_fairness}. We first present the studies focusing on fairness for job seekers and then the papers addressing fairness issues for job postings. An overview of the approaches that address fairness issues in e-recruitment recommendation systems is presented in Fig.~\ref{fig_fairness}.

To provide \textbf{fair recommendations concerning job seekers}, Geyik et al. \cite{546/geyik2019fairness} proposed a fairness-aware framework for ranking job seekers as used in search and recommending job seekers. Four deterministic reranking algorithms were proposed to mitigate biased prediction towards any sensitive group. Islam et al. \cite{531/islam2021debias} addressed the gender bias in job recommedation by proposing a neural fair collaborative filtering model (NFCF). Job seeker embeddings were pre-trained from non-e-recruitment recommendation data (e.g., movie recommendation) and then debiased with a similar technique of debiasing word vectors so that the gender component is removed from each job seeker embedding. Next, the debiased job seeker embeddings were used in the fine-tuning stage for job recommendation to ensure that sensitive attributes do not affect the outputs of the system.

To provide \textbf{fairness for job postings}, Chen et al. \cite{64/chen2019correcting} tackled the recency bias in job recommendation. They considered the recency bias as a type of selection bias imposed by the job seekers and designed an unbiased loss using inverse propensity weighting in a neural collaborative filtering model. 

\begin {figure*}
\centering
\begin{adjustbox}{width=\linewidth}
\begin{tikzpicture}[
every node/.style = {draw, rounded corners, text width=27mm, align=center, anchor=north},
level distance = 18mm,
sibling distance = 62mm,
edge from parent fork down
                        ]
 \node {Large scale}
    child{ node {Training phase}
             child{ node {Item-based methods \cite{44/ShalabyAKPAQZ17}}}
             child{ node {Scalable algorithms (e.g. parallel methods) \cite{540/zhang2016glmix}}}
        }             
    child{ node {Inference phase}
             child{ node {Two-stage methods \cite{132/zhao2021embedding,533/borisyuk2016casmos}}
        }    
    }
    child{ node {Both phases}
             child{ node {Big data computations \cite{501/boukari2020huntalent}}
        }    
             child{ node {Reducing the number of individual jobs/job seekers (e.g., by clustering) \cite{506/ChenZDGHWW18,509/DongLZBL17,37/MhamdiMGAM20}}
        }    
    }    
      ;
\end{tikzpicture}
\end{adjustbox}
\caption{An overview of the \textit{large scale} challenge}
\label{fig_large_scale}
\end{figure*}
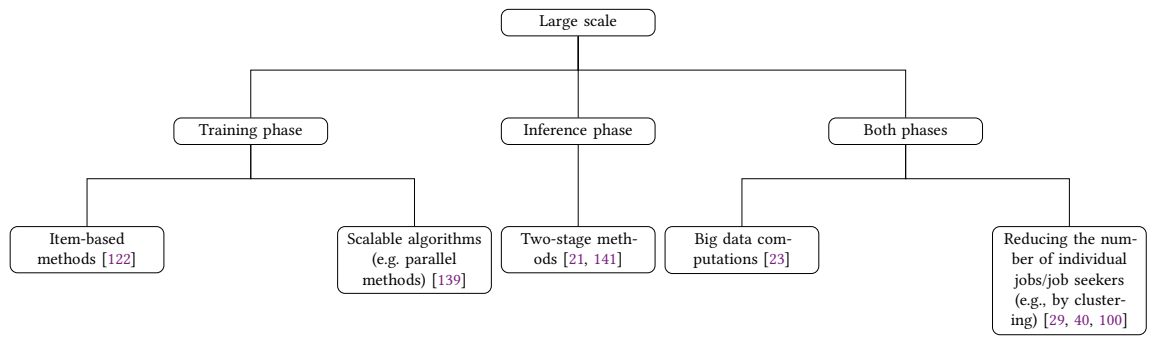

\subsection{Large scale}\label{sect:large scale}
Real-world job recommendation systems have to deal with millions of job seekers and job postings. Hence, recommending at large scale needs to be considered in online job market platforms. We briefly discuss the papers dealing with large scale issues described in Section~\ref{problem_item_large_scale}, which include reducing execution time and consumed storage/memory in training and inference phases. An overview of the approaches that address large scale issues in e-recruitment recommendation systems is presented in Fig.~\ref{fig_large_scale}.

To deal with the execution time and consumed storage/memory issues during the \textbf{training phase}, a study from CareerBuilder\footnote{\url{https://www.careerbuilder.com/}} \cite{44/ShalabyAKPAQZ17} created an item-based graph of jobs with edges representing job similarities based on behavioral and content-based signals. An item-based graph of jobs with different similarity scores was used rather than a user-based (job seeker based) or user-item (job-job seeker) graph for scalability. A subgraph of this job graph was selected by a job seeker's resume or past clicks, and the recommendations were generated by applying PageRank algorithm to this subgraph. In a study at LinkedIn \cite{540/zhang2016glmix}, a scalable algorithm (a parallel block-wise coordinate descent algorithm) was designed for learning the GLMix model to predict the user response.

To deal with the response time in the \textbf{inference phase} a two-stage architecture is often used by the industry leaders, where the first stage selects a pool of candidates from a large number of items using a computationally inexpensive model, and the second stage reranks the results using a more expensive model. One example of the two-stage architectures was designed for recommendation at CareerBuilder \cite{132/zhao2021embedding}. The first stage was designed to select hundreds of candidates from millions using FAISS \cite{johnson2019billion} to find the nearest neighbors of an entity in the embedding space. The embeddings were calculated through three components; a deep neural network to learn from the textual data, a representation framework to learn from three graphs constructed from jobs and skills \cite{74/dave2018combined}, and a geolocation embedding calculator \cite{65/liu2019tripartite}. The second stage was designed to rerank the candidates using a weighted linear combination of the first stage scores and context-based scores. 
In \cite{533/borisyuk2016casmos}, a candidate selection model, CasMoS, was proposed as the first stage in the two-stage recommendation framework at LinkedIn. CasMoS is the framework that learns the first stage model, candidate selection, using the Weighted AND (WAND) query operator \cite{broder2003efficient}.

From another perspective, to deal with scalability issues both in the \textbf{training and inference phases}, Boukari et al. \cite{501/boukari2020huntalent} employed Apache Spark, a tool to process big data, to recommend jobs to job seekers using a content-based algorithm. Another proposed approach to deal with big data and the large number of entities is to cluster jobs and/or job seekers \cite{506/ChenZDGHWW18,509/DongLZBL17,37/MhamdiMGAM20}.

\subsection{Papers not included in previous sections}\label{sect:remaining papers}
Some of the collected papers are not included in the previous sections because they did not directly address any of the challenges discussed in this survey \cite{542/almalis2014content,74/dave2018combined,522/patel2017capar,114/zhao2021summer,48/FengJWYL21,88/kenthapadi2017personalized,63/jiang2019user,32/HongZWS13,9/RivasCGCC19,20/Gonzalez-Briones18,18/HeggoA18,21/Al-OtaibiY17,67/lacic2019should,91/DomeniconiMPPP16,94/CarpiEKSCPQ16,95/LeksinO16,96/MishraR16,97/PacukSWWW16,98/PessemierVM16,99/PolatoA16,100/XiaoXLMW16,101/Zibriczky16,142/Fernandez-Reyes19,171/LinLAL16,178/HeapKWBC14,273/BituinAE20,332/chenni2015content,390/eitle2021impact,518/almalis2014acontent}. However, some papers tackle a specific challenge in e-recruitment recommendation systems, such as dealing with missing features \cite{63/jiang2019user} or applying different recommendation strategies for different groups of job seekers \cite{32/HongZWS13,63/jiang2019user}. We did not discuss such challenges in these papers since either there were not many papers dealing with the same issues or these issues were considered to be of lesser practical significance as compared to the challenges highlighted in the present survey. Practical challenges and lessons learned from the e-recruitment recommendation system at LinkedIn are also discussed in two talks \cite{86/KenthapadiLV17,530/geyik2018talent}. 





\section{Conclusion}\label{sec_conclusion}
In this section we provide our final remarks. We first provide a summary of this survey in Section~\ref{sec_conclusion_summary}. Next, we discuss the limitations of this survey in Section~\ref{sec_limitations}. Finally, open challenges and future research directions of recommendation in e-recruitment are discussed in Section~\ref{sec_future_directions}.

\subsection{Summary}\label{sec_conclusion_summary}
E-recruitment recommendation includes recommending jobs to job seekers and job seekers to jobs. We identified eight challenges that have been studied in the past decade for recommendation in e-recruitment. Since the available data for training an e-recruitment recommendation model include the interactions between job seekers and job positions together with their features and textual contents, several studies have addressed \textbf{data quality} issues.

Job seekers' and jobs' data usually include textual content, location, categorical features, etc., which could also be enriched by external data sources. Moreover, there are many interaction types, such as click, apply, invite, chat, interview, etc., in e-recruitment platforms. Therefore, dealing with \textbf{heterogeneous data, and multiple interaction types and data sources} is another challenge in e-recruitment.

Since job positions with the same content are often represented as different entities in e-recruitment recommendation systems (different job entities with distinct IDs may have the same title/content), \textbf{cold start} problem needs more attention in e-recruitment recommendation compared to the traditional recommenders. The availability of many facets in e-recruitment domain could help alleviate the cold start problem.

Traditional recommendation systems mainly consider user preferences for generating the recommendations, while e-recruitment recommendation systems have to match job seekers with jobs based on the job seekers' skills and jobs' required skills as well. Hence, e-recruitment recommendation systems should consider \textbf{user preferences as well as suitability}.

Explainable recommendations in general help users make better decisions. Nonetheless,  \textbf{interpretability and explainability} are even more important in e-recruitment recommendation systems since e-recruitment recommendation has a great influence on job seekers' future careers and also on the employers of companies.

Recommendation systems in a specific domain could have \textbf{specific objectives}. In e-recruitment, the goal is usually to satisfy multiple stakeholders, including job seekers, recruiters, and service providers. Moreover, e-recruitment recommendation systems should consider the fact that each job seeker could be employed for one or a few job positions and vice versa, which can introduce new objectives for recommendation systems.

\textbf{Bias and fairness} issues are challenging for most recommendation systems. In e-recruitment, it is even more critical to provide fair recommendations due to the possible high-stakes involved for both job seekers and employers.

Finally, \textbf{large scale} issues are unignorable in designing real-world recommendation systems. Since e-recruitment recommendation systems usually have to provide services for thousands/millions of job seekers and job positions, they have to consider the large scale aspects of the recommendation system.


\subsection{Limitations of the survey}\label{sec_limitations}
We have selected and elaborated the main challenges in the e-recruitment recommendation from our point of view, but there could be other challenges in this domain. For example, extracting features from textual data with different granularity could also be considered as another challenge, albeit not specific to the e-recruitment domain. Identifying more challenges and categorizing papers based on their approaches to address them remain for the future.

Since e-recruitment recommendation could be a reciprocal recommendation task (recommending jobs to job seekers and vice versa), reviewing the challenges in other reciprocal recommendation systems (e.g., online dating) could also be useful for designing e-recruitment recommendation systems. We omitted papers from other reciprocal recommendation domains to limit the scope of this survey.

\subsection{Open challenges and future research directions}\label{sec_future_directions}

While there has been much useful work in addressing certain aspects of e-recruitment recommendation systems, there are still some open challenges in this domain that could be investigated in future research works. Some of such challenges that we personally consider promising include:
\begin{itemize}
    \item \textbf{One worker, one job} (OWOJ aspect \ref{sect:e-recruitment_overview}.\ref{job_vs_trad_matching}). Since each job seeker can only be employed for one or a few jobs and a job can be assigned to one or a few candidates, balancing the recommendations in a way that job postings do not receive too many or too few applications is of great importance. Moreover, each job/job seeker should receive recommendations with a high chance of success. This would require the recommendation system to consider the relative probability of matching, that is, how likely one's recommended jobs would be successfully matched with other job seekers. Although some aspects of these issues have been addressed in a few papers (see Section~\ref{sect:specific obj}), this challenge still needs further investigation for more insights and new solutions.
    \item \textbf{Career path recommendation}. Some job seekers choose their next jobs in a way that helps them reach their dream jobs in the future. This problem has been addressed by a few career path recommendation systems, which is to recommend intermediate jobs to reach the final career goal \cite{ghosh2020skill}. This line of research could be investigated in future studies.
    
    \item \textbf{Domain adaptation}. Domain adaptation techniques can improve model performance with limited labeled data, but the application of such techniques in e-recruitment recommendation has not been well investigated except for in a few studies such as \cite{484/bian2019domain}. Methods for domain adaptation between different job sectors, languages, platforms, countries, etc., would be worth investigating to improve the performance of e-recruitment recommendation systems.
    
    \item \textbf{Multi-linguality}. Many platforms/countries have resumes and job postings in multiple languages. Hence, e-recruitment recommendation systems in those platforms/countries should support multiple languages and cross-matching resumes and job postings with different languages. Although some papers have addressed this problem (see Section~\ref{sect:data quality}), further investigations are still in need to provide better support for multi-lingual platforms.

    \item \textbf{Conversational}. Conversational recommendation systems perform multi-turn dialogue with users to achieve recommendation related goals \cite{jannach2021survey}. Although conversational recommendation systems have become more popular in recent years \cite{gao2021advances}, few studies have explored conversational settings in the e-recruitment domain \cite{49/BiancofioreNSNP21,51/BiancofioreNSNP21,52/Mentec0HR21,58/bellini2020guapp}. Conversational recommendation can elicit the current user's preference, provide explanations, make use of the explicit feedback, etc., which makes it valuable to e-recruitment and worthwhile for future studies. \citep{gao2021advances}.

    \item \textbf{Specific job seekers}. Some groups of job seekers may need special attention by e-recruitment recommendation systems. First, user interfaces need to be designed specifically for certain user groups to enhance their interactions with the system (e.g., for people with special needs). This aspect should also be considered for some groups of recruiters. Moreover, some groups of job seekers might be fit for some specific jobs. For example, adults with autism are among the most under employed demographics \citep{115/bills2021looking}. However, they have special skills to contribute to the workplace if applied to the right job \citep{115/bills2021looking}. Although there have been some job recommenders designed for \textit{specific job seekers} such as students and new graduates \cite{88/kenthapadi2017personalized,48/FengJWYL21,77/LiuROX17,90/LiuORSTX16,114/zhao2021summer,160/SchmittGCS17,402/lu2013recommender,522/patel2017capar}, the elderly \cite{156/arita2017gber}, and people with special needs \cite{10/ShishehchiB19,115/bills2021looking}, exploring the needs of more subgroups of job seekers could greatly benefit the e-recruitment field. More specifically, designing a taxonomy of different groups of job seekers with their characteristics and needs would be a good starting point, which could further encourage collecting data for designing recommendation methods that can take the differences between different groups of job seekers into consideration.
        
    \item \textbf{Fairness}. Fair recommendation in e-recruitment is even more important than that in other recommendation systems because people's career choices are influenced by their recommended jobs and the recommendation may also have a long-term impact on the labor market (HS aspect \ref{sect:e-recruitment_overview}.\ref{job_vs_trad_high_stakes}). Although there has been growing attention to fairness issues in general recommendation settings, not many papers specifically address these issues in e-recruitment recommendation systems (as shown in Section~\ref{sect:bias}). One reason could be that the fairness issues are more complicated than the other recommendation systems due to the reciprocal nature and multiple stakeholders involved in e-recruitment. Another reason might be that there are relatively few open datasets for this specific field, as elaborated below.

\end{itemize}

Another challenge in research for e-recruitment recommendation systems is that few public datasets are available. As far as we know, there are only two public datasets: \emph{CareerBuilder 2012 dataset}\footnote{\url{https://www.kaggle.com/c/job-recommendation}} on Kaggle\footnote{\url{https://www.kaggle.com/}} from the e-recruitment platform CareerBuilder\footnote{\url{https://www.careerbuilder.com/}} and \emph{Zhilian dataset}\footnote{\url{https://tianchi.aliyun.com/dataset/dataDetail?dataId=31623}} from a Chinese e-recruitment platform Zhilian\footnote{\url{https://www.zhaopin.com}}. The two datasets for the RecSys challenges 2016 \cite{abel2016recsys} and 2017 \cite{abel2017recsys} provided by the e-recruitment platform Xing\footnote{\url{https://www.xing.com}}, although used in some related studies, are not publicly available. Advances in e-recruitment recommendation systems from academic research depend on the availability of public datasets: more publicly available data could help to establish stronger benchmarks; a larger datasets of variety  could also facilitate new ideas to appear in the field.

\begin{acks}
The research leading to these results has received funding from the European Research Council under the European Union's Seventh Framework Programme (FP7/2007-2013) (ERC Grant Agreement no. 615517), and under the European Union’s Horizon 2020 research and innovation programme (ERC Grant Agreement no. 963924), from the Special Research Fund (BOF) of Ghent University (BOF20/IBF/117), from the Flemish Government under the ``Onderzoeksprogramma Artificiële Intelligentie (AI) Vlaanderen'' programme, and from the FWO (project no. G0F9816N, 3G042220).
\end{acks}

\bibliographystyle{ACM-Reference-Format}
\bibliography{bibliography}


\begin{thebibliography}{144}


\ifx \showCODEN    \undefined \def \showCODEN     #1{\unskip}     \fi
\ifx \showDOI      \undefined \def \showDOI       #1{#1}\fi
\ifx \showISBNx    \undefined \def \showISBNx     #1{\unskip}     \fi
\ifx \showISBNxiii \undefined \def \showISBNxiii  #1{\unskip}     \fi
\ifx \showISSN     \undefined \def \showISSN      #1{\unskip}     \fi
\ifx \showLCCN     \undefined \def \showLCCN      #1{\unskip}     \fi
\ifx \shownote     \undefined \def \shownote      #1{#1}          \fi
\ifx \showarticletitle \undefined \def \showarticletitle #1{#1}   \fi
\ifx \showURL      \undefined \def \showURL       {\relax}        \fi
\providecommand\bibfield[2]{#2}
\providecommand\bibinfo[2]{#2}
\providecommand\natexlab[1]{#1}
\providecommand\showeprint[2][]{arXiv:#2}

\bibitem[Abdollahpouri et~al\mbox{.}(2020a)]%
        {abdollahpouri2020multistakeholder}
\bibfield{author}{\bibinfo{person}{Himan Abdollahpouri},
  \bibinfo{person}{Gediminas Adomavicius}, \bibinfo{person}{Robin Burke},
  \bibinfo{person}{Ido Guy}, \bibinfo{person}{Dietmar Jannach},
  \bibinfo{person}{Toshihiro Kamishima}, \bibinfo{person}{Jan Krasnodebski},
  {and} \bibinfo{person}{Luiz Pizzato}.} \bibinfo{year}{2020}\natexlab{a}.
\newblock \showarticletitle{Multistakeholder recommendation: Survey and
  research directions}.
\newblock \bibinfo{journal}{\emph{User Modeling and User-Adapted Interaction}}
  \bibinfo{volume}{30}, \bibinfo{number}{1} (\bibinfo{year}{2020}),
  \bibinfo{pages}{127--158}.
\newblock


\bibitem[Abdollahpouri et~al\mbox{.}(2020b)]%
        {abdollahpouri2020addressing}
\bibfield{author}{\bibinfo{person}{Himan Abdollahpouri},
  \bibinfo{person}{Masoud Mansoury}, \bibinfo{person}{Robin Burke}, {and}
  \bibinfo{person}{Bamshad Mobasher}.} \bibinfo{year}{2020}\natexlab{b}.
\newblock \showarticletitle{Addressing the multistakeholder impact of
  popularity bias in recommendation through calibration}.
\newblock \bibinfo{journal}{\emph{arXiv preprint arXiv:2007.12230}}
  (\bibinfo{year}{2020}).
\newblock


\bibitem[Abel et~al\mbox{.}(2016)]%
        {abel2016recsys}
\bibfield{author}{\bibinfo{person}{Fabian Abel}, \bibinfo{person}{Andr{\'a}s
  Bencz{\'u}r}, \bibinfo{person}{Daniel Kohlsdorf}, \bibinfo{person}{Martha
  Larson}, {and} \bibinfo{person}{R{\'o}bert P{\'a}lovics}.}
  \bibinfo{year}{2016}\natexlab{}.
\newblock \showarticletitle{Recsys challenge 2016: Job recommendations}. In
  \bibinfo{booktitle}{\emph{Proceedings of the 10th ACM conference on
  recommender systems}}. \bibinfo{pages}{425--426}.
\newblock


\bibitem[Abel et~al\mbox{.}(2017)]%
        {abel2017recsys}
\bibfield{author}{\bibinfo{person}{Fabian Abel}, \bibinfo{person}{Yashar
  Deldjoo}, \bibinfo{person}{Mehdi Elahi}, {and} \bibinfo{person}{Daniel
  Kohlsdorf}.} \bibinfo{year}{2017}\natexlab{}.
\newblock \showarticletitle{Recsys challenge 2017: Offline and online
  evaluation}. In \bibinfo{booktitle}{\emph{Proceedings of the eleventh acm
  conference on recommender systems}}. \bibinfo{pages}{372--373}.
\newblock


\bibitem[Al-Otaibi and Ykhlef(2017)]%
        {21/Al-OtaibiY17}
\bibfield{author}{\bibinfo{person}{Shaha Al-Otaibi} {and}
  \bibinfo{person}{Mourad Ykhlef}.} \bibinfo{year}{2017}\natexlab{}.
\newblock \showarticletitle{Hybrid immunizing solution for job recommender
  system}.
\newblock \bibinfo{journal}{\emph{Frontiers of Computer Science}}
  \bibinfo{volume}{11}, \bibinfo{number}{3} (\bibinfo{year}{2017}),
  \bibinfo{pages}{511--527}.
\newblock


\bibitem[Almalis et~al\mbox{.}(2014a)]%
        {542/almalis2014content}
\bibfield{author}{\bibinfo{person}{Nikolaos~D Almalis},
  \bibinfo{person}{George~A Tsihrintzis}, {and} \bibinfo{person}{Nikolaos
  Karagiannis}.} \bibinfo{year}{2014}\natexlab{a}.
\newblock \showarticletitle{A content based approach for recommending personnel
  for job positions}. In \bibinfo{booktitle}{\emph{IISA 2014, The 5th
  International Conference on Information, Intelligence, Systems and
  Applications}}. IEEE, \bibinfo{pages}{45--49}.
\newblock


\bibitem[Almalis et~al\mbox{.}(2014b)]%
        {518/almalis2014acontent}
\bibfield{author}{\bibinfo{person}{Nikolaos~D Almalis},
  \bibinfo{person}{George~A Tsihrintzis}, {and} \bibinfo{person}{Nikolaos
  Karagiannis}.} \bibinfo{year}{2014}\natexlab{b}.
\newblock \showarticletitle{A content based approach for recommending personnel
  for job positions}. In \bibinfo{booktitle}{\emph{IISA 2014, The 5th
  International Conference on Information, Intelligence, Systems and
  Applications}}. IEEE, \bibinfo{pages}{45--49}.
\newblock


\bibitem[Almalis et~al\mbox{.}(2015)]%
        {46/AlmalisTKS15}
\bibfield{author}{\bibinfo{person}{Nikolaos~D Almalis},
  \bibinfo{person}{George~A Tsihrintzis}, \bibinfo{person}{Nikolaos
  Karagiannis}, {and} \bibinfo{person}{Aggeliki~D Strati}.}
  \bibinfo{year}{2015}\natexlab{}.
\newblock \showarticletitle{FoDRA - {A} new content-based job recommendation
  algorithm for job seeking and recruiting}. In \bibinfo{booktitle}{\emph{2015
  6th International Conference on Information, Intelligence, Systems and
  Applications (IISA)}}. IEEE, \bibinfo{pages}{1--7}.
\newblock


\bibitem[Apaza et~al\mbox{.}(2021)]%
        {517/apaza2021job}
\bibfield{author}{\bibinfo{person}{Honorio Apaza},
  \bibinfo{person}{Am{\'e}rico~Ariel Rubin~de Celis~Vidal}, {and}
  \bibinfo{person}{Josimar~Edinson Chire~Saire}.}
  \bibinfo{year}{2021}\natexlab{}.
\newblock \showarticletitle{Job Recommendation Based on Curriculum Vitae Using
  Text Mining}. In \bibinfo{booktitle}{\emph{Future of Information and
  Communication Conference}}. Springer, \bibinfo{pages}{1051--1059}.
\newblock


\bibitem[Arita et~al\mbox{.}(2017)]%
        {156/arita2017gber}
\bibfield{author}{\bibinfo{person}{Shoma Arita}, \bibinfo{person}{Atsushi
  Hiyama}, {and} \bibinfo{person}{Michitaka Hirose}.}
  \bibinfo{year}{2017}\natexlab{}.
\newblock \showarticletitle{Gber: A social matching app which utilizes time,
  place, and skills of workers and jobs}. In
  \bibinfo{booktitle}{\emph{Companion of the 2017 ACM Conference on Computer
  Supported Cooperative Work and Social Computing}}. \bibinfo{pages}{127--130}.
\newblock


\bibitem[Bansal et~al\mbox{.}(2017)]%
        {84/BansalSA17}
\bibfield{author}{\bibinfo{person}{Shivam Bansal}, \bibinfo{person}{Aman
  Srivastava}, {and} \bibinfo{person}{Anuja Arora}.}
  \bibinfo{year}{2017}\natexlab{}.
\newblock \showarticletitle{Topic modeling driven content based jobs
  recommendation engine for recruitment industry}.
\newblock \bibinfo{journal}{\emph{Procedia computer science}}
  \bibinfo{volume}{122} (\bibinfo{year}{2017}), \bibinfo{pages}{865--872}.
\newblock


\bibitem[Bellini et~al\mbox{.}(2020)]%
        {58/bellini2020guapp}
\bibfield{author}{\bibinfo{person}{Vito Bellini},
  \bibinfo{person}{Giovanni~Maria Biancofiore}, \bibinfo{person}{Tommaso
  Di~Noia}, \bibinfo{person}{Eugenio Di~Sciascio}, \bibinfo{person}{Fedelucio
  Narducci}, {and} \bibinfo{person}{Claudio Pomo}.}
  \bibinfo{year}{2020}\natexlab{}.
\newblock \showarticletitle{GUapp: a conversational agent for job
  recommendation for the Italian public administration}. In
  \bibinfo{booktitle}{\emph{2020 IEEE Conference on Evolving and Adaptive
  Intelligent Systems (EAIS)}}. IEEE, \bibinfo{pages}{1--7}.
\newblock


\bibitem[Bian et~al\mbox{.}(2020)]%
        {527/bian2020learning}
\bibfield{author}{\bibinfo{person}{Shuqing Bian}, \bibinfo{person}{Xu Chen},
  \bibinfo{person}{Wayne~Xin Zhao}, \bibinfo{person}{Kun Zhou},
  \bibinfo{person}{Yupeng Hou}, \bibinfo{person}{Yang Song},
  \bibinfo{person}{Tao Zhang}, {and} \bibinfo{person}{Ji-Rong Wen}.}
  \bibinfo{year}{2020}\natexlab{}.
\newblock \showarticletitle{Learning to match jobs with resumes from sparse
  interaction data using multi-view co-teaching network}. In
  \bibinfo{booktitle}{\emph{Proceedings of the 29th ACM International
  Conference on Information \& Knowledge Management}}. \bibinfo{pages}{65--74}.
\newblock


\bibitem[Bian et~al\mbox{.}(2019)]%
        {484/bian2019domain}
\bibfield{author}{\bibinfo{person}{Shuqing Bian}, \bibinfo{person}{Wayne~Xin
  Zhao}, \bibinfo{person}{Yang Song}, \bibinfo{person}{Tao Zhang}, {and}
  \bibinfo{person}{Ji-Rong Wen}.} \bibinfo{year}{2019}\natexlab{}.
\newblock \showarticletitle{Domain adaptation for person-job fit with
  transferable deep global match network}. In
  \bibinfo{booktitle}{\emph{Proceedings of the 2019 Conference on Empirical
  Methods in Natural Language Processing and the 9th International Joint
  Conference on Natural Language Processing (EMNLP-IJCNLP)}}.
  \bibinfo{pages}{4810--4820}.
\newblock


\bibitem[Bianchi et~al\mbox{.}(2017)]%
        {538/bianchi2017content}
\bibfield{author}{\bibinfo{person}{Mattia Bianchi}, \bibinfo{person}{Federico
  Cesaro}, \bibinfo{person}{Filippo Ciceri}, \bibinfo{person}{Mattia Dagrada},
  \bibinfo{person}{Alberto Gasparin}, \bibinfo{person}{Daniele Grattarola},
  \bibinfo{person}{Ilyas Inajjar}, \bibinfo{person}{Alberto~Maria Metelli},
  {and} \bibinfo{person}{Leonardo Cella}.} \bibinfo{year}{2017}\natexlab{}.
\newblock \showarticletitle{Content-based approaches for cold-start job
  recommendations}.
\newblock In \bibinfo{booktitle}{\emph{Proceedings of the Recommender Systems
  Challenge 2017}}. \bibinfo{pages}{1--5}.
\newblock


\bibitem[Biancofiore et~al\mbox{.}(2021a)]%
        {51/BiancofioreNSNP21}
\bibfield{author}{\bibinfo{person}{Giovanni~Maria Biancofiore},
  \bibinfo{person}{Tommaso Di~Noia}, \bibinfo{person}{Eugenio Di~Sciascio},
  \bibinfo{person}{Fedelucio Narducci}, {and} \bibinfo{person}{Paolo Pastore}.}
  \bibinfo{year}{2021}\natexlab{a}.
\newblock \showarticletitle{Guapp: a knowledge-aware conversational agent for
  job recommendation}. In \bibinfo{booktitle}{\emph{Proceedings of the Joint
  KaRS \& ComplexRec Workshop. CEUR-WS}}.
\newblock


\bibitem[Biancofiore et~al\mbox{.}(2021b)]%
        {49/BiancofioreNSNP21}
\bibfield{author}{\bibinfo{person}{Giovanni~Maria Biancofiore},
  \bibinfo{person}{Tommaso Di~Noia}, \bibinfo{person}{Eugenio Di~Sciascio},
  \bibinfo{person}{Fedelucio Narducci}, {and} \bibinfo{person}{Paolo Pastore}.}
  \bibinfo{year}{2021}\natexlab{b}.
\newblock \showarticletitle{GUapp: Enhancing Job Recommendations with Knowledge
  Graphs.}. In \bibinfo{booktitle}{\emph{Proceedings of the 11th Italian
  Information Retrieval Workshop. CEUR-WS}}.
\newblock


\bibitem[Bills and Ng(2021)]%
        {115/bills2021looking}
\bibfield{author}{\bibinfo{person}{Joseph Bills} {and}
  \bibinfo{person}{Yiu-kai~Dennis Ng}.} \bibinfo{year}{2021}\natexlab{}.
\newblock \showarticletitle{Looking for Jobs? Matching Adults with Autism with
  Potential Employers for Job Opportunities}. In \bibinfo{booktitle}{\emph{25th
  International Database Engineering \& Applications Symposium}}.
  \bibinfo{pages}{212--221}.
\newblock


\bibitem[Bituin et~al\mbox{.}(2020)]%
        {273/BituinAE20}
\bibfield{author}{\bibinfo{person}{Ronie~C Bituin}, \bibinfo{person}{Ronielle~B
  Antonio}, {and} \bibinfo{person}{James~A Esquivel}.}
  \bibinfo{year}{2020}\natexlab{}.
\newblock \showarticletitle{Harmonic Means between TF-IDF and Angle of
  Similarity to Identify Prospective Applicants in a Recruitment Setting}. In
  \bibinfo{booktitle}{\emph{2020 3rd International Conference on Algorithms,
  Computing and Artificial Intelligence}}. \bibinfo{pages}{1--5}.
\newblock


\bibitem[Bollinger et~al\mbox{.}(2012)]%
        {184/BollingerHM12}
\bibfield{author}{\bibinfo{person}{Jacob Bollinger}, \bibinfo{person}{David
  Hardtke}, {and} \bibinfo{person}{Ben Martin}.}
  \bibinfo{year}{2012}\natexlab{}.
\newblock \showarticletitle{Using social data for resume job matching}. In
  \bibinfo{booktitle}{\emph{Proceedings of the 2012 workshop on Data-driven
  user behavioral modelling and mining from social media}}.
  \bibinfo{pages}{27--30}.
\newblock


\bibitem[Borisyuk et~al\mbox{.}(2016)]%
        {533/borisyuk2016casmos}
\bibfield{author}{\bibinfo{person}{Fedor Borisyuk}, \bibinfo{person}{Krishnaram
  Kenthapadi}, \bibinfo{person}{David Stein}, {and} \bibinfo{person}{Bo Zhao}.}
  \bibinfo{year}{2016}\natexlab{}.
\newblock \showarticletitle{CaSMoS: A framework for learning candidate
  selection models over structured queries and documents}. In
  \bibinfo{booktitle}{\emph{Proceedings of the 22nd ACM SIGKDD International
  Conference on Knowledge Discovery and Data Mining}}.
  \bibinfo{pages}{441--450}.
\newblock


\bibitem[Borisyuk et~al\mbox{.}(2017)]%
        {541/borisyuk2017lijar}
\bibfield{author}{\bibinfo{person}{Fedor Borisyuk}, \bibinfo{person}{Liang
  Zhang}, {and} \bibinfo{person}{Krishnaram Kenthapadi}.}
  \bibinfo{year}{2017}\natexlab{}.
\newblock \showarticletitle{LiJAR: A system for job application redistribution
  towards efficient career marketplace}. In
  \bibinfo{booktitle}{\emph{Proceedings of the 23rd ACM SIGKDD International
  Conference on Knowledge Discovery and Data Mining}}.
  \bibinfo{pages}{1397--1406}.
\newblock


\bibitem[Boukari et~al\mbox{.}(2020)]%
        {501/boukari2020huntalent}
\bibfield{author}{\bibinfo{person}{Shayma Boukari}, \bibinfo{person}{Sondes
  Fayech}, {and} \bibinfo{person}{Rim Faiz}.} \bibinfo{year}{2020}\natexlab{}.
\newblock \showarticletitle{Huntalent: A candidates recommendation system for
  automatic recruitment via LinkedIn}. In \bibinfo{booktitle}{\emph{2020
  Seventh International Conference on Social Networks Analysis, Management and
  Security (SNAMS)}}. IEEE, \bibinfo{pages}{1--7}.
\newblock


\bibitem[Broder et~al\mbox{.}(2003)]%
        {broder2003efficient}
\bibfield{author}{\bibinfo{person}{Andrei~Z Broder}, \bibinfo{person}{David
  Carmel}, \bibinfo{person}{Michael Herscovici}, \bibinfo{person}{Aya Soffer},
  {and} \bibinfo{person}{Jason Zien}.} \bibinfo{year}{2003}\natexlab{}.
\newblock \showarticletitle{Efficient query evaluation using a two-level
  retrieval process}. In \bibinfo{booktitle}{\emph{Proceedings of the twelfth
  international conference on Information and knowledge management}}.
  \bibinfo{pages}{426--434}.
\newblock


\bibitem[Cardoso et~al\mbox{.}(2021)]%
        {528/cardoso2021matching}
\bibfield{author}{\bibinfo{person}{Alan Cardoso}, \bibinfo{person}{Fernando
  Mour{\~a}o}, {and} \bibinfo{person}{Leonardo Rocha}.}
  \bibinfo{year}{2021}\natexlab{}.
\newblock \showarticletitle{The matching scarcity problem: When recommenders do
  not connect the edges in recruitment services}.
\newblock \bibinfo{journal}{\emph{Expert Systems with Applications}}
  \bibinfo{volume}{175} (\bibinfo{year}{2021}), \bibinfo{pages}{114764}.
\newblock


\bibitem[Carpi et~al\mbox{.}(2016)]%
        {94/CarpiEKSCPQ16}
\bibfield{author}{\bibinfo{person}{Tommaso Carpi}, \bibinfo{person}{Marco
  Edemanti}, \bibinfo{person}{Ervin Kamberoski}, \bibinfo{person}{Elena
  Sacchi}, \bibinfo{person}{Paolo Cremonesi}, \bibinfo{person}{Roberto Pagano},
  {and} \bibinfo{person}{Massimo Quadrana}.} \bibinfo{year}{2016}\natexlab{}.
\newblock \showarticletitle{Multi-stack ensemble for job recommendation}.
\newblock In \bibinfo{booktitle}{\emph{Proceedings of the Recommender Systems
  Challenge}}. \bibinfo{pages}{1--4}.
\newblock


\bibitem[Chala and Fathi(2017)]%
        {159/ChalaF17}
\bibfield{author}{\bibinfo{person}{Sisay Chala} {and} \bibinfo{person}{Madjid
  Fathi}.} \bibinfo{year}{2017}\natexlab{}.
\newblock \showarticletitle{Job seeker to vacancy matching using social network
  analysis}. In \bibinfo{booktitle}{\emph{2017 IEEE International Conference on
  Industrial Technology (ICIT)}}. IEEE, \bibinfo{pages}{1250--1255}.
\newblock


\bibitem[Chen et~al\mbox{.}(2019)]%
        {64/chen2019correcting}
\bibfield{author}{\bibinfo{person}{Ruey-Cheng Chen}, \bibinfo{person}{Qingyao
  Ai}, \bibinfo{person}{Gaya Jayasinghe}, {and} \bibinfo{person}{W~Bruce
  Croft}.} \bibinfo{year}{2019}\natexlab{}.
\newblock \showarticletitle{Correcting for recency bias in job recommendation}.
  In \bibinfo{booktitle}{\emph{Proceedings of the 28th ACM International
  Conference on Information and Knowledge Management}}.
  \bibinfo{pages}{2185--2188}.
\newblock


\bibitem[Chen et~al\mbox{.}(2018)]%
        {506/ChenZDGHWW18}
\bibfield{author}{\bibinfo{person}{Wenbo Chen}, \bibinfo{person}{Pan Zhou},
  \bibinfo{person}{Shaokang Dong}, \bibinfo{person}{Shimin Gong},
  \bibinfo{person}{Menglan Hu}, \bibinfo{person}{Kehao Wang}, {and}
  \bibinfo{person}{Dapeng Wu}.} \bibinfo{year}{2018}\natexlab{}.
\newblock \showarticletitle{Tree-based contextual learning for online job or
  candidate recommendation with big data support in professional social
  networks}.
\newblock \bibinfo{journal}{\emph{IEEE Access}}  \bibinfo{volume}{6}
  (\bibinfo{year}{2018}), \bibinfo{pages}{77725--77739}.
\newblock


\bibitem[Chenni et~al\mbox{.}(2015)]%
        {332/chenni2015content}
\bibfield{author}{\bibinfo{person}{Oualid Chenni}, \bibinfo{person}{Yanis
  Bouda}, \bibinfo{person}{Hamid Benachour}, {and} \bibinfo{person}{Chahnez
  Zakaria}.} \bibinfo{year}{2015}\natexlab{}.
\newblock \showarticletitle{A content-based recommendation approach using
  semantic user profile in e-recruitment}. In
  \bibinfo{booktitle}{\emph{International Conference on Theory and Practice of
  Natural Computing}}. Springer, \bibinfo{pages}{23--32}.
\newblock


\bibitem[Coelho et~al\mbox{.}(2015)]%
        {106/coelho2015hyred}
\bibfield{author}{\bibinfo{person}{Bruno Coelho}, \bibinfo{person}{Fernando
  Costa}, {and} \bibinfo{person}{Gil~M Gon{\c{c}}alves}.}
  \bibinfo{year}{2015}\natexlab{}.
\newblock \showarticletitle{Hyred: hybrid job recommendation system}. In
  \bibinfo{booktitle}{\emph{2015 12th International Joint Conference on
  e-Business and Telecommunications (ICETE)}}, Vol.~\bibinfo{volume}{2}. IEEE,
  \bibinfo{pages}{29--38}.
\newblock


\bibitem[{Council of European Union}(2022)]%
        {eu-aiact-2022}
\bibfield{author}{\bibinfo{person}{{Council of European Union}}.}
  \bibinfo{year}{2022}\natexlab{}.
\newblock \bibinfo{title}{Proposal for a REGULATION OF THE EUROPEAN PARLIAMENT
  AND OF THE COUNCIL LAYING DOWN HARMONISED RULES ON ARTIFICIAL INTELLIGENCE
  (ARTIFICIAL INTELLIGENCE ACT) AND AMENDING CERTAIN UNION LEGISLATIVE
  ACTS2014}.
\newblock
\newblock
\urldef\tempurl%
\url{https://eur-lex.europa.eu/legal-content/EN/TXT/?qid=1623335154975&uri=CELEX%3A52021PC0206}
\showURL{%
\tempurl}


\bibitem[Dave et~al\mbox{.}(2018)]%
        {74/dave2018combined}
\bibfield{author}{\bibinfo{person}{Vachik~S Dave}, \bibinfo{person}{Baichuan
  Zhang}, \bibinfo{person}{Mohammad Al~Hasan}, \bibinfo{person}{Khalifeh
  AlJadda}, {and} \bibinfo{person}{Mohammed Korayem}.}
  \bibinfo{year}{2018}\natexlab{}.
\newblock \showarticletitle{A combined representation learning approach for
  better job and skill recommendation}. In
  \bibinfo{booktitle}{\emph{Proceedings of the 27th ACM International
  Conference on Information and Knowledge Management}}.
  \bibinfo{pages}{1997--2005}.
\newblock


\bibitem[De~Pessemier et~al\mbox{.}(2016)]%
        {98/PessemierVM16}
\bibfield{author}{\bibinfo{person}{Toon De~Pessemier}, \bibinfo{person}{Kris
  Vanhecke}, {and} \bibinfo{person}{Luc Martens}.}
  \bibinfo{year}{2016}\natexlab{}.
\newblock \showarticletitle{A scalable, high-performance Algorithm for hybrid
  job recommendations}.
\newblock In \bibinfo{booktitle}{\emph{Proceedings of the Recommender Systems
  Challenge}}. \bibinfo{pages}{1--4}.
\newblock


\bibitem[de~Ruijt and Bhulai(2021)]%
        {7/deRuijt2021}
\bibfield{author}{\bibinfo{person}{Corn{\'e} de Ruijt} {and}
  \bibinfo{person}{Sandjai Bhulai}.} \bibinfo{year}{2021}\natexlab{}.
\newblock \showarticletitle{Job recommender systems: A review}.
\newblock \bibinfo{journal}{\emph{arXiv preprint arXiv:2111.13576}}
  (\bibinfo{year}{2021}).
\newblock


\bibitem[Diaby and Viennet(2014)]%
        {29/DiabyV14}
\bibfield{author}{\bibinfo{person}{Mamadou Diaby} {and}
  \bibinfo{person}{Emmanuel Viennet}.} \bibinfo{year}{2014}\natexlab{}.
\newblock \showarticletitle{Taxonomy-based job recommender systems on Facebook
  and LinkedIn profiles}. In \bibinfo{booktitle}{\emph{2014 IEEE Eighth
  International Conference on Research Challenges in Information Science
  (RCIS)}}. IEEE, \bibinfo{pages}{1--6}.
\newblock


\bibitem[Diaby et~al\mbox{.}(2013)]%
        {454/diaby2013toward}
\bibfield{author}{\bibinfo{person}{Mamadou Diaby}, \bibinfo{person}{Emmanuel
  Viennet}, {and} \bibinfo{person}{Tristan Launay}.}
  \bibinfo{year}{2013}\natexlab{}.
\newblock \showarticletitle{Toward the next generation of recruitment tools: an
  online social network-based job recommender system}. In
  \bibinfo{booktitle}{\emph{2013 IEEE/ACM International Conference on Advances
  in Social Networks Analysis and Mining (ASONAM 2013)}}. IEEE,
  \bibinfo{pages}{821--828}.
\newblock


\bibitem[Diaby et~al\mbox{.}(2014)]%
        {27/DiabyVL14}
\bibfield{author}{\bibinfo{person}{Mamadou Diaby}, \bibinfo{person}{Emmanuel
  Viennet}, {and} \bibinfo{person}{Tristan Launay}.}
  \bibinfo{year}{2014}\natexlab{}.
\newblock \showarticletitle{Exploration of methodologies to improve job
  recommender systems on social networks}.
\newblock \bibinfo{journal}{\emph{Social Network Analysis and Mining}}
  \bibinfo{volume}{4}, \bibinfo{number}{1} (\bibinfo{year}{2014}),
  \bibinfo{pages}{1--17}.
\newblock


\bibitem[Domeniconi et~al\mbox{.}(2016)]%
        {91/DomeniconiMPPP16}
\bibfield{author}{\bibinfo{person}{Giacomo Domeniconi},
  \bibinfo{person}{Gianluca Moro}, \bibinfo{person}{Andrea Pagliarani},
  \bibinfo{person}{Karin Pasini}, {and} \bibinfo{person}{Roberto Pasolini}.}
  \bibinfo{year}{2016}\natexlab{}.
\newblock \showarticletitle{Job recommendation from semantic similarity of
  linkedin users’ skills}. In \bibinfo{booktitle}{\emph{International
  Conference on Pattern Recognition Applications and Methods}},
  Vol.~\bibinfo{volume}{2}. SciTePress, \bibinfo{pages}{270--277}.
\newblock


\bibitem[Dong et~al\mbox{.}(2017)]%
        {509/DongLZBL17}
\bibfield{author}{\bibinfo{person}{Shaokang Dong}, \bibinfo{person}{Zijian
  Lei}, \bibinfo{person}{Pan Zhou}, \bibinfo{person}{Kaigui Bian}, {and}
  \bibinfo{person}{Guanghui Liu}.} \bibinfo{year}{2017}\natexlab{}.
\newblock \showarticletitle{Job and candidate recommendation with big data
  support: a contextual online learning approach}. In
  \bibinfo{booktitle}{\emph{GLOBECOM 2017-2017 IEEE Global Communications
  Conference}}. IEEE, \bibinfo{pages}{1--7}.
\newblock


\bibitem[Eitle et~al\mbox{.}(2021)]%
        {390/eitle2021impact}
\bibfield{author}{\bibinfo{person}{Verena Eitle}, \bibinfo{person}{Felix
  Peters}, \bibinfo{person}{Andreas Welsch}, {and} \bibinfo{person}{Peter
  Buxmann}.} \bibinfo{year}{2021}\natexlab{}.
\newblock \showarticletitle{The Impact of CV Recommender Systems on Procedural
  Justice in Recruiting: An Experiment in Candidate Selection}.
\newblock  (\bibinfo{year}{2021}).
\newblock


\bibitem[Elgammal et~al\mbox{.}(2021)]%
        {129/elgammal2021matching}
\bibfield{author}{\bibinfo{person}{Ziad Elgammal}, \bibinfo{person}{Abdullah
  Barmu}, \bibinfo{person}{Hamza Hassan}, \bibinfo{person}{Khaled Elgammal},
  \bibinfo{person}{Tansel {\"O}zyer}, {and} \bibinfo{person}{Reda Alhajj}.}
  \bibinfo{year}{2021}\natexlab{}.
\newblock \showarticletitle{Matching Applicants with Positions for Better
  Allocation of Employees in the Job Market}. In \bibinfo{booktitle}{\emph{2021
  22nd International Arab Conference on Information Technology (ACIT)}}. IEEE,
  \bibinfo{pages}{1--5}.
\newblock


\bibitem[Elsafty et~al\mbox{.}(2018)]%
        {19/ElsaftyRB18}
\bibfield{author}{\bibinfo{person}{Ahmed Elsafty}, \bibinfo{person}{Martin
  Riedl}, {and} \bibinfo{person}{Chris Biemann}.}
  \bibinfo{year}{2018}\natexlab{}.
\newblock \showarticletitle{Document-based recommender system for job postings
  using dense representations}. In \bibinfo{booktitle}{\emph{Proceedings of the
  2018 Conference of the North American Chapter of the Association for
  Computational Linguistics: Human Language Technologies, Volume 3 (Industry
  Papers)}}. \bibinfo{pages}{216--224}.
\newblock


\bibitem[Faliagka et~al\mbox{.}(2014)]%
        {211/FaliagkaIKRSTT14}
\bibfield{author}{\bibinfo{person}{Evanthia Faliagka}, \bibinfo{person}{Lazaros
  Iliadis}, \bibinfo{person}{Ioannis Karydis}, \bibinfo{person}{Maria Rigou},
  \bibinfo{person}{Spyros Sioutas}, \bibinfo{person}{Athanasios Tsakalidis},
  {and} \bibinfo{person}{Giannis Tzimas}.} \bibinfo{year}{2014}\natexlab{}.
\newblock \showarticletitle{On-line consistent ranking on e-recruitment:
  seeking the truth behind a well-formed CV}.
\newblock \bibinfo{journal}{\emph{Artificial Intelligence Review}}
  \bibinfo{volume}{42}, \bibinfo{number}{3} (\bibinfo{year}{2014}),
  \bibinfo{pages}{515--528}.
\newblock


\bibitem[Faliagka et~al\mbox{.}(2012)]%
        {212/FaliagkaTT12}
\bibfield{author}{\bibinfo{person}{Evanthia Faliagka},
  \bibinfo{person}{Athanasios Tsakalidis}, {and} \bibinfo{person}{Giannis
  Tzimas}.} \bibinfo{year}{2012}\natexlab{}.
\newblock \showarticletitle{An integrated e-recruitment system for automated
  personality mining and applicant ranking}.
\newblock \bibinfo{journal}{\emph{Internet research}} (\bibinfo{year}{2012}).
\newblock


\bibitem[Feng et~al\mbox{.}(2021)]%
        {48/FengJWYL21}
\bibfield{author}{\bibinfo{person}{Peini Feng}, \bibinfo{person}{Charles
  Jiahao~Jiang}, \bibinfo{person}{Jiale Wang}, \bibinfo{person}{Sunny Yeung},
  {and} \bibinfo{person}{Xijie Li}.} \bibinfo{year}{2021}\natexlab{}.
\newblock \showarticletitle{Job Recommendation System Based on Analytic
  Hierarchy Process and K-means Clustering}. In \bibinfo{booktitle}{\emph{2021
  The 13th International Conference on Computer Modeling and Simulation}}.
  \bibinfo{pages}{104--113}.
\newblock


\bibitem[Fern{\'a}ndez-Reyes and Shinde(2019)]%
        {142/Fernandez-Reyes19}
\bibfield{author}{\bibinfo{person}{Francis~C Fern{\'a}ndez-Reyes} {and}
  \bibinfo{person}{Suraj Shinde}.} \bibinfo{year}{2019}\natexlab{}.
\newblock \showarticletitle{CV Retrieval System based on job description
  matching using hybrid word embeddings}.
\newblock \bibinfo{journal}{\emph{Computer Speech \& Language}}
  \bibinfo{volume}{56} (\bibinfo{year}{2019}), \bibinfo{pages}{73--79}.
\newblock


\bibitem[Freire and Castro(2020)]%
        {277/freire2020framework}
\bibfield{author}{\bibinfo{person}{Mauricio~Noris Freire} {and}
  \bibinfo{person}{Leandro Nunes~de Castro}.} \bibinfo{year}{2020}\natexlab{}.
\newblock \showarticletitle{A Framework for e-Recruitment Recommender Systems}.
  In \bibinfo{booktitle}{\emph{International Conference on Artificial
  Intelligence and Soft Computing}}. Springer, \bibinfo{pages}{165--175}.
\newblock


\bibitem[Freire and de~Castro(2021)]%
        {259/freire2021recruitment}
\bibfield{author}{\bibinfo{person}{Mauricio~Noris Freire} {and}
  \bibinfo{person}{Leandro~Nunes de Castro}.} \bibinfo{year}{2021}\natexlab{}.
\newblock \showarticletitle{e-Recruitment recommender systems: a systematic
  review}.
\newblock \bibinfo{journal}{\emph{Knowledge and Information Systems}}
  \bibinfo{volume}{63}, \bibinfo{number}{1} (\bibinfo{year}{2021}),
  \bibinfo{pages}{1--20}.
\newblock


\bibitem[Fu et~al\mbox{.}(2021)]%
        {131/fu2021beyond}
\bibfield{author}{\bibinfo{person}{Bin Fu}, \bibinfo{person}{Hongzhi Liu},
  \bibinfo{person}{Yao Zhu}, \bibinfo{person}{Yang Song}, \bibinfo{person}{Tao
  Zhang}, {and} \bibinfo{person}{Zhonghai Wu}.}
  \bibinfo{year}{2021}\natexlab{}.
\newblock \showarticletitle{Beyond Matching: Modeling Two-Sided
  Multi-Behavioral Sequences for Dynamic Person-Job Fit}. In
  \bibinfo{booktitle}{\emph{International Conference on Database Systems for
  Advanced Applications}}. Springer, \bibinfo{pages}{359--375}.
\newblock


\bibitem[Gale and Shapley(2013)]%
        {gale2013college}
\bibfield{author}{\bibinfo{person}{David Gale} {and} \bibinfo{person}{Lloyd~S
  Shapley}.} \bibinfo{year}{2013}\natexlab{}.
\newblock \showarticletitle{College admissions and the stability of marriage}.
\newblock \bibinfo{journal}{\emph{The American Mathematical Monthly}}
  \bibinfo{volume}{120}, \bibinfo{number}{5} (\bibinfo{year}{2013}),
  \bibinfo{pages}{386--391}.
\newblock


\bibitem[Gao et~al\mbox{.}(2021)]%
        {gao2021advances}
\bibfield{author}{\bibinfo{person}{Chongming Gao}, \bibinfo{person}{Wenqiang
  Lei}, \bibinfo{person}{Xiangnan He}, \bibinfo{person}{Maarten de Rijke},
  {and} \bibinfo{person}{Tat-Seng Chua}.} \bibinfo{year}{2021}\natexlab{}.
\newblock \showarticletitle{Advances and challenges in conversational
  recommender systems: A survey}.
\newblock \bibinfo{journal}{\emph{AI Open}}  \bibinfo{volume}{2}
  (\bibinfo{year}{2021}), \bibinfo{pages}{100--126}.
\newblock


\bibitem[Geyik et~al\mbox{.}(2019)]%
        {546/geyik2019fairness}
\bibfield{author}{\bibinfo{person}{Sahin~Cem Geyik}, \bibinfo{person}{Stuart
  Ambler}, {and} \bibinfo{person}{Krishnaram Kenthapadi}.}
  \bibinfo{year}{2019}\natexlab{}.
\newblock \showarticletitle{Fairness-aware ranking in search \& recommendation
  systems with application to linkedin talent search}. In
  \bibinfo{booktitle}{\emph{Proceedings of the 25th acm sigkdd international
  conference on knowledge discovery \& data mining}}.
  \bibinfo{pages}{2221--2231}.
\newblock


\bibitem[Geyik et~al\mbox{.}(2018)]%
        {530/geyik2018talent}
\bibfield{author}{\bibinfo{person}{Sahin~Cem Geyik}, \bibinfo{person}{Qi Guo},
  \bibinfo{person}{Bo Hu}, \bibinfo{person}{Cagri Ozcaglar},
  \bibinfo{person}{Ketan Thakkar}, \bibinfo{person}{Xianren Wu}, {and}
  \bibinfo{person}{Krishnaram Kenthapadi}.} \bibinfo{year}{2018}\natexlab{}.
\newblock \showarticletitle{Talent search and recommendation systems at
  LinkedIn: Practical challenges and lessons learned}. In
  \bibinfo{booktitle}{\emph{The 41st International ACM SIGIR Conference on
  Research \& Development in Information Retrieval}}.
  \bibinfo{pages}{1353--1354}.
\newblock


\bibitem[Ghosh et~al\mbox{.}(2020)]%
        {ghosh2020skill}
\bibfield{author}{\bibinfo{person}{Aritra Ghosh}, \bibinfo{person}{Beverly
  Woolf}, \bibinfo{person}{Shlomo Zilberstein}, {and} \bibinfo{person}{Andrew
  Lan}.} \bibinfo{year}{2020}\natexlab{}.
\newblock \showarticletitle{Skill-based Career Path Modeling and
  Recommendation}. In \bibinfo{booktitle}{\emph{2020 IEEE International
  Conference on Big Data (Big Data)}}. IEEE, \bibinfo{pages}{1156--1165}.
\newblock


\bibitem[Gonz{\'a}lez-Briones et~al\mbox{.}(2018)]%
        {20/Gonzalez-Briones18}
\bibfield{author}{\bibinfo{person}{Alfonso Gonz{\'a}lez-Briones},
  \bibinfo{person}{Alberto Rivas}, \bibinfo{person}{Pablo Chamoso},
  \bibinfo{person}{Roberto Casado-Vara}, {and} \bibinfo{person}{Juan~Manuel
  Corchado}.} \bibinfo{year}{2018}\natexlab{}.
\newblock \showarticletitle{Case-based reasoning and agent based job offer
  recommender system}. In \bibinfo{booktitle}{\emph{The 13th International
  Conference on Soft Computing Models in Industrial and Environmental
  Applications}}. Springer, \bibinfo{pages}{21--33}.
\newblock


\bibitem[Gugnani and Misra(2020)]%
        {57/GugnaniM20}
\bibfield{author}{\bibinfo{person}{Akshay Gugnani} {and}
  \bibinfo{person}{Hemant Misra}.} \bibinfo{year}{2020}\natexlab{}.
\newblock \showarticletitle{Implicit skills extraction using document embedding
  and its use in job recommendation}. In \bibinfo{booktitle}{\emph{Proceedings
  of the AAAI Conference on Artificial Intelligence}},
  Vol.~\bibinfo{volume}{34}. \bibinfo{pages}{13286--13293}.
\newblock


\bibitem[Guo et~al\mbox{.}(2017)]%
        {535/guo2017howinte}
\bibfield{author}{\bibinfo{person}{Cheng Guo}, \bibinfo{person}{Hongyu Lu},
  \bibinfo{person}{Shaoyun Shi}, \bibinfo{person}{Bin Hao},
  \bibinfo{person}{Bin Liu}, \bibinfo{person}{Min Zhang},
  \bibinfo{person}{Yiqun Liu}, {and} \bibinfo{person}{Shaoping Ma}.}
  \bibinfo{year}{2017}\natexlab{}.
\newblock \showarticletitle{How integration helps on cold-start
  recommendations}.
\newblock In \bibinfo{booktitle}{\emph{Proceedings of the Recommender Systems
  Challenge 2017}}. \bibinfo{pages}{1--6}.
\newblock


\bibitem[Guo et~al\mbox{.}(2016)]%
        {164/GuoAH16}
\bibfield{author}{\bibinfo{person}{Shiqiang Guo}, \bibinfo{person}{Folami
  Alamudun}, {and} \bibinfo{person}{Tracy Hammond}.}
  \bibinfo{year}{2016}\natexlab{}.
\newblock \showarticletitle{R{\'e}suMatcher: A personalized r{\'e}sum{\'e}-job
  matching system}.
\newblock \bibinfo{journal}{\emph{Expert Systems with Applications}}
  \bibinfo{volume}{60} (\bibinfo{year}{2016}), \bibinfo{pages}{169--182}.
\newblock


\bibitem[Gupta and Garg(2014)]%
        {3/GuptaG14}
\bibfield{author}{\bibinfo{person}{Anika Gupta} {and} \bibinfo{person}{Deepak
  Garg}.} \bibinfo{year}{2014}\natexlab{}.
\newblock \showarticletitle{Applying data mining techniques in job recommender
  system for considering candidate job preferences}. In
  \bibinfo{booktitle}{\emph{2014 International Conference on Advances in
  Computing, Communications and Informatics (ICACCI)}}. IEEE,
  \bibinfo{pages}{1458--1465}.
\newblock


\bibitem[Guti{\'e}rrez et~al\mbox{.}(2019)]%
        {39/GutierrezCCHGV19}
\bibfield{author}{\bibinfo{person}{Francisco Guti{\'e}rrez},
  \bibinfo{person}{Sven Charleer}, \bibinfo{person}{Robin De~Croon},
  \bibinfo{person}{Nyi~Nyi Htun}, \bibinfo{person}{Gerd Goetschalckx}, {and}
  \bibinfo{person}{Katrien Verbert}.} \bibinfo{year}{2019}\natexlab{}.
\newblock \showarticletitle{Explaining and exploring job recommendations: a
  user-driven approach for interacting with knowledge-based job recommender
  systems}. In \bibinfo{booktitle}{\emph{Proceedings of the 13th ACM Conference
  on Recommender Systems}}. \bibinfo{pages}{60--68}.
\newblock


\bibitem[Habous and Nfaoui(2021)]%
        {126/habous2021fuzzy}
\bibfield{author}{\bibinfo{person}{Amine Habous} {and}
  \bibinfo{person}{El~Habib Nfaoui}.} \bibinfo{year}{2021}\natexlab{}.
\newblock \showarticletitle{A fuzzy logic and ontology-based approach for
  improving the CV and job offer matching in recruitment process}.
\newblock \bibinfo{journal}{\emph{International Journal of Metadata, Semantics
  and Ontologies}} \bibinfo{volume}{15}, \bibinfo{number}{2}
  (\bibinfo{year}{2021}), \bibinfo{pages}{104--120}.
\newblock


\bibitem[Hauff and Gousios(2015)]%
        {174/HauffG15}
\bibfield{author}{\bibinfo{person}{Claudia Hauff} {and}
  \bibinfo{person}{Georgios Gousios}.} \bibinfo{year}{2015}\natexlab{}.
\newblock \showarticletitle{Matching GitHub developer profiles to job
  advertisements}. In \bibinfo{booktitle}{\emph{2015 IEEE/ACM 12th Working
  Conference on Mining Software Repositories}}. IEEE,
  \bibinfo{pages}{362--366}.
\newblock


\bibitem[He et~al\mbox{.}(2021a)]%
        {525/he2021self}
\bibfield{author}{\bibinfo{person}{Miao He}, \bibinfo{person}{Dayong Shen},
  \bibinfo{person}{Tao Wang}, \bibinfo{person}{Hua Zhao},
  \bibinfo{person}{Zhongshan Zhang}, {and} \bibinfo{person}{Renjie He}.}
  \bibinfo{year}{2021}\natexlab{a}.
\newblock \showarticletitle{Self-Attentional Multi-Field Features
  Representation and Interaction Learning for Person-Job Fit}.
\newblock \bibinfo{journal}{\emph{IEEE Transactions on Computational Social
  Systems}} (\bibinfo{year}{2021}).
\newblock


\bibitem[He et~al\mbox{.}(2021b)]%
        {526/he2021finn}
\bibfield{author}{\bibinfo{person}{Miao He}, \bibinfo{person}{Tao Wang},
  \bibinfo{person}{Yuanyuan Zhu}, \bibinfo{person}{Yingguo Chen},
  \bibinfo{person}{Feng Yao}, {and} \bibinfo{person}{Ning Wang}.}
  \bibinfo{year}{2021}\natexlab{b}.
\newblock \showarticletitle{FINN: Feature Interaction Neural Network for
  Person-Job Fit}. In \bibinfo{booktitle}{\emph{2021 7th International
  Conference on Big Data and Information Analytics (BigDIA)}}. IEEE,
  \bibinfo{pages}{123--130}.
\newblock


\bibitem[Heap et~al\mbox{.}(2014)]%
        {178/HeapKWBC14}
\bibfield{author}{\bibinfo{person}{Bradford Heap}, \bibinfo{person}{Alfred
  Krzywicki}, \bibinfo{person}{Wayne Wobcke}, \bibinfo{person}{Mike Bain},
  {and} \bibinfo{person}{Paul Compton}.} \bibinfo{year}{2014}\natexlab{}.
\newblock \showarticletitle{Combining career progression and profile matching
  in a job recommender system}. In \bibinfo{booktitle}{\emph{Pacific Rim
  International Conference on Artificial Intelligence}}. Springer,
  \bibinfo{pages}{396--408}.
\newblock


\bibitem[Heggo and Abdelbaki(2018)]%
        {18/HeggoA18}
\bibfield{author}{\bibinfo{person}{Islam~A Heggo} {and} \bibinfo{person}{Nashwa
  Abdelbaki}.} \bibinfo{year}{2018}\natexlab{}.
\newblock \showarticletitle{Hybrid information filtering engine for
  personalized job recommender system}. In
  \bibinfo{booktitle}{\emph{International Conference on Advanced Machine
  Learning Technologies and Applications}}. Springer,
  \bibinfo{pages}{553--563}.
\newblock


\bibitem[Hong et~al\mbox{.}(2013a)]%
        {519/hong2013dynamic}
\bibfield{author}{\bibinfo{person}{Wenxing Hong}, \bibinfo{person}{Siting
  Zheng}, {and} \bibinfo{person}{Huan Wang}.} \bibinfo{year}{2013}\natexlab{a}.
\newblock \showarticletitle{Dynamic user profile-based job recommender system}.
  In \bibinfo{booktitle}{\emph{2013 8th International Conference on Computer
  Science \& Education}}. IEEE, \bibinfo{pages}{1499--1503}.
\newblock


\bibitem[Hong et~al\mbox{.}(2013b)]%
        {32/HongZWS13}
\bibfield{author}{\bibinfo{person}{Wenxing Hong}, \bibinfo{person}{Siting
  Zheng}, \bibinfo{person}{Huan Wang}, {and} \bibinfo{person}{Jianchao Shi}.}
  \bibinfo{year}{2013}\natexlab{b}.
\newblock \showarticletitle{A job recommender system based on user clustering.}
\newblock \bibinfo{journal}{\emph{J. Comput.}} \bibinfo{volume}{8},
  \bibinfo{number}{8} (\bibinfo{year}{2013}), \bibinfo{pages}{1960--1967}.
\newblock


\bibitem[Islam et~al\mbox{.}(2021)]%
        {531/islam2021debias}
\bibfield{author}{\bibinfo{person}{Rashidul Islam},
  \bibinfo{person}{Kamrun~Naher Keya}, \bibinfo{person}{Ziqian Zeng},
  \bibinfo{person}{Shimei Pan}, {and} \bibinfo{person}{James Foulds}.}
  \bibinfo{year}{2021}\natexlab{}.
\newblock \showarticletitle{Debiasing career recommendations with neural fair
  collaborative filtering}. In \bibinfo{booktitle}{\emph{Proceedings of the Web
  Conference 2021}}. \bibinfo{pages}{3779--3790}.
\newblock


\bibitem[Jannach et~al\mbox{.}(2021)]%
        {jannach2021survey}
\bibfield{author}{\bibinfo{person}{Dietmar Jannach}, \bibinfo{person}{Ahtsham
  Manzoor}, \bibinfo{person}{Wanling Cai}, {and} \bibinfo{person}{Li Chen}.}
  \bibinfo{year}{2021}\natexlab{}.
\newblock \showarticletitle{A survey on conversational recommender systems}.
\newblock \bibinfo{journal}{\emph{ACM Computing Surveys (CSUR)}}
  \bibinfo{volume}{54}, \bibinfo{number}{5} (\bibinfo{year}{2021}),
  \bibinfo{pages}{1--36}.
\newblock


\bibitem[Jiang et~al\mbox{.}(2020)]%
        {479/jiang2020learning}
\bibfield{author}{\bibinfo{person}{Junshu Jiang}, \bibinfo{person}{Songyun Ye},
  \bibinfo{person}{Wei Wang}, \bibinfo{person}{Jingran Xu}, {and}
  \bibinfo{person}{Xiaosheng Luo}.} \bibinfo{year}{2020}\natexlab{}.
\newblock \showarticletitle{Learning effective representations for person-job
  fit by feature fusion}. In \bibinfo{booktitle}{\emph{Proceedings of the 29th
  ACM International Conference on Information \& Knowledge Management}}.
  \bibinfo{pages}{2549--2556}.
\newblock


\bibitem[Jiang et~al\mbox{.}(2019)]%
        {63/jiang2019user}
\bibfield{author}{\bibinfo{person}{Miao Jiang}, \bibinfo{person}{Yi Fang},
  \bibinfo{person}{Huangming Xie}, \bibinfo{person}{Jike Chong}, {and}
  \bibinfo{person}{Meng Meng}.} \bibinfo{year}{2019}\natexlab{}.
\newblock \showarticletitle{User click prediction for personalized job
  recommendation}.
\newblock \bibinfo{journal}{\emph{World Wide Web}} \bibinfo{volume}{22},
  \bibinfo{number}{1} (\bibinfo{year}{2019}), \bibinfo{pages}{325--345}.
\newblock


\bibitem[Johnson et~al\mbox{.}(2019)]%
        {johnson2019billion}
\bibfield{author}{\bibinfo{person}{Jeff Johnson}, \bibinfo{person}{Matthijs
  Douze}, {and} \bibinfo{person}{Herv{\'e} J{\'e}gou}.}
  \bibinfo{year}{2019}\natexlab{}.
\newblock \showarticletitle{Billion-scale similarity search with gpus}.
\newblock \bibinfo{journal}{\emph{IEEE Transactions on Big Data}}
  \bibinfo{volume}{7}, \bibinfo{number}{3} (\bibinfo{year}{2019}),
  \bibinfo{pages}{535--547}.
\newblock


\bibitem[Kenthapadi et~al\mbox{.}(2017a)]%
        {88/kenthapadi2017personalized}
\bibfield{author}{\bibinfo{person}{Krishnaram Kenthapadi},
  \bibinfo{person}{Benjamin Le}, {and} \bibinfo{person}{Ganesh Venkataraman}.}
  \bibinfo{year}{2017}\natexlab{a}.
\newblock \showarticletitle{Personalized job recommendation system at linkedin:
  Practical challenges and lessons learned}. In
  \bibinfo{booktitle}{\emph{Proceedings of the eleventh ACM conference on
  recommender systems}}. \bibinfo{pages}{346--347}.
\newblock


\bibitem[Kenthapadi et~al\mbox{.}(2017b)]%
        {86/KenthapadiLV17}
\bibfield{author}{\bibinfo{person}{Krishnaram Kenthapadi},
  \bibinfo{person}{Benjamin Le}, {and} \bibinfo{person}{Ganesh Venkataraman}.}
  \bibinfo{year}{2017}\natexlab{b}.
\newblock \showarticletitle{Personalized job recommendation system at linkedin:
  Practical challenges and lessons learned}. In
  \bibinfo{booktitle}{\emph{Proceedings of the eleventh ACM conference on
  recommender systems}}. \bibinfo{pages}{346--347}.
\newblock


\bibitem[Khatua and Nejdl(2020)]%
        {137/khatua2020matching}
\bibfield{author}{\bibinfo{person}{Aparup Khatua} {and}
  \bibinfo{person}{Wolfgang Nejdl}.} \bibinfo{year}{2020}\natexlab{}.
\newblock \showarticletitle{Matching recruiters and jobseekers on twitter}. In
  \bibinfo{booktitle}{\emph{2020 IEEE/ACM International Conference on Advances
  in Social Networks Analysis and Mining (ASONAM)}}. IEEE,
  \bibinfo{pages}{266--269}.
\newblock


\bibitem[Lacic et~al\mbox{.}(2019)]%
        {67/lacic2019should}
\bibfield{author}{\bibinfo{person}{Emanuel Lacic}, \bibinfo{person}{Markus
  Reiter-Haas}, \bibinfo{person}{Tomislav Duricic}, \bibinfo{person}{Valentin
  Slawicek}, {and} \bibinfo{person}{Elisabeth Lex}.}
  \bibinfo{year}{2019}\natexlab{}.
\newblock \showarticletitle{Should we embed? A study on the online performance
  of utilizing embeddings for real-time job recommendations}. In
  \bibinfo{booktitle}{\emph{Proceedings of the 13th ACM Conference on
  Recommender Systems}}. \bibinfo{pages}{496--500}.
\newblock


\bibitem[Lacic et~al\mbox{.}(2020)]%
        {56/LacicRKDCL20}
\bibfield{author}{\bibinfo{person}{Emanuel Lacic}, \bibinfo{person}{Markus
  Reiter-Haas}, \bibinfo{person}{Dominik Kowald}, \bibinfo{person}{Manoj
  Reddy~Dareddy}, \bibinfo{person}{Junghoo Cho}, {and}
  \bibinfo{person}{Elisabeth Lex}.} \bibinfo{year}{2020}\natexlab{}.
\newblock \showarticletitle{Using autoencoders for session-based job
  recommendations}.
\newblock \bibinfo{journal}{\emph{User Modeling and User-Adapted Interaction}}
  \bibinfo{volume}{30}, \bibinfo{number}{4} (\bibinfo{year}{2020}),
  \bibinfo{pages}{617--658}.
\newblock


\bibitem[Lavi et~al\mbox{.}(2021)]%
        {116/lavi2021consultantbert}
\bibfield{author}{\bibinfo{person}{Dor Lavi}, \bibinfo{person}{Volodymyr
  Medentsiy}, {and} \bibinfo{person}{David Graus}.}
  \bibinfo{year}{2021}\natexlab{}.
\newblock \showarticletitle{conSultantBERT: Fine-tuned Siamese Sentence-BERT
  for Matching Jobs and Job Seekers}.
\newblock \bibinfo{journal}{\emph{arXiv preprint arXiv:2109.06501}}
  (\bibinfo{year}{2021}).
\newblock


\bibitem[Le et~al\mbox{.}(2019)]%
        {483/le2019towards}
\bibfield{author}{\bibinfo{person}{Ran Le}, \bibinfo{person}{Wenpeng Hu},
  \bibinfo{person}{Yang Song}, \bibinfo{person}{Tao Zhang},
  \bibinfo{person}{Dongyan Zhao}, {and} \bibinfo{person}{Rui Yan}.}
  \bibinfo{year}{2019}\natexlab{}.
\newblock \showarticletitle{Towards effective and interpretable person-job
  fitting}. In \bibinfo{booktitle}{\emph{Proceedings of the 28th ACM
  International Conference on Information and Knowledge Management}}.
  \bibinfo{pages}{1883--1892}.
\newblock


\bibitem[Lee et~al\mbox{.}(2016)]%
        {102/LeeHK16}
\bibfield{author}{\bibinfo{person}{Yeon-Chang Lee}, \bibinfo{person}{Jiwon
  Hong}, {and} \bibinfo{person}{Sang-Wook Kim}.}
  \bibinfo{year}{2016}\natexlab{}.
\newblock \showarticletitle{Job recommendation in askstory: experiences,
  methods, and evaluation}. In \bibinfo{booktitle}{\emph{Proceedings of the
  31st Annual ACM Symposium on Applied Computing}}. \bibinfo{pages}{780--786}.
\newblock


\bibitem[Leksin and Ostapets(2016)]%
        {95/LeksinO16}
\bibfield{author}{\bibinfo{person}{Vasily Leksin} {and} \bibinfo{person}{Andrey
  Ostapets}.} \bibinfo{year}{2016}\natexlab{}.
\newblock \showarticletitle{Job recommendation based on factorization machine
  and topic modelling}.
\newblock In \bibinfo{booktitle}{\emph{Proceedings of the Recommender Systems
  Challenge}}. \bibinfo{pages}{1--4}.
\newblock


\bibitem[Li et~al\mbox{.}(2020)]%
        {138/LiFTPHC20}
\bibfield{author}{\bibinfo{person}{Changmao Li}, \bibinfo{person}{Elaine
  Fisher}, \bibinfo{person}{Rebecca Thomas}, \bibinfo{person}{Steve Pittard},
  \bibinfo{person}{Vicki Hertzberg}, {and} \bibinfo{person}{Jinho~D Choi}.}
  \bibinfo{year}{2020}\natexlab{}.
\newblock \showarticletitle{Competence-level prediction and resume \& job
  description matching using context-aware transformer models}.
\newblock \bibinfo{journal}{\emph{arXiv preprint arXiv:2011.02998}}
  (\bibinfo{year}{2020}).
\newblock


\bibitem[Li et~al\mbox{.}(2022)]%
        {li2022fairness}
\bibfield{author}{\bibinfo{person}{Yunqi Li}, \bibinfo{person}{Hanxiong Chen},
  \bibinfo{person}{Shuyuan Xu}, \bibinfo{person}{Yingqiang Ge},
  \bibinfo{person}{Juntao Tan}, \bibinfo{person}{Shuchang Liu}, {and}
  \bibinfo{person}{Yongfeng Zhang}.} \bibinfo{year}{2022}\natexlab{}.
\newblock \showarticletitle{Fairness in Recommendation: A Survey}.
\newblock \bibinfo{journal}{\emph{arXiv preprint arXiv:2205.13619}}
  (\bibinfo{year}{2022}).
\newblock


\bibitem[Lian et~al\mbox{.}(2017)]%
        {537/lian2017practical}
\bibfield{author}{\bibinfo{person}{Jianxun Lian}, \bibinfo{person}{Fuzheng
  Zhang}, \bibinfo{person}{Min Hou}, \bibinfo{person}{Hongwei Wang},
  \bibinfo{person}{Xing Xie}, {and} \bibinfo{person}{Guangzhong Sun}.}
  \bibinfo{year}{2017}\natexlab{}.
\newblock \showarticletitle{Practical lessons for job recommendations in the
  cold-start scenario}.
\newblock In \bibinfo{booktitle}{\emph{Proceedings of the Recommender Systems
  Challenge 2017}}. \bibinfo{pages}{1--6}.
\newblock


\bibitem[Lin et~al\mbox{.}(2016)]%
        {171/LinLAL16}
\bibfield{author}{\bibinfo{person}{Yiou Lin}, \bibinfo{person}{Hang Lei},
  \bibinfo{person}{Prince~Clement Addo}, {and} \bibinfo{person}{Xiaoyu Li}.}
  \bibinfo{year}{2016}\natexlab{}.
\newblock \showarticletitle{Machine learned resume-job matching solution}.
\newblock \bibinfo{journal}{\emph{arXiv preprint arXiv:1607.07657}}
  (\bibinfo{year}{2016}).
\newblock


\bibitem[Liu et~al\mbox{.}(2016b)]%
        {23/LiuSKZN16}
\bibfield{author}{\bibinfo{person}{Kuan Liu}, \bibinfo{person}{Xing Shi},
  \bibinfo{person}{Anoop Kumar}, \bibinfo{person}{Linhong Zhu}, {and}
  \bibinfo{person}{Prem Natarajan}.} \bibinfo{year}{2016}\natexlab{b}.
\newblock \showarticletitle{Temporal learning and sequence modeling for a job
  recommender system}.
\newblock In \bibinfo{booktitle}{\emph{Proceedings of the Recommender Systems
  Challenge}}. \bibinfo{pages}{1--4}.
\newblock


\bibitem[Liu et~al\mbox{.}(2019)]%
        {65/liu2019tripartite}
\bibfield{author}{\bibinfo{person}{Mengshu Liu}, \bibinfo{person}{Jingya Wang},
  \bibinfo{person}{Kareem Abdelfatah}, {and} \bibinfo{person}{Mohammed
  Korayem}.} \bibinfo{year}{2019}\natexlab{}.
\newblock \showarticletitle{Tripartite vector representations for better job
  recommendation}.
\newblock \bibinfo{journal}{\emph{arXiv preprint arXiv:1907.12379}}
  (\bibinfo{year}{2019}).
\newblock


\bibitem[Liu et~al\mbox{.}(2016a)]%
        {90/LiuORSTX16}
\bibfield{author}{\bibinfo{person}{Rui Liu}, \bibinfo{person}{Yuanxin Ouyang},
  \bibinfo{person}{Wenge Rong}, \bibinfo{person}{Xin Song},
  \bibinfo{person}{Cui Tang}, {and} \bibinfo{person}{Zhang Xiong}.}
  \bibinfo{year}{2016}\natexlab{a}.
\newblock \showarticletitle{Rating prediction based job recommendation service
  for college students}. In \bibinfo{booktitle}{\emph{International conference
  on computational science and its applications}}. Springer,
  \bibinfo{pages}{453--467}.
\newblock


\bibitem[Liu et~al\mbox{.}(2017)]%
        {77/LiuROX17}
\bibfield{author}{\bibinfo{person}{Rui Liu}, \bibinfo{person}{Wenge Rong},
  \bibinfo{person}{Yuanxin Ouyang}, {and} \bibinfo{person}{Zhang Xiong}.}
  \bibinfo{year}{2017}\natexlab{}.
\newblock \showarticletitle{A hierarchical similarity based job recommendation
  service framework for university students}.
\newblock \bibinfo{journal}{\emph{Frontiers of Computer Science}}
  \bibinfo{volume}{11}, \bibinfo{number}{5} (\bibinfo{year}{2017}),
  \bibinfo{pages}{912--922}.
\newblock


\bibitem[Lu et~al\mbox{.}(2013)]%
        {402/lu2013recommender}
\bibfield{author}{\bibinfo{person}{Yao Lu}, \bibinfo{person}{Sandy El~Helou},
  {and} \bibinfo{person}{Denis Gillet}.} \bibinfo{year}{2013}\natexlab{}.
\newblock \showarticletitle{A recommender system for job seeking and recruiting
  website}. In \bibinfo{booktitle}{\emph{Proceedings of the 22nd International
  Conference on World Wide Web}}. \bibinfo{pages}{963--966}.
\newblock


\bibitem[Luo et~al\mbox{.}(2019)]%
        {523/luo2019resumegan}
\bibfield{author}{\bibinfo{person}{Yong Luo}, \bibinfo{person}{Huaizheng
  Zhang}, \bibinfo{person}{Yonggang Wen}, {and} \bibinfo{person}{Xinwen
  Zhang}.} \bibinfo{year}{2019}\natexlab{}.
\newblock \showarticletitle{Resumegan: an optimized deep representation
  learning framework for talent-job fit via adversarial learning}. In
  \bibinfo{booktitle}{\emph{Proceedings of the 28th ACM international
  conference on information and knowledge management}}.
  \bibinfo{pages}{1101--1110}.
\newblock


\bibitem[Maheshwary and Misra(2018)]%
        {153/MaheshwaryM18}
\bibfield{author}{\bibinfo{person}{Saket Maheshwary} {and}
  \bibinfo{person}{Hemant Misra}.} \bibinfo{year}{2018}\natexlab{}.
\newblock \showarticletitle{Matching resumes to jobs via deep siamese network}.
  In \bibinfo{booktitle}{\emph{Companion Proceedings of the The Web Conference
  2018}}. \bibinfo{pages}{87--88}.
\newblock


\bibitem[Malherbe et~al\mbox{.}(2014)]%
        {110/malherbe2014field}
\bibfield{author}{\bibinfo{person}{Emmanuel Malherbe}, \bibinfo{person}{Mamadou
  Diaby}, \bibinfo{person}{Mario Cataldi}, \bibinfo{person}{Emmanuel Viennet},
  {and} \bibinfo{person}{Marie-Aude Aufaure}.} \bibinfo{year}{2014}\natexlab{}.
\newblock \showarticletitle{Field selection for job categorization and
  recommendation to social network users}. In \bibinfo{booktitle}{\emph{2014
  IEEE/ACM International Conference on Advances in Social Networks Analysis and
  Mining (ASONAM 2014)}}. IEEE, \bibinfo{pages}{588--595}.
\newblock


\bibitem[Maree et~al\mbox{.}(2019)]%
        {288/MareeKB19}
\bibfield{author}{\bibinfo{person}{Mohammed Maree}, \bibinfo{person}{Aseel~B
  Kmail}, {and} \bibinfo{person}{Mohammed Belkhatir}.}
  \bibinfo{year}{2019}\natexlab{}.
\newblock \showarticletitle{Analysis and shortcomings of e-recruitment systems:
  Towards a semantics-based approach addressing knowledge incompleteness and
  limited domain coverage}.
\newblock \bibinfo{journal}{\emph{Journal of Information Science}}
  \bibinfo{volume}{45}, \bibinfo{number}{6} (\bibinfo{year}{2019}),
  \bibinfo{pages}{713--735}.
\newblock


\bibitem[Martinez-Gil et~al\mbox{.}(2018)]%
        {75/martinez2018recommendation}
\bibfield{author}{\bibinfo{person}{Jorge Martinez-Gil},
  \bibinfo{person}{Bernhard Freudenthaler}, {and} \bibinfo{person}{Thomas
  Natschl{\"a}ger}.} \bibinfo{year}{2018}\natexlab{}.
\newblock \showarticletitle{Recommendation of job offers using random forests
  and support vector machines}. In \bibinfo{booktitle}{\emph{Proceedings of the
  of the EDBT/ICDT joint conference}}.
\newblock


\bibitem[Menacer et~al\mbox{.}(2021)]%
        {477/menacer2021interpretable}
\bibfield{author}{\bibinfo{person}{Mohamed~Amine Menacer},
  \bibinfo{person}{Fatma~Ben Hamda}, \bibinfo{person}{Ghada Mighri},
  \bibinfo{person}{Sabeur~Ben Hamidene}, {and} \bibinfo{person}{Maxime
  Cariou}.} \bibinfo{year}{2021}\natexlab{}.
\newblock \showarticletitle{An interpretable person-job fitting approach based
  on classification and ranking}. In \bibinfo{booktitle}{\emph{Proceedings of
  The Fourth International Conference on Natural Language and Speech Processing
  (ICNLSP 2021)}}. \bibinfo{pages}{130--138}.
\newblock


\bibitem[Mentec et~al\mbox{.}(2021)]%
        {52/Mentec0HR21}
\bibfield{author}{\bibinfo{person}{Fran{\c{c}}ois Mentec},
  \bibinfo{person}{Zolt{\'a}n Mikl{\'o}s}, \bibinfo{person}{S{\'e}bastien
  Hervieu}, {and} \bibinfo{person}{Thierry Roger}.}
  \bibinfo{year}{2021}\natexlab{}.
\newblock \showarticletitle{Conversational recommendations for job recruiters}.
  In \bibinfo{booktitle}{\emph{Knowledge-aware and Conversational Recommender
  Systems}}.
\newblock


\bibitem[Mhamdi et~al\mbox{.}(2020)]%
        {37/MhamdiMGAM20}
\bibfield{author}{\bibinfo{person}{D Mhamdi}, \bibinfo{person}{Reda Moulouki},
  \bibinfo{person}{Mohammed~Yassine El~Ghoumari}, \bibinfo{person}{M Azzouazi},
  {and} \bibinfo{person}{L Moussaid}.} \bibinfo{year}{2020}\natexlab{}.
\newblock \showarticletitle{Job recommendation based on job profile clustering
  and job seeker behavior}.
\newblock \bibinfo{journal}{\emph{Procedia Computer Science}}
  \bibinfo{volume}{175} (\bibinfo{year}{2020}), \bibinfo{pages}{695--699}.
\newblock


\bibitem[Mine et~al\mbox{.}(2013)]%
        {181/MineKO13}
\bibfield{author}{\bibinfo{person}{Tsunenori Mine}, \bibinfo{person}{Tomoyuki
  Kakuta}, {and} \bibinfo{person}{Akira Ono}.} \bibinfo{year}{2013}\natexlab{}.
\newblock \showarticletitle{Reciprocal recommendation for job matching with
  bidirectional feedback}. In \bibinfo{booktitle}{\emph{2013 Second IIAI
  International Conference on Advanced Applied Informatics}}. IEEE,
  \bibinfo{pages}{39--44}.
\newblock


\bibitem[Mishra and Reddy(2016)]%
        {96/MishraR16}
\bibfield{author}{\bibinfo{person}{Sonu~K Mishra} {and} \bibinfo{person}{Manoj
  Reddy}.} \bibinfo{year}{2016}\natexlab{}.
\newblock \showarticletitle{A bottom-up approach to job recommendation system}.
\newblock In \bibinfo{booktitle}{\emph{Proceedings of the Recommender Systems
  Challenge}}. \bibinfo{pages}{1--4}.
\newblock


\bibitem[Mughaid et~al\mbox{.}(2019)]%
        {16/MughaidOHAA19}
\bibfield{author}{\bibinfo{person}{Ala Mughaid}, \bibinfo{person}{Ibrahim
  Obeidat}, \bibinfo{person}{Bilal Hawashin}, \bibinfo{person}{Shadi AlZu'bi},
  {and} \bibinfo{person}{Darah Aqel}.} \bibinfo{year}{2019}\natexlab{}.
\newblock \showarticletitle{A smart geo-location job recommender system based
  on social media posts}. In \bibinfo{booktitle}{\emph{2019 Sixth International
  Conference on Social Networks Analysis, Management and Security (SNAMS)}}.
  IEEE, \bibinfo{pages}{505--510}.
\newblock


\bibitem[Nigam et~al\mbox{.}(2019)]%
        {38/NigamRSW19}
\bibfield{author}{\bibinfo{person}{Amber Nigam}, \bibinfo{person}{Aakash Roy},
  \bibinfo{person}{Hartaran Singh}, {and} \bibinfo{person}{Harsimran Waila}.}
  \bibinfo{year}{2019}\natexlab{}.
\newblock \showarticletitle{Job recommendation through progression of job
  selection}. In \bibinfo{booktitle}{\emph{2019 IEEE 6th International
  Conference on Cloud Computing and Intelligence Systems (CCIS)}}. IEEE,
  \bibinfo{pages}{212--216}.
\newblock


\bibitem[Pacuk et~al\mbox{.}(2016)]%
        {97/PacukSWWW16}
\bibfield{author}{\bibinfo{person}{Andrzej Pacuk}, \bibinfo{person}{Piotr
  Sankowski}, \bibinfo{person}{Karol Wegrzycki}, \bibinfo{person}{Adam
  Witkowski}, {and} \bibinfo{person}{Piotr Wygocki}.}
  \bibinfo{year}{2016}\natexlab{}.
\newblock \showarticletitle{RecSys Challenge 2016: Job recommendations based on
  preselection of offers and gradient boosting}.
\newblock In \bibinfo{booktitle}{\emph{Proceedings of the Recommender Systems
  Challenge}}. \bibinfo{pages}{1--4}.
\newblock


\bibitem[Paparrizos et~al\mbox{.}(2011)]%
        {paparrizos2011machine}
\bibfield{author}{\bibinfo{person}{Ioannis Paparrizos},
  \bibinfo{person}{B~Barla Cambazoglu}, {and} \bibinfo{person}{Aristides
  Gionis}.} \bibinfo{year}{2011}\natexlab{}.
\newblock \showarticletitle{Machine learned job recommendation}. In
  \bibinfo{booktitle}{\emph{Proceedings of the fifth ACM Conference on
  Recommender Systems}}. \bibinfo{pages}{325--328}.
\newblock


\bibitem[Patel et~al\mbox{.}(2017)]%
        {522/patel2017capar}
\bibfield{author}{\bibinfo{person}{Bharat Patel}, \bibinfo{person}{Varun
  Kakuste}, {and} \bibinfo{person}{Magdalini Eirinaki}.}
  \bibinfo{year}{2017}\natexlab{}.
\newblock \showarticletitle{CaPaR: a career path recommendation framework}. In
  \bibinfo{booktitle}{\emph{2017 IEEE Third International Conference on Big
  Data Computing Service and Applications (BigDataService)}}. IEEE,
  \bibinfo{pages}{23--30}.
\newblock


\bibitem[Polato and Aiolli(2016)]%
        {99/PolatoA16}
\bibfield{author}{\bibinfo{person}{Mirko Polato} {and} \bibinfo{person}{Fabio
  Aiolli}.} \bibinfo{year}{2016}\natexlab{}.
\newblock \showarticletitle{A preliminary study on a recommender system for the
  job recommendation challenge}.
\newblock In \bibinfo{booktitle}{\emph{Proceedings of the Recommender Systems
  Challenge}}. \bibinfo{pages}{1--4}.
\newblock


\bibitem[Qin et~al\mbox{.}(2018)]%
        {485/qin2018enhancing}
\bibfield{author}{\bibinfo{person}{Chuan Qin}, \bibinfo{person}{Hengshu Zhu},
  \bibinfo{person}{Tong Xu}, \bibinfo{person}{Chen Zhu}, \bibinfo{person}{Liang
  Jiang}, \bibinfo{person}{Enhong Chen}, {and} \bibinfo{person}{Hui Xiong}.}
  \bibinfo{year}{2018}\natexlab{}.
\newblock \showarticletitle{Enhancing person-job fit for talent recruitment: An
  ability-aware neural network approach}. In \bibinfo{booktitle}{\emph{The 41st
  international ACM SIGIR conference on research \& development in information
  retrieval}}. \bibinfo{pages}{25--34}.
\newblock


\bibitem[Qin et~al\mbox{.}(2020)]%
        {478/qin2020enhanced}
\bibfield{author}{\bibinfo{person}{Chuan Qin}, \bibinfo{person}{Hengshu Zhu},
  \bibinfo{person}{Tong Xu}, \bibinfo{person}{Chen Zhu}, \bibinfo{person}{Chao
  Ma}, \bibinfo{person}{Enhong Chen}, {and} \bibinfo{person}{Hui Xiong}.}
  \bibinfo{year}{2020}\natexlab{}.
\newblock \showarticletitle{An enhanced neural network approach to person-job
  fit in talent recruitment}.
\newblock \bibinfo{journal}{\emph{ACM Transactions on Information Systems
  (TOIS)}} \bibinfo{volume}{38}, \bibinfo{number}{2} (\bibinfo{year}{2020}),
  \bibinfo{pages}{1--33}.
\newblock


\bibitem[R{\'a}cz et~al\mbox{.}(2016)]%
        {166/RaczSS16}
\bibfield{author}{\bibinfo{person}{G{\'a}bor R{\'a}cz}, \bibinfo{person}{Attila
  Sali}, {and} \bibinfo{person}{Klaus-Dieter Schewe}.}
  \bibinfo{year}{2016}\natexlab{}.
\newblock \showarticletitle{Semantic matching strategies for job recruitment: A
  comparison of new and known approaches}. In
  \bibinfo{booktitle}{\emph{FoIKS}}. Springer, \bibinfo{pages}{149--168}.
\newblock


\bibitem[Raghavan et~al\mbox{.}(2020)]%
        {544/raghavan2020miti}
\bibfield{author}{\bibinfo{person}{Manish Raghavan}, \bibinfo{person}{Solon
  Barocas}, \bibinfo{person}{Jon Kleinberg}, {and} \bibinfo{person}{Karen
  Levy}.} \bibinfo{year}{2020}\natexlab{}.
\newblock \showarticletitle{Mitigating bias in algorithmic hiring: Evaluating
  claims and practices}. In \bibinfo{booktitle}{\emph{Proceedings of the 2020
  conference on fairness, accountability, and transparency}}.
  \bibinfo{pages}{469--481}.
\newblock


\bibitem[Reusens et~al\mbox{.}(2017)]%
        {76/ReusensLBS17}
\bibfield{author}{\bibinfo{person}{Michael Reusens}, \bibinfo{person}{Wilfried
  Lemahieu}, \bibinfo{person}{Bart Baesens}, {and} \bibinfo{person}{Luc Sels}.}
  \bibinfo{year}{2017}\natexlab{}.
\newblock \showarticletitle{A note on explicit versus implicit information for
  job recommendation}.
\newblock \bibinfo{journal}{\emph{Decision Support Systems}}
  \bibinfo{volume}{98} (\bibinfo{year}{2017}), \bibinfo{pages}{26--35}.
\newblock


\bibitem[Rivas et~al\mbox{.}(2019)]%
        {9/RivasCGCC19}
\bibfield{author}{\bibinfo{person}{Alberto Rivas}, \bibinfo{person}{Pablo
  Chamoso}, \bibinfo{person}{Alfonso Gonz{\'a}lez-Briones},
  \bibinfo{person}{Roberto Casado-Vara}, {and} \bibinfo{person}{Juan~Manuel
  Corchado}.} \bibinfo{year}{2019}\natexlab{}.
\newblock \showarticletitle{Hybrid job offer recommender system in a social
  network}.
\newblock \bibinfo{journal}{\emph{Expert Systems}} \bibinfo{volume}{36},
  \bibinfo{number}{4} (\bibinfo{year}{2019}), \bibinfo{pages}{e12416}.
\newblock


\bibitem[Rodriguez and Chavez(2019)]%
        {117/RodriguezC19}
\bibfield{author}{\bibinfo{person}{Leah~G Rodriguez} {and}
  \bibinfo{person}{Enrico~P Chavez}.} \bibinfo{year}{2019}\natexlab{}.
\newblock \showarticletitle{Feature selection for job matching application
  using profile matching model}. In \bibinfo{booktitle}{\emph{2019 IEEE 4th
  International Conference on Computer and Communication Systems (ICCCS)}}.
  IEEE, \bibinfo{pages}{263--266}.
\newblock


\bibitem[Roy et~al\mbox{.}(2020)]%
        {521/roy2020machine}
\bibfield{author}{\bibinfo{person}{Pradeep~Kumar Roy},
  \bibinfo{person}{Sarabjeet~Singh Chowdhary}, {and} \bibinfo{person}{Rocky
  Bhatia}.} \bibinfo{year}{2020}\natexlab{}.
\newblock \showarticletitle{A Machine Learning approach for automation of
  Resume Recommendation system}.
\newblock \bibinfo{journal}{\emph{Procedia Computer Science}}
  \bibinfo{volume}{167} (\bibinfo{year}{2020}), \bibinfo{pages}{2318--2327}.
\newblock


\bibitem[Salazar et~al\mbox{.}(2015)]%
        {331/salazar2015case}
\bibfield{author}{\bibinfo{person}{Oscar~M Salazar}, \bibinfo{person}{Juan~C
  Jaramillo}, \bibinfo{person}{Demetrio~A Ovalle}, {and}
  \bibinfo{person}{Jaime~A Guzm{\'a}n}.} \bibinfo{year}{2015}\natexlab{}.
\newblock \showarticletitle{A case-based multi-agent and recommendation
  environment to improve the e-recruitment process}. In
  \bibinfo{booktitle}{\emph{International Conference on Practical Applications
  of Agents and Multi-Agent Systems}}. Springer, \bibinfo{pages}{389--397}.
\newblock


\bibitem[S{\'a}nchez-Monedero et~al\mbox{.}(2020)]%
        {545/sanchez2020whatdoes}
\bibfield{author}{\bibinfo{person}{Javier S{\'a}nchez-Monedero},
  \bibinfo{person}{Lina Dencik}, {and} \bibinfo{person}{Lilian Edwards}.}
  \bibinfo{year}{2020}\natexlab{}.
\newblock \showarticletitle{What does it mean to'solve'the problem of
  discrimination in hiring? Social, technical and legal perspectives from the
  UK on automated hiring systems}. In \bibinfo{booktitle}{\emph{Proceedings of
  the 2020 conference on fairness, accountability, and transparency}}.
  \bibinfo{pages}{458--468}.
\newblock


\bibitem[Sato et~al\mbox{.}(2017)]%
        {520/sato2017explor}
\bibfield{author}{\bibinfo{person}{Masahiro Sato}, \bibinfo{person}{Koki
  Nagatani}, {and} \bibinfo{person}{Takuji Tahara}.}
  \bibinfo{year}{2017}\natexlab{}.
\newblock \showarticletitle{Exploring an optimal online model for new job
  recommendation: Solution for recsys challenge 2017}.
\newblock In \bibinfo{booktitle}{\emph{Proceedings of the Recommender Systems
  Challenge 2017}}. \bibinfo{pages}{1--5}.
\newblock


\bibitem[Schmitt et~al\mbox{.}(2016)]%
        {167/SchmittCS16}
\bibfield{author}{\bibinfo{person}{Thomas Schmitt}, \bibinfo{person}{Philippe
  Caillou}, {and} \bibinfo{person}{Michele Sebag}.}
  \bibinfo{year}{2016}\natexlab{}.
\newblock \showarticletitle{Matching jobs and resumes: a deep collaborative
  filtering task}. In \bibinfo{booktitle}{\emph{GCAI 2016-2nd Global Conference
  on Artificial Intelligence}}, Vol.~\bibinfo{volume}{41}.
\newblock


\bibitem[Schmitt et~al\mbox{.}(2017)]%
        {160/SchmittGCS17}
\bibfield{author}{\bibinfo{person}{Thomas Schmitt},
  \bibinfo{person}{Fran{\c{c}}ois Gonard}, \bibinfo{person}{Philippe Caillou},
  {and} \bibinfo{person}{Mich{\`e}le Sebag}.} \bibinfo{year}{2017}\natexlab{}.
\newblock \showarticletitle{Language modelling for collaborative filtering:
  Application to job applicant matching}. In \bibinfo{booktitle}{\emph{2017
  IEEE 29th International Conference on Tools with Artificial Intelligence
  (ICTAI)}}. IEEE, \bibinfo{pages}{1226--1233}.
\newblock


\bibitem[Shalaby et~al\mbox{.}(2017)]%
        {44/ShalabyAKPAQZ17}
\bibfield{author}{\bibinfo{person}{Walid Shalaby}, \bibinfo{person}{BahaaEddin
  AlAila}, \bibinfo{person}{Mohammed Korayem}, \bibinfo{person}{Layla
  Pournajaf}, \bibinfo{person}{Khalifeh AlJadda}, \bibinfo{person}{Shannon
  Quinn}, {and} \bibinfo{person}{Wlodek Zadrozny}.}
  \bibinfo{year}{2017}\natexlab{}.
\newblock \showarticletitle{Help me find a job: A graph-based approach for job
  recommendation at scale}. In \bibinfo{booktitle}{\emph{2017 IEEE
  international conference on big data (big data)}}. IEEE,
  \bibinfo{pages}{1544--1553}.
\newblock


\bibitem[Shi et~al\mbox{.}(2020)]%
        {524/shi2020salience}
\bibfield{author}{\bibinfo{person}{Baoxu Shi}, \bibinfo{person}{Jaewon Yang},
  \bibinfo{person}{Feng Guo}, {and} \bibinfo{person}{Qi He}.}
  \bibinfo{year}{2020}\natexlab{}.
\newblock \showarticletitle{Salience and market-aware skill extraction for job
  targeting}. In \bibinfo{booktitle}{\emph{Proceedings of the 26th ACM SIGKDD
  International Conference on Knowledge Discovery \& Data Mining}}.
  \bibinfo{pages}{2871--2879}.
\newblock


\bibitem[Shishehchi and Banihashem(2019)]%
        {10/ShishehchiB19}
\bibfield{author}{\bibinfo{person}{Saman Shishehchi} {and}
  \bibinfo{person}{Seyed~Yashar Banihashem}.} \bibinfo{year}{2019}\natexlab{}.
\newblock \showarticletitle{Jrdp: a job recommender system based on ontology
  for disabled people}.
\newblock \bibinfo{journal}{\emph{International Journal of Technology and Human
  Interaction (IJTHI)}} \bibinfo{volume}{15}, \bibinfo{number}{1}
  (\bibinfo{year}{2019}), \bibinfo{pages}{85--99}.
\newblock


\bibitem[Slama and Darmon(2021)]%
        {495/slama2021novel}
\bibfield{author}{\bibinfo{person}{Olfa Slama} {and} \bibinfo{person}{Patrice
  Darmon}.} \bibinfo{year}{2021}\natexlab{}.
\newblock \showarticletitle{A Novel Personalized Preference-based Approach for
  Job/Candidate Recommendation}. In \bibinfo{booktitle}{\emph{International
  Conference on Research Challenges in Information Science}}. Springer,
  \bibinfo{pages}{418--434}.
\newblock


\bibitem[Smith et~al\mbox{.}(2021)]%
        {130/smith2021skill}
\bibfield{author}{\bibinfo{person}{Ellery Smith}, \bibinfo{person}{Andreas
  Weiler}, {and} \bibinfo{person}{Martin Braschler}.}
  \bibinfo{year}{2021}\natexlab{}.
\newblock \showarticletitle{Skill Extraction for Domain-Specific Text Retrieval
  in a Job-Matching Platform}. In \bibinfo{booktitle}{\emph{International
  Conference of the Cross-Language Evaluation Forum for European Languages}}.
  Springer, \bibinfo{pages}{116--128}.
\newblock


\bibitem[Upadhyay et~al\mbox{.}(2021)]%
        {53/UpadhyayA0F21}
\bibfield{author}{\bibinfo{person}{Chirayu Upadhyay}, \bibinfo{person}{Hasan
  Abu-Rasheed}, \bibinfo{person}{Christian Weber}, {and}
  \bibinfo{person}{Madjid Fathi}.} \bibinfo{year}{2021}\natexlab{}.
\newblock \showarticletitle{Explainable Job-Posting Recommendations Using
  Knowledge Graphs and Named Entity Recognition}. In
  \bibinfo{booktitle}{\emph{2021 IEEE International Conference on Systems, Man,
  and Cybernetics (SMC)}}. IEEE, \bibinfo{pages}{3291--3296}.
\newblock


\bibitem[Valverde-Rebaza et~al\mbox{.}(2018)]%
        {41/Valverde-Rebaza18}
\bibfield{author}{\bibinfo{person}{Jorge~Carlos Valverde-Rebaza},
  \bibinfo{person}{Ricardo Puma}, \bibinfo{person}{Paul Bustios}, {and}
  \bibinfo{person}{Nathalia~C Silva}.} \bibinfo{year}{2018}\natexlab{}.
\newblock \showarticletitle{Job Recommendation Based on Job Seeker Skills: An
  Empirical Study.}. In \bibinfo{booktitle}{\emph{Text2Story@ ECIR}}.
  \bibinfo{pages}{47--51}.
\newblock


\bibitem[Volkovs et~al\mbox{.}(2017)]%
        {539/volkovs2017content}
\bibfield{author}{\bibinfo{person}{Maksims Volkovs}, \bibinfo{person}{Guang~Wei
  Yu}, {and} \bibinfo{person}{Tomi Poutanen}.} \bibinfo{year}{2017}\natexlab{}.
\newblock \showarticletitle{Content-based neighbor models for cold start in
  recommender systems}.
\newblock In \bibinfo{booktitle}{\emph{Proceedings of the Recommender Systems
  Challenge 2017}}. \bibinfo{pages}{1--6}.
\newblock


\bibitem[Wang et~al\mbox{.}(2022)]%
        {543/wang2022dohumans}
\bibfield{author}{\bibinfo{person}{Clarice Wang}, \bibinfo{person}{Kathryn
  Wang}, \bibinfo{person}{Andrew Bian}, \bibinfo{person}{Rashidul Islam},
  \bibinfo{person}{Kamrun~Naher Keya}, \bibinfo{person}{James Foulds}, {and}
  \bibinfo{person}{Shimei Pan}.} \bibinfo{year}{2022}\natexlab{}.
\newblock \showarticletitle{Do Humans Prefer Debiased AI Algorithms? A Case
  Study in Career Recommendation}. In \bibinfo{booktitle}{\emph{27th
  International Conference on Intelligent User Interfaces}}.
  \bibinfo{pages}{134--147}.
\newblock


\bibitem[Wang et~al\mbox{.}(2021)]%
        {124/WangJP21}
\bibfield{author}{\bibinfo{person}{Xiaowei Wang}, \bibinfo{person}{Zhenhong
  Jiang}, {and} \bibinfo{person}{Lingxi Peng}.}
  \bibinfo{year}{2021}\natexlab{}.
\newblock \showarticletitle{A Deep-Learning-Inspired Person-Job Matching Model
  Based on Sentence Vectors and Subject-Term Graphs}.
\newblock \bibinfo{journal}{\emph{Complexity}}  \bibinfo{volume}{2021}
  (\bibinfo{year}{2021}).
\newblock


\bibitem[Wang et~al\mbox{.}(2020)]%
        {60/wang2020session}
\bibfield{author}{\bibinfo{person}{Yusen Wang}, \bibinfo{person}{Kaize Shi},
  {and} \bibinfo{person}{Zhendong Niu}.} \bibinfo{year}{2020}\natexlab{}.
\newblock \showarticletitle{A Session-based Job Recommendation System Combining
  Area Knowledge and Interest Graph Neural Networks.}. In
  \bibinfo{booktitle}{\emph{SEKE}}. \bibinfo{pages}{489--492}.
\newblock


\bibitem[Xiao et~al\mbox{.}(2016)]%
        {100/XiaoXLMW16}
\bibfield{author}{\bibinfo{person}{Wenming Xiao}, \bibinfo{person}{Xiao Xu},
  \bibinfo{person}{Kang Liang}, \bibinfo{person}{Junkang Mao}, {and}
  \bibinfo{person}{Jun Wang}.} \bibinfo{year}{2016}\natexlab{}.
\newblock \showarticletitle{Job recommendation with hawkes process: an
  effective solution for recsys challenge 2016}.
\newblock In \bibinfo{booktitle}{\emph{Proceedings of the recommender systems
  challenge}}. \bibinfo{pages}{1--4}.
\newblock


\bibitem[Xu and Barbosa(2018)]%
        {150/xu2018matching}
\bibfield{author}{\bibinfo{person}{Peng Xu} {and} \bibinfo{person}{Denilson
  Barbosa}.} \bibinfo{year}{2018}\natexlab{}.
\newblock \showarticletitle{Matching r{\'e}sum{\'e}s to job descriptions with
  stacked models}. In \bibinfo{booktitle}{\emph{Canadian Conference on
  Artificial Intelligence}}. Springer, \bibinfo{pages}{304--309}.
\newblock


\bibitem[Yagci and Gurgen(2017)]%
        {516/yagci2017aranker}
\bibfield{author}{\bibinfo{person}{Murat Yagci} {and} \bibinfo{person}{Fikret
  Gurgen}.} \bibinfo{year}{2017}\natexlab{}.
\newblock \showarticletitle{A ranker ensemble for multi-objective job
  recommendation in an item cold start setting}.
\newblock In \bibinfo{booktitle}{\emph{Proceedings of the Recommender Systems
  Challenge 2017}}. \bibinfo{pages}{1--4}.
\newblock


\bibitem[Yan et~al\mbox{.}(2019)]%
        {145/YanLSZZ019}
\bibfield{author}{\bibinfo{person}{Rui Yan}, \bibinfo{person}{Ran Le},
  \bibinfo{person}{Yang Song}, \bibinfo{person}{Tao Zhang},
  \bibinfo{person}{Xiangliang Zhang}, {and} \bibinfo{person}{Dongyan Zhao}.}
  \bibinfo{year}{2019}\natexlab{}.
\newblock \showarticletitle{Interview choice reveals your preference on the
  market: To improve job-resume matching through profiling memories}. In
  \bibinfo{booktitle}{\emph{Proceedings of the 25th ACM SIGKDD International
  Conference on Knowledge Discovery \& Data Mining}}.
  \bibinfo{pages}{914--922}.
\newblock


\bibitem[Yang et~al\mbox{.}(2017)]%
        {78/yang2017combining}
\bibfield{author}{\bibinfo{person}{Shuo Yang}, \bibinfo{person}{Mohammed
  Korayem}, \bibinfo{person}{Khalifeh AlJadda}, \bibinfo{person}{Trey
  Grainger}, {and} \bibinfo{person}{Sriraam Natarajan}.}
  \bibinfo{year}{2017}\natexlab{}.
\newblock \showarticletitle{Combining content-based and collaborative filtering
  for job recommendation system: A cost-sensitive Statistical Relational
  Learning approach}.
\newblock \bibinfo{journal}{\emph{Knowledge-Based Systems}}
  \bibinfo{volume}{136} (\bibinfo{year}{2017}), \bibinfo{pages}{37--45}.
\newblock


\bibitem[Zhang and Cheng(2016)]%
        {25/ZhangC16}
\bibfield{author}{\bibinfo{person}{Chenrui Zhang} {and} \bibinfo{person}{Xueqi
  Cheng}.} \bibinfo{year}{2016}\natexlab{}.
\newblock \showarticletitle{An ensemble method for job recommender systems}.
\newblock In \bibinfo{booktitle}{\emph{Proceedings of the Recommender Systems
  Challenge}}. \bibinfo{pages}{1--4}.
\newblock


\bibitem[Zhang et~al\mbox{.}(2016)]%
        {540/zhang2016glmix}
\bibfield{author}{\bibinfo{person}{XianXing Zhang}, \bibinfo{person}{Yitong
  Zhou}, \bibinfo{person}{Yiming Ma}, \bibinfo{person}{Bee-Chung Chen},
  \bibinfo{person}{Liang Zhang}, {and} \bibinfo{person}{Deepak Agarwal}.}
  \bibinfo{year}{2016}\natexlab{}.
\newblock \showarticletitle{Glmix: Generalized linear mixed models for
  large-scale response prediction}. In \bibinfo{booktitle}{\emph{Proceedings of
  the 22nd ACM SIGKDD international conference on knowledge discovery and data
  mining}}. \bibinfo{pages}{363--372}.
\newblock


\bibitem[Zhang et~al\mbox{.}(2021)]%
        {475/zhang2021explainable}
\bibfield{author}{\bibinfo{person}{Yunchong Zhang}, \bibinfo{person}{Baisong
  Liu}, \bibinfo{person}{Jiangbo Qian}, \bibinfo{person}{Jiangcheng Qin},
  \bibinfo{person}{Xueyuan Zhang}, {and} \bibinfo{person}{Xueyong Jiang}.}
  \bibinfo{year}{2021}\natexlab{}.
\newblock \showarticletitle{An Explainable Person-Job Fit Model Incorporating
  Structured Information}. In \bibinfo{booktitle}{\emph{2021 IEEE International
  Conference on Big Data (Big Data)}}. IEEE, \bibinfo{pages}{3571--3579}.
\newblock


\bibitem[Zhao et~al\mbox{.}(2021a)]%
        {132/zhao2021embedding}
\bibfield{author}{\bibinfo{person}{Jing Zhao}, \bibinfo{person}{Jingya Wang},
  \bibinfo{person}{Madhav Sigdel}, \bibinfo{person}{Bopeng Zhang},
  \bibinfo{person}{Phuong Hoang}, \bibinfo{person}{Mengshu Liu}, {and}
  \bibinfo{person}{Mohammed Korayem}.} \bibinfo{year}{2021}\natexlab{a}.
\newblock \showarticletitle{Embedding-based Recommender System for Job to
  Candidate Matching on Scale}.
\newblock \bibinfo{journal}{\emph{arXiv preprint arXiv:2107.00221}}
  (\bibinfo{year}{2021}).
\newblock


\bibitem[Zhao et~al\mbox{.}(2021b)]%
        {114/zhao2021summer}
\bibfield{author}{\bibinfo{person}{Tianhua Zhao}, \bibinfo{person}{Cheng Wuyu},
  {and} \bibinfo{person}{Chen Zhixiang}.} \bibinfo{year}{2021}\natexlab{b}.
\newblock \showarticletitle{Summer Job Selection Model Based on Job Matching
  and Comprehensive Evaluation Algorithm}. In \bibinfo{booktitle}{\emph{2021
  2nd International Conference on Artificial Intelligence and Information
  Systems}}. \bibinfo{pages}{1--5}.
\newblock


\bibitem[Zhu et~al\mbox{.}(2018)]%
        {469/zhu2018person}
\bibfield{author}{\bibinfo{person}{Chen Zhu}, \bibinfo{person}{Hengshu Zhu},
  \bibinfo{person}{Hui Xiong}, \bibinfo{person}{Chao Ma}, \bibinfo{person}{Fang
  Xie}, \bibinfo{person}{Pengliang Ding}, {and} \bibinfo{person}{Pan Li}.}
  \bibinfo{year}{2018}\natexlab{}.
\newblock \showarticletitle{Person-job fit: Adapting the right talent for the
  right job with joint representation learning}.
\newblock \bibinfo{journal}{\emph{ACM Transactions on Management Information
  Systems (TMIS)}} \bibinfo{volume}{9}, \bibinfo{number}{3}
  (\bibinfo{year}{2018}), \bibinfo{pages}{1--17}.
\newblock


\bibitem[Zibriczky(2016)]%
        {101/Zibriczky16}
\bibfield{author}{\bibinfo{person}{D{\'a}vid Zibriczky}.}
  \bibinfo{year}{2016}\natexlab{}.
\newblock \showarticletitle{A combination of simple models by forward predictor
  selection for job recommendation}.
\newblock In \bibinfo{booktitle}{\emph{Proceedings of the Recommender Systems
  Challenge}}. \bibinfo{pages}{1--4}.
\newblock


\end{thebibliography}

\clearpage

\appendix
\section{Supplementary Materials}\label{appendix}
Table~\ref{tab:all_papers} gives an overview of all the papers that have been collected with the literature search methodology in Section~\ref{sect:methodology}.


\setlength\tabcolsep{1pt}
\begin{center}
\linespread{0.5}
\begin{small}
\renewcommand*{\arraystretch}{1}
  \begin{longtable}{?C{0.045\linewidth}?C{0.045\linewidth}?C{0.06\linewidth}|C{0.06\linewidth}|C{0.06\linewidth}?C{0.04\linewidth}|C{0.04\linewidth}|C{0.04\linewidth}|C{0.04\linewidth}?C{0.04\linewidth}|C{0.08\linewidth}|C{0.04\linewidth}|C{0.08\linewidth}|C{0.04\linewidth}|C{0.04\linewidth}|C{0.04\linewidth}|C{0.04\linewidth}?}
\caption{An overview of e-recruitment recommendation systems is presented. Regarding the recommended entities, although some papers could be reciprocal in design, we did not report them as reciprocal since they did not claim to be reciprocal and also they only experimented with the job or job seeker recommendation task. The methods cover a broad range of content based (CB), collaborative filtering (CF), knowledge based (KB), and hybrid/other methods. Some papers focus on preprocessing, postprocessing or re-ranking, and do not mention the recommendation method type in detail. Hence, we also do not report the recommendation method type for those papers. The papers are sorted based on their publication year.}\label{tab:all_papers}\\
\hline \multirow{2}{*}{Paper} & \multirow{2}{*}{Year} & \multicolumn{3}{c?}{Recommended entities} & %
    \multicolumn{4}{c?}{Method} & \multicolumn{8}{c?}{Challenge}\\
\cline{3-17} &  & \rot{Job} & \rot{Job seeker} & \rot{Reciprocal} & CB & CF & KB & \rot{Hybrid/Other} & \rot{\ref{sect:data quality} Data quality} & \rot{\parbox{22mm}{\ref{sect:hetero data} Heterogenous\\data, multiple int-\\eraction types\\and data sources}} & \rot{\ref{sect:cold start} Cold start} & \rot{\parbox{22mm}{\ref{sect:user preference} User prefer-\\ences as well as\\suitability}} & \rot{\parbox{22mm}{\ref{sect:interpretability} Interpretability\\and explainability}} & \rot{\parbox{22mm}{\ref{sect:specific obj} Specific\\objectives}} & \rot{\parbox{22mm}{\ref{sect:bias} Bias\\and fairness}} & \rot{\ref{sect:large scale} Large scale}\\
\hline
\endhead

\cite{184/BollingerHM12} &2012 & \fullmoon & \fullmoon & \newmoon & \fullmoon & \fullmoon & \fullmoon & \newmoon & \fullmoon & \newmoon & \fullmoon & \fullmoon & \fullmoon & \newmoon & \fullmoon & \fullmoon \\
\cite{212/FaliagkaTT12} &2012 & \fullmoon & \newmoon & \fullmoon & \newmoon & \fullmoon & \fullmoon & \fullmoon & \fullmoon & \newmoon & \fullmoon & \fullmoon & \fullmoon & \fullmoon & \fullmoon & \fullmoon \\
\cite{32/HongZWS13} &2013 & \newmoon & \fullmoon & \fullmoon & \fullmoon & \fullmoon & \fullmoon & \newmoon & \fullmoon & \fullmoon & \fullmoon & \fullmoon & \fullmoon & \fullmoon & \fullmoon & \fullmoon \\
\cite{181/MineKO13} &2013 & \fullmoon & \fullmoon & \newmoon & \fullmoon & \fullmoon & \fullmoon & \newmoon & \fullmoon & \fullmoon & \fullmoon & \fullmoon & \fullmoon & \newmoon & \fullmoon & \fullmoon \\
\cite{402/lu2013recommender} &2013 & \fullmoon & \fullmoon & \newmoon & \fullmoon & \fullmoon & \fullmoon & \newmoon & \fullmoon & \newmoon & \fullmoon & \fullmoon & \fullmoon & \newmoon & \fullmoon & \fullmoon \\
\cite{454/diaby2013toward} &2013 & \newmoon & \fullmoon & \fullmoon & \newmoon & \fullmoon & \fullmoon & \fullmoon & \newmoon & \fullmoon & \fullmoon & \fullmoon & \fullmoon & \fullmoon & \fullmoon & \fullmoon \\
\cite{519/hong2013dynamic} &2013 & \newmoon & \fullmoon & \fullmoon & \fullmoon & \fullmoon & \fullmoon & \newmoon & \fullmoon & \fullmoon & \newmoon & \fullmoon & \fullmoon & \fullmoon & \fullmoon & \fullmoon \\
\cite{3/GuptaG14} &2014 & \newmoon & \fullmoon & \fullmoon & \fullmoon & \fullmoon & \fullmoon & \newmoon & \fullmoon & \fullmoon & \newmoon & \newmoon & \fullmoon & \fullmoon & \fullmoon & \fullmoon \\
\cite{27/DiabyVL14} &2014 & \newmoon & \fullmoon & \fullmoon & \fullmoon & \fullmoon & \fullmoon & \newmoon & \newmoon & \newmoon & \fullmoon & \fullmoon & \fullmoon & \fullmoon & \fullmoon & \fullmoon \\
\cite{29/DiabyV14} &2014 & \newmoon & \fullmoon & \fullmoon & \fullmoon & \fullmoon & \fullmoon & \newmoon & \newmoon & \fullmoon & \fullmoon & \fullmoon & \fullmoon & \fullmoon & \fullmoon & \fullmoon \\
\cite{110/malherbe2014field} &2014 & \newmoon & \fullmoon & \fullmoon & \newmoon & \fullmoon & \fullmoon & \fullmoon & \fullmoon & \newmoon & \fullmoon & \fullmoon & \fullmoon & \fullmoon & \fullmoon & \fullmoon \\
\cite{178/HeapKWBC14} &2014 & \newmoon & \fullmoon & \fullmoon & \newmoon & \fullmoon & \fullmoon & \fullmoon & \fullmoon & \fullmoon & \fullmoon & \fullmoon & \fullmoon & \fullmoon & \fullmoon & \fullmoon \\
\cite{211/FaliagkaIKRSTT14} &2014 & \fullmoon & \newmoon & \fullmoon & \fullmoon & \fullmoon & \fullmoon & \newmoon & \newmoon & \newmoon & \fullmoon & \fullmoon & \fullmoon & \fullmoon & \fullmoon & \fullmoon \\
\cite{518/almalis2014acontent} &2014 & \fullmoon & \newmoon & \fullmoon & \newmoon & \fullmoon & \fullmoon & \fullmoon & \fullmoon & \fullmoon & \fullmoon & \fullmoon & \fullmoon & \fullmoon & \fullmoon & \fullmoon \\
\cite{542/almalis2014content} &2014 & \fullmoon & \newmoon & \fullmoon & \newmoon & \fullmoon & \fullmoon & \fullmoon & \fullmoon & \fullmoon & \fullmoon & \fullmoon & \fullmoon & \fullmoon & \fullmoon & \fullmoon \\
\cite{46/AlmalisTKS15} &2015 & \fullmoon & \fullmoon & \newmoon & \newmoon & \fullmoon & \fullmoon & \fullmoon & \fullmoon & \fullmoon & \fullmoon & \fullmoon & \fullmoon & \newmoon & \fullmoon & \fullmoon \\
\cite{106/coelho2015hyred} &2015 & \fullmoon & \fullmoon & \newmoon & \fullmoon & \fullmoon & \fullmoon & \newmoon & \fullmoon & \newmoon & \fullmoon & \fullmoon & \fullmoon & \newmoon & \fullmoon & \fullmoon \\
\cite{174/HauffG15} &2015 & \fullmoon & \fullmoon & \newmoon & \fullmoon & \fullmoon & \newmoon & \fullmoon & \newmoon & \fullmoon & \fullmoon & \fullmoon & \fullmoon & \newmoon & \fullmoon & \fullmoon \\
\cite{331/salazar2015case} &2015 & \fullmoon & \fullmoon & \newmoon & \fullmoon & \fullmoon & \fullmoon & \newmoon & \fullmoon & \fullmoon & \fullmoon & \fullmoon & \fullmoon & \newmoon & \fullmoon & \fullmoon \\
\cite{332/chenni2015content} &2015 & \newmoon & \fullmoon & \fullmoon & \newmoon & \fullmoon & \fullmoon & \fullmoon & \fullmoon & \fullmoon & \fullmoon & \fullmoon & \fullmoon & \fullmoon & \fullmoon & \fullmoon \\
\cite{23/LiuSKZN16} &2016 & \newmoon & \fullmoon & \fullmoon & \fullmoon & \fullmoon & \fullmoon & \newmoon & \fullmoon & \fullmoon & \fullmoon & \newmoon & \fullmoon & \fullmoon & \fullmoon & \fullmoon \\
\cite{25/ZhangC16} &2016 & \newmoon & \fullmoon & \fullmoon & \fullmoon & \fullmoon & \fullmoon & \newmoon & \fullmoon & \newmoon & \newmoon & \fullmoon & \fullmoon & \fullmoon & \fullmoon & \fullmoon \\
\cite{90/LiuORSTX16} &2016 & \newmoon & \fullmoon & \fullmoon & \fullmoon & \fullmoon & \fullmoon & \newmoon & \fullmoon & \newmoon & \newmoon & \fullmoon & \fullmoon & \fullmoon & \fullmoon & \fullmoon \\
\cite{91/DomeniconiMPPP16} &2016 & \newmoon & \fullmoon & \fullmoon & \newmoon & \fullmoon & \fullmoon & \fullmoon & \fullmoon & \fullmoon & \fullmoon & \fullmoon & \fullmoon & \fullmoon & \fullmoon & \fullmoon \\
\cite{94/CarpiEKSCPQ16} &2016 & \newmoon & \fullmoon & \fullmoon & \fullmoon & \fullmoon & \fullmoon & \newmoon & \fullmoon & \fullmoon & \fullmoon & \fullmoon & \fullmoon & \fullmoon & \fullmoon & \fullmoon \\
\cite{95/LeksinO16} &2016 & \newmoon & \fullmoon & \fullmoon & \fullmoon & \fullmoon & \fullmoon & \newmoon & \fullmoon & \fullmoon & \fullmoon & \fullmoon & \fullmoon & \fullmoon & \fullmoon & \fullmoon \\
\cite{96/MishraR16} &2016 & \newmoon & \fullmoon & \fullmoon & \fullmoon & \fullmoon & \fullmoon & \newmoon & \fullmoon & \fullmoon & \fullmoon & \fullmoon & \fullmoon & \fullmoon & \fullmoon & \fullmoon \\
\cite{97/PacukSWWW16} &2016 & \newmoon & \fullmoon & \fullmoon & \fullmoon & \fullmoon & \fullmoon & \newmoon & \fullmoon & \fullmoon & \fullmoon & \fullmoon & \fullmoon & \fullmoon & \fullmoon & \fullmoon \\
\cite{98/PessemierVM16} &2016 & \newmoon & \fullmoon & \fullmoon & \fullmoon & \fullmoon & \fullmoon & \newmoon & \fullmoon & \fullmoon & \fullmoon & \fullmoon & \fullmoon & \fullmoon & \fullmoon & \fullmoon \\
\cite{99/PolatoA16} &2016 & \newmoon & \fullmoon & \fullmoon & \fullmoon & \fullmoon & \fullmoon & \newmoon & \fullmoon & \fullmoon & \fullmoon & \fullmoon & \fullmoon & \fullmoon & \fullmoon & \fullmoon \\
\cite{100/XiaoXLMW16} &2016 & \newmoon & \fullmoon & \fullmoon & \fullmoon & \fullmoon & \fullmoon & \newmoon & \fullmoon & \fullmoon & \fullmoon & \fullmoon & \fullmoon & \fullmoon & \fullmoon & \fullmoon \\
\cite{101/Zibriczky16} &2016 & \newmoon & \fullmoon & \fullmoon & \fullmoon & \fullmoon & \fullmoon & \newmoon & \fullmoon & \fullmoon & \fullmoon & \fullmoon & \fullmoon & \fullmoon & \fullmoon & \fullmoon \\
\cite{102/LeeHK16} &2016 & \newmoon & \fullmoon & \fullmoon & \fullmoon & \fullmoon & \fullmoon & \newmoon & \newmoon & \fullmoon & \newmoon & \fullmoon & \fullmoon & \fullmoon & \fullmoon & \fullmoon \\
\cite{164/GuoAH16} &2016 & \newmoon & \fullmoon & \fullmoon & \fullmoon & \fullmoon & \fullmoon & \newmoon & \newmoon & \newmoon & \fullmoon & \fullmoon & \fullmoon & \fullmoon & \fullmoon & \fullmoon \\
\cite{166/RaczSS16} &2016 & \fullmoon & \fullmoon & \newmoon & \fullmoon & \fullmoon & \fullmoon & \newmoon & \fullmoon & \fullmoon & \fullmoon & \fullmoon & \fullmoon & \newmoon & \fullmoon & \fullmoon \\
\cite{167/SchmittCS16} &2016 & \newmoon & \fullmoon & \fullmoon & \fullmoon & \fullmoon & \fullmoon & \newmoon & \newmoon & \fullmoon & \newmoon & \fullmoon & \fullmoon & \fullmoon & \fullmoon & \fullmoon \\
\cite{171/LinLAL16} &2016 & \newmoon & \fullmoon & \fullmoon & \fullmoon & \fullmoon & \fullmoon & \newmoon & \fullmoon & \fullmoon & \fullmoon & \fullmoon & \fullmoon & \fullmoon & \fullmoon & \fullmoon \\
\cite{533/borisyuk2016casmos} &2016 & \newmoon & \fullmoon & \fullmoon & \newmoon & \fullmoon & \fullmoon & \fullmoon & \fullmoon & \fullmoon & \fullmoon & \fullmoon & \fullmoon & \fullmoon & \fullmoon & \newmoon \\
\cite{540/zhang2016glmix} &2016 & \newmoon & \fullmoon & \fullmoon & \fullmoon & \fullmoon & \fullmoon & \newmoon & \fullmoon & \fullmoon & \fullmoon & \fullmoon & \fullmoon & \fullmoon & \fullmoon & \newmoon \\
\cite{21/Al-OtaibiY17} &2017 & \newmoon & \fullmoon & \fullmoon & \fullmoon & \fullmoon & \fullmoon & \newmoon & \fullmoon & \fullmoon & \fullmoon & \fullmoon & \fullmoon & \fullmoon & \fullmoon & \fullmoon \\
\cite{44/ShalabyAKPAQZ17} &2017 & \newmoon & \fullmoon & \fullmoon & \fullmoon & \fullmoon & \fullmoon & \newmoon & \newmoon & \fullmoon & \newmoon & \fullmoon & \fullmoon & \fullmoon & \fullmoon & \newmoon \\
\cite{76/ReusensLBS17} &2017 & \newmoon & \fullmoon & \fullmoon & \fullmoon & \newmoon & \fullmoon & \fullmoon & \fullmoon & \fullmoon & \fullmoon & \fullmoon & \fullmoon & \fullmoon & \fullmoon & \fullmoon \\
\cite{77/LiuROX17} &2017 & \newmoon & \fullmoon & \fullmoon & \fullmoon & \fullmoon & \fullmoon & \newmoon & \fullmoon & \newmoon & \newmoon & \fullmoon & \fullmoon & \fullmoon & \fullmoon & \fullmoon \\
\cite{78/yang2017combining} &2017 & \newmoon & \fullmoon & \fullmoon & \fullmoon & \fullmoon & \fullmoon & \newmoon & \fullmoon & \fullmoon & \newmoon & \fullmoon & \fullmoon & \newmoon & \fullmoon & \fullmoon \\
\cite{84/BansalSA17} &2017 & \newmoon & \fullmoon & \fullmoon & \newmoon & \fullmoon & \fullmoon & \fullmoon & \newmoon & \fullmoon & \fullmoon & \fullmoon & \fullmoon & \fullmoon & \fullmoon & \fullmoon \\
\cite{88/kenthapadi2017personalized} &2017 & \newmoon & \fullmoon & \fullmoon & \newmoon & \fullmoon & \fullmoon & \fullmoon & \fullmoon & \fullmoon & \fullmoon & \fullmoon & \fullmoon & \fullmoon & \fullmoon & \fullmoon \\
\cite{156/arita2017gber} &2017 & \fullmoon & \fullmoon & \newmoon & \newmoon & \fullmoon & \fullmoon & \fullmoon & \fullmoon & \fullmoon & \fullmoon & \fullmoon & \fullmoon & \newmoon & \fullmoon & \fullmoon \\
\cite{159/ChalaF17} &2017 & \fullmoon & \fullmoon & \newmoon & \newmoon & \fullmoon & \fullmoon & \fullmoon & \fullmoon & \newmoon & \fullmoon & \fullmoon & \fullmoon & \newmoon & \fullmoon & \fullmoon \\
\cite{160/SchmittGCS17} &2017 & \newmoon & \fullmoon & \fullmoon & \fullmoon & \fullmoon & \fullmoon & \newmoon & \fullmoon & \fullmoon & \newmoon & \fullmoon & \fullmoon & \fullmoon & \fullmoon & \fullmoon \\
\cite{509/DongLZBL17} &2017 & \newmoon & \fullmoon & \fullmoon & \fullmoon & \fullmoon & \fullmoon & \newmoon & \newmoon & \fullmoon & \fullmoon & \fullmoon & \fullmoon & \fullmoon & \fullmoon & \newmoon \\
\cite{516/yagci2017aranker} &2017 & \fullmoon & \fullmoon & \newmoon & \fullmoon & \fullmoon & \fullmoon & \newmoon & \fullmoon & \fullmoon & \newmoon & \fullmoon & \fullmoon & \newmoon & \fullmoon & \fullmoon \\
\cite{520/sato2017explor} &2017 & \fullmoon & \fullmoon & \newmoon & \newmoon & \fullmoon & \fullmoon & \fullmoon & \fullmoon & \fullmoon & \newmoon & \fullmoon & \fullmoon & \newmoon & \fullmoon & \fullmoon \\
\cite{522/patel2017capar} &2017 & \newmoon & \fullmoon & \fullmoon & \fullmoon & \fullmoon & \fullmoon & \newmoon & \fullmoon & \fullmoon & \fullmoon & \fullmoon & \fullmoon & \fullmoon & \fullmoon & \fullmoon \\
\cite{535/guo2017howinte} &2017 & \fullmoon & \fullmoon & \newmoon & \newmoon & \fullmoon & \fullmoon & \fullmoon & \fullmoon & \fullmoon & \newmoon & \fullmoon & \fullmoon & \newmoon & \fullmoon & \fullmoon \\
\cite{537/lian2017practical} &2017 & \fullmoon & \fullmoon & \newmoon & \newmoon & \fullmoon & \fullmoon & \fullmoon & \fullmoon & \fullmoon & \newmoon & \fullmoon & \fullmoon & \newmoon & \fullmoon & \fullmoon \\
\cite{538/bianchi2017content} &2017 & \fullmoon & \fullmoon & \newmoon & \fullmoon & \fullmoon & \fullmoon & \newmoon & \fullmoon & \fullmoon & \newmoon & \fullmoon & \fullmoon & \newmoon & \fullmoon & \fullmoon \\
\cite{539/volkovs2017content} &2017 & \fullmoon & \fullmoon & \newmoon & \fullmoon & \fullmoon & \fullmoon & \newmoon & \fullmoon & \newmoon & \newmoon & \fullmoon & \fullmoon & \newmoon & \fullmoon & \fullmoon \\
\cite{541/borisyuk2017lijar} &2017 & \newmoon & \fullmoon & \fullmoon & \fullmoon & \fullmoon & \fullmoon & \fullmoon & \fullmoon & \fullmoon & \fullmoon & \fullmoon & \fullmoon & \newmoon & \fullmoon & \fullmoon \\
\cite{18/HeggoA18} &2018 & \newmoon & \fullmoon & \fullmoon & \fullmoon & \fullmoon & \fullmoon & \newmoon & \fullmoon & \fullmoon & \fullmoon & \fullmoon & \fullmoon & \fullmoon & \fullmoon & \fullmoon \\
\cite{19/ElsaftyRB18} &2018 & \newmoon & \fullmoon & \fullmoon & \newmoon & \fullmoon & \fullmoon & \fullmoon & \newmoon & \fullmoon & \fullmoon & \fullmoon & \fullmoon & \fullmoon & \fullmoon & \fullmoon \\
\cite{20/Gonzalez-Briones18} &2018 & \newmoon & \fullmoon & \fullmoon & \fullmoon & \fullmoon & \fullmoon & \newmoon & \fullmoon & \fullmoon & \fullmoon & \fullmoon & \fullmoon & \fullmoon & \fullmoon & \fullmoon \\
\cite{41/Valverde-Rebaza18} &2018 & \newmoon & \fullmoon & \fullmoon & \newmoon & \fullmoon & \fullmoon & \fullmoon & \newmoon & \fullmoon & \fullmoon & \fullmoon & \fullmoon & \fullmoon & \fullmoon & \fullmoon \\
\cite{74/dave2018combined} &2018 & \newmoon & \fullmoon & \fullmoon & \fullmoon & \fullmoon & \fullmoon & \newmoon & \fullmoon & \fullmoon & \fullmoon & \fullmoon & \fullmoon & \fullmoon & \fullmoon & \fullmoon \\
\cite{75/martinez2018recommendation} &2018 & \newmoon & \fullmoon & \fullmoon & \newmoon & \fullmoon & \fullmoon & \fullmoon & \fullmoon & \fullmoon & \fullmoon & \fullmoon & \newmoon & \fullmoon & \fullmoon & \fullmoon \\
\cite{150/xu2018matching} &2018 & \fullmoon & \newmoon & \fullmoon & \fullmoon & \fullmoon & \fullmoon & \newmoon & \newmoon & \fullmoon & \fullmoon & \fullmoon & \fullmoon & \fullmoon & \fullmoon & \fullmoon \\
\cite{153/MaheshwaryM18} &2018 & \fullmoon & \fullmoon & \newmoon & \newmoon & \fullmoon & \fullmoon & \fullmoon & \fullmoon & \fullmoon & \fullmoon & \fullmoon & \fullmoon & \newmoon & \fullmoon & \fullmoon \\
\cite{469/zhu2018person} &2018 & \fullmoon & \fullmoon & \newmoon & \newmoon & \fullmoon & \fullmoon & \fullmoon & \fullmoon & \fullmoon & \fullmoon & \fullmoon & \newmoon & \newmoon & \fullmoon & \fullmoon \\
\cite{485/qin2018enhancing} &2018 & \fullmoon & \fullmoon & \newmoon & \newmoon & \fullmoon & \fullmoon & \fullmoon & \fullmoon & \fullmoon & \fullmoon & \fullmoon & \newmoon & \newmoon & \fullmoon & \fullmoon \\
\cite{506/ChenZDGHWW18} &2018 & \newmoon & \newmoon & \fullmoon & \fullmoon & \fullmoon & \fullmoon & \newmoon & \newmoon & \fullmoon & \newmoon & \fullmoon & \fullmoon & \fullmoon & \fullmoon & \newmoon \\
\cite{9/RivasCGCC19} &2019 & \newmoon & \fullmoon & \fullmoon & \fullmoon & \fullmoon & \fullmoon & \newmoon & \fullmoon & \fullmoon & \fullmoon & \fullmoon & \fullmoon & \fullmoon & \fullmoon & \fullmoon \\
\cite{10/ShishehchiB19} &2019 & \newmoon & \fullmoon & \fullmoon & \fullmoon & \fullmoon & \newmoon & \fullmoon & \newmoon & \fullmoon & \fullmoon & \fullmoon & \fullmoon & \fullmoon & \fullmoon & \fullmoon \\
\cite{16/MughaidOHAA19} &2019 & \newmoon & \fullmoon & \fullmoon & \newmoon & \fullmoon & \fullmoon & \fullmoon & \newmoon & \fullmoon & \fullmoon & \fullmoon & \fullmoon & \fullmoon & \fullmoon & \fullmoon \\
\cite{38/NigamRSW19} &2019 & \newmoon & \fullmoon & \fullmoon & \fullmoon & \fullmoon & \fullmoon & \newmoon & \fullmoon & \fullmoon & \newmoon & \newmoon & \fullmoon & \fullmoon & \fullmoon & \fullmoon \\
\cite{39/GutierrezCCHGV19} &2019 & \newmoon & \fullmoon & \fullmoon & \fullmoon & \fullmoon & \newmoon & \fullmoon & \fullmoon & \fullmoon & \fullmoon & \newmoon & \newmoon & \fullmoon & \fullmoon & \fullmoon \\
\cite{63/jiang2019user} &2019 & \newmoon & \fullmoon & \fullmoon & \fullmoon & \fullmoon & \fullmoon & \newmoon & \fullmoon & \fullmoon & \fullmoon & \fullmoon & \fullmoon & \fullmoon & \fullmoon & \fullmoon \\
\cite{64/chen2019correcting} &2019 & \newmoon & \fullmoon & \fullmoon & \fullmoon & \newmoon & \fullmoon & \fullmoon & \fullmoon & \fullmoon & \fullmoon & \fullmoon & \fullmoon & \fullmoon & \newmoon & \fullmoon \\
\cite{65/liu2019tripartite} &2019 & \fullmoon & \fullmoon & \newmoon & \fullmoon & \fullmoon & \fullmoon & \newmoon & \fullmoon & \fullmoon & \fullmoon & \fullmoon & \fullmoon & \newmoon & \fullmoon & \fullmoon \\
\cite{67/lacic2019should} &2019 & \newmoon & \fullmoon & \fullmoon & \newmoon & \fullmoon & \fullmoon & \fullmoon & \fullmoon & \fullmoon & \fullmoon & \fullmoon & \fullmoon & \fullmoon & \fullmoon & \fullmoon \\
\cite{117/RodriguezC19} &2019 & \fullmoon & \newmoon & \fullmoon & \newmoon & \fullmoon & \fullmoon & \fullmoon & \newmoon & \newmoon & \fullmoon & \fullmoon & \fullmoon & \fullmoon & \fullmoon & \fullmoon \\
\cite{142/Fernandez-Reyes19} &2019 & \fullmoon & \newmoon & \fullmoon & \newmoon & \fullmoon & \fullmoon & \fullmoon & \fullmoon & \fullmoon & \fullmoon & \fullmoon & \fullmoon & \fullmoon & \fullmoon & \fullmoon \\
\cite{145/YanLSZZ019} &2019 & \fullmoon & \fullmoon & \newmoon & \fullmoon & \fullmoon & \fullmoon & \newmoon & \fullmoon & \fullmoon & \fullmoon & \fullmoon & \fullmoon & \newmoon & \fullmoon & \fullmoon \\
\cite{288/MareeKB19} &2019 & \fullmoon & \fullmoon & \newmoon & \fullmoon & \fullmoon & \newmoon & \fullmoon & \newmoon & \fullmoon & \fullmoon & \fullmoon & \fullmoon & \newmoon & \fullmoon & \fullmoon \\
\cite{483/le2019towards} &2019 & \fullmoon & \fullmoon & \newmoon & \fullmoon & \fullmoon & \fullmoon & \newmoon & \fullmoon & \fullmoon & \fullmoon & \fullmoon & \newmoon & \newmoon & \fullmoon & \fullmoon \\
\cite{484/bian2019domain} &2019 & \fullmoon & \fullmoon & \newmoon & \newmoon & \fullmoon & \fullmoon & \fullmoon & \fullmoon & \fullmoon & \fullmoon & \fullmoon & \fullmoon & \newmoon & \fullmoon & \fullmoon \\
\cite{523/luo2019resumegan} &2019 & \fullmoon & \fullmoon & \fullmoon & \newmoon & \fullmoon & \fullmoon & \fullmoon & \fullmoon & \newmoon & \fullmoon & \fullmoon & \fullmoon & \newmoon & \fullmoon & \fullmoon \\
\cite{546/geyik2019fairness} &2019 & \fullmoon & \newmoon & \fullmoon & \fullmoon & \fullmoon & \fullmoon & \fullmoon & \fullmoon & \fullmoon & \fullmoon & \fullmoon & \fullmoon & \fullmoon & \newmoon & \fullmoon \\
\cite{37/MhamdiMGAM20} &2020 & \newmoon & \fullmoon & \fullmoon & \fullmoon & \fullmoon & \fullmoon & \newmoon & \fullmoon & \fullmoon & \fullmoon & \fullmoon & \fullmoon & \fullmoon & \fullmoon & \newmoon \\
\cite{56/LacicRKDCL20} &2020 & \newmoon & \fullmoon & \fullmoon & \fullmoon & \fullmoon & \fullmoon & \newmoon & \fullmoon & \fullmoon & \fullmoon & \fullmoon & \fullmoon & \fullmoon & \fullmoon & \fullmoon \\
\cite{57/GugnaniM20} &2020 & \newmoon & \fullmoon & \fullmoon & \newmoon & \fullmoon & \fullmoon & \fullmoon & \newmoon & \fullmoon & \fullmoon & \fullmoon & \fullmoon & \fullmoon & \fullmoon & \fullmoon \\
\cite{58/bellini2020guapp} &2020 & \newmoon & \fullmoon & \fullmoon & \newmoon & \fullmoon & \fullmoon & \fullmoon & \newmoon & \fullmoon & \fullmoon & \newmoon & \fullmoon & \fullmoon & \fullmoon & \fullmoon \\
\cite{60/wang2020session} &2020 & \newmoon & \fullmoon & \fullmoon & \fullmoon & \fullmoon & \fullmoon & \newmoon & \fullmoon & \fullmoon & \fullmoon & \fullmoon & \fullmoon & \fullmoon & \fullmoon & \fullmoon \\
\cite{137/khatua2020matching} &2020 & \fullmoon & \fullmoon & \newmoon & \newmoon & \fullmoon & \fullmoon & \fullmoon & \newmoon & \fullmoon & \fullmoon & \fullmoon & \fullmoon & \newmoon & \fullmoon & \fullmoon \\
\cite{138/LiFTPHC20} &2020 & \fullmoon & \fullmoon & \newmoon & \newmoon & \fullmoon & \fullmoon & \fullmoon & \fullmoon & \fullmoon & \fullmoon & \fullmoon & \fullmoon & \newmoon & \fullmoon & \fullmoon \\
\cite{273/BituinAE20} &2020 & \fullmoon & \newmoon & \fullmoon & \newmoon & \fullmoon & \fullmoon & \fullmoon & \fullmoon & \fullmoon & \fullmoon & \fullmoon & \fullmoon & \fullmoon & \fullmoon & \fullmoon \\
\cite{277/freire2020framework} &2020 & \fullmoon & \fullmoon & \newmoon & \fullmoon & \fullmoon & \fullmoon & \newmoon & \fullmoon & \fullmoon & \fullmoon & \fullmoon & \fullmoon & \newmoon & \fullmoon & \fullmoon \\
\cite{478/qin2020enhanced} &2020 & \fullmoon & \fullmoon & \newmoon & \fullmoon & \fullmoon & \fullmoon & \newmoon & \fullmoon & \fullmoon & \fullmoon & \fullmoon & \newmoon & \newmoon & \fullmoon & \fullmoon \\
\cite{479/jiang2020learning} &2020 & \fullmoon & \fullmoon & \newmoon & \newmoon & \fullmoon & \fullmoon & \fullmoon & \fullmoon & \fullmoon & \fullmoon & \fullmoon & \fullmoon & \newmoon & \fullmoon & \fullmoon \\
\cite{501/boukari2020huntalent} &2020 & \fullmoon & \newmoon & \fullmoon & \newmoon & \fullmoon & \fullmoon & \fullmoon & \newmoon & \fullmoon & \fullmoon & \fullmoon & \fullmoon & \fullmoon & \fullmoon & \newmoon \\
\cite{521/roy2020machine} &2020 & \fullmoon & \newmoon & \fullmoon & \newmoon & \fullmoon & \fullmoon & \fullmoon & \newmoon & \fullmoon & \fullmoon & \fullmoon & \fullmoon & \fullmoon & \fullmoon & \fullmoon \\
\cite{524/shi2020salience} &2020 & \newmoon & \fullmoon & \fullmoon & \newmoon & \fullmoon & \fullmoon & \fullmoon & \newmoon & \fullmoon & \fullmoon & \fullmoon & \fullmoon & \fullmoon & \fullmoon & \fullmoon \\
\cite{527/bian2020learning} &2020 & \fullmoon & \fullmoon & \newmoon & \fullmoon & \fullmoon & \fullmoon & \newmoon & \fullmoon & \fullmoon & \fullmoon & \fullmoon & \fullmoon & \newmoon & \fullmoon & \fullmoon \\
\cite{48/FengJWYL21} &2021 & \newmoon & \fullmoon & \fullmoon & \fullmoon & \fullmoon & \fullmoon & \newmoon & \fullmoon & \fullmoon & \fullmoon & \fullmoon & \fullmoon & \fullmoon & \fullmoon & \fullmoon \\
\cite{49/BiancofioreNSNP21} &2021 & \newmoon & \fullmoon & \fullmoon & \fullmoon & \fullmoon & \fullmoon & \newmoon & \fullmoon & \fullmoon & \fullmoon & \newmoon & \fullmoon & \fullmoon & \fullmoon & \fullmoon \\
\cite{51/BiancofioreNSNP21} &2021 & \newmoon & \fullmoon & \fullmoon & \fullmoon & \fullmoon & \fullmoon & \newmoon & \fullmoon & \fullmoon & \fullmoon & \newmoon & \fullmoon & \fullmoon & \fullmoon & \fullmoon \\
\cite{52/Mentec0HR21} &2021 & \fullmoon & \newmoon & \fullmoon & \fullmoon & \fullmoon & \fullmoon & \newmoon & \newmoon & \fullmoon & \fullmoon & \fullmoon & \newmoon & \fullmoon & \fullmoon & \fullmoon \\
\cite{53/UpadhyayA0F21} &2021 & \newmoon & \fullmoon & \fullmoon & \fullmoon & \fullmoon & \fullmoon & \newmoon & \fullmoon & \fullmoon & \fullmoon & \fullmoon & \newmoon & \fullmoon & \fullmoon & \fullmoon \\
\cite{114/zhao2021summer} &2021 & \newmoon & \fullmoon & \fullmoon & \newmoon & \fullmoon & \fullmoon & \fullmoon & \fullmoon & \fullmoon & \fullmoon & \fullmoon & \fullmoon & \fullmoon & \fullmoon & \fullmoon \\
\cite{115/bills2021looking} &2021 & \fullmoon & \fullmoon & \newmoon & \fullmoon & \fullmoon & \fullmoon & \newmoon & \fullmoon & \fullmoon & \fullmoon & \newmoon & \fullmoon & \newmoon & \fullmoon & \fullmoon \\
\cite{116/lavi2021consultantbert} &2021 & \fullmoon & \fullmoon & \newmoon & \newmoon & \fullmoon & \fullmoon & \fullmoon & \newmoon & \fullmoon & \fullmoon & \fullmoon & \fullmoon & \newmoon & \fullmoon & \fullmoon \\
\cite{124/WangJP21} &2021 & \fullmoon & \fullmoon & \newmoon & \fullmoon & \fullmoon & \fullmoon & \newmoon & \fullmoon & \fullmoon & \fullmoon & \fullmoon & \fullmoon & \newmoon & \fullmoon & \fullmoon \\
\cite{126/habous2021fuzzy} &2021 & \fullmoon & \fullmoon & \newmoon & \fullmoon & \fullmoon & \fullmoon & \newmoon & \newmoon & \fullmoon & \fullmoon & \fullmoon & \fullmoon & \newmoon & \fullmoon & \fullmoon \\
\cite{129/elgammal2021matching} &2021 & \fullmoon & \fullmoon & \newmoon & \newmoon & \fullmoon & \fullmoon & \fullmoon & \fullmoon & \newmoon & \fullmoon & \fullmoon & \fullmoon & \newmoon & \fullmoon & \fullmoon \\
\cite{130/smith2021skill} &2021 & \fullmoon & \fullmoon & \newmoon & \newmoon & \fullmoon & \fullmoon & \fullmoon & \newmoon & \fullmoon & \fullmoon & \fullmoon & \fullmoon & \newmoon & \fullmoon & \fullmoon \\
\cite{131/fu2021beyond} &2021 & \fullmoon & \fullmoon & \newmoon & \fullmoon & \fullmoon & \fullmoon & \newmoon & \fullmoon & \newmoon & \fullmoon & \newmoon & \fullmoon & \newmoon & \fullmoon & \fullmoon \\
\cite{132/zhao2021embedding} &2021 & \fullmoon & \fullmoon & \newmoon & \fullmoon & \fullmoon & \fullmoon & \newmoon & \fullmoon & \newmoon & \fullmoon & \fullmoon & \fullmoon & \newmoon & \fullmoon & \newmoon \\
\cite{475/zhang2021explainable} &2021 & \fullmoon & \fullmoon & \newmoon & \newmoon & \fullmoon & \fullmoon & \fullmoon & \fullmoon & \fullmoon & \fullmoon & \fullmoon & \newmoon & \newmoon & \fullmoon & \fullmoon \\
\cite{477/menacer2021interpretable} &2021 & \fullmoon & \fullmoon & \newmoon & \newmoon & \fullmoon & \fullmoon & \fullmoon & \newmoon & \fullmoon & \fullmoon & \fullmoon & \newmoon & \newmoon & \fullmoon & \fullmoon \\
\cite{495/slama2021novel} &2021 & \fullmoon & \fullmoon & \newmoon & \newmoon & \fullmoon & \fullmoon & \fullmoon & \fullmoon & \fullmoon & \fullmoon & \newmoon & \fullmoon & \fullmoon & \fullmoon & \fullmoon \\
\cite{517/apaza2021job} &2021 & \newmoon & \fullmoon & \fullmoon & \newmoon & \fullmoon & \fullmoon & \fullmoon & \newmoon & \fullmoon & \fullmoon & \fullmoon & \fullmoon & \fullmoon & \fullmoon & \fullmoon \\
\cite{525/he2021self} &2021 & \fullmoon & \fullmoon & \newmoon & \newmoon & \fullmoon & \fullmoon & \fullmoon & \fullmoon & \newmoon & \fullmoon & \fullmoon & \fullmoon & \newmoon & \fullmoon & \fullmoon \\
\cite{526/he2021finn} &2021 & \fullmoon & \fullmoon & \newmoon & \newmoon & \fullmoon & \fullmoon & \fullmoon & \fullmoon & \newmoon & \fullmoon & \fullmoon & \fullmoon & \newmoon & \fullmoon & \fullmoon \\
\cite{528/cardoso2021matching} &2021 & \fullmoon & \fullmoon & \newmoon & \fullmoon & \fullmoon & \fullmoon & \fullmoon & \fullmoon & \fullmoon & \fullmoon & \fullmoon & \fullmoon & \newmoon & \fullmoon & \fullmoon \\
\cite{531/islam2021debias} &2021 & \newmoon & \fullmoon & \fullmoon & \fullmoon & \newmoon & \fullmoon & \fullmoon & \fullmoon & \newmoon & \fullmoon & \fullmoon & \fullmoon & \fullmoon & \newmoon & \fullmoon \\

\hline
  \end{longtable}
\end{small}
\end{center}

\end{document}